\documentclass[a4paper,11pt]{article}

\usepackage[utf8]{inputenc}
\usepackage{fullpage}
\usepackage{amsmath, amssymb}
\usepackage{mathrsfs}
\usepackage{cite}
\usepackage{float}
\usepackage{color}
\usepackage{xcolor}
\usepackage{multirow}
\usepackage{graphicx,caption,subcaption}
\usepackage{bm}
\usepackage{braket}
\usepackage{dsfont}
\usepackage{cancel}
\usepackage{latexsym}
\usepackage{slashed}
\usepackage[space]{grffile}
\usepackage{tikz}

\definecolor{darkblue}{rgb}{0.1,0.1,.8}
\usepackage[breaklinks=true,linktocpage=true,
colorlinks=true,urlcolor=darkblue,linkcolor=darkblue,
citecolor=darkblue,pdfpagelabels=true,hypertexnames=true,
plainpages=false,naturalnames=false,]{hyperref}

\numberwithin{equation}{section}

\newcommand{\ga}{{\gamma}}
\newcommand{\de}{{\delta}}
\newcommand{\ep}{{\epsilon}}

\newcommand{\lm}{{\lambda}}

\newcommand{\te}{{\theta}}
\newcommand{\zt}{{\zeta}}
\newcommand{\vep}{{\varepsilon}}
\newcommand{\vphi}{{\varphi}}

\newcommand{\De}{{\Delta}}

\newcommand{\cA}{{\mathcal{A}}}
\newcommand{\cD}{{\mathcal{D}}}
\newcommand{\cH}{{\mathcal{H}}}
\newcommand{\cL}{{\mathcal{L}}}

\newcommand{\pd}{{\partial}}

\newcommand{\nn}{{\nonumber}}
\newcommand{\Pl}{{\mathrm{Pl}}}

\newcommand{\ol}[1]{{\overline{#1}}}

\newcommand{\bbra}{{\langle\kern-2.5pt\langle}}
\newcommand{\kket}{{\rangle\kern-2.5pt\rangle}}
\newcommand{\Bbra}{{\Big\langle\kern-4.5pt\Big\langle}}
\newcommand{\Kket}{{\Big\rangle\kern-4.5pt\Big\rangle}}

\begin{document}

\renewcommand{\baselinestretch}{1.2}
\thispagestyle{empty}

\mbox{}

\vskip3.0cm
\begin{center}

{\LARGE\bf New leading contributions to non-gaussianity in single field inflation}

\vglue.3in

\hspace{-2pt}Ignatios Antoniadis,${}^{a,b,}$\footnote{antoniad@lpthe.jussieu.fr} Jules Cunat,${}^{b,}$\footnote{jcunat@lpthe.jussieu.fr} Auttakit Chatrabhuti,${}^{a,}$\footnote{auttakit.c@chula.ac.th} Hiroshi Isono${}^{a,}$\footnote{hiroshi.i@chula.ac.th}
\vglue.1in

${}^a$~{\it Department of Physics, Faculty of Science, Chulalongkorn University,\\ Phayathai Road, Pathumwan, Bangkok 10330, Thailand}\\
${}^b$~{\it Laboratoire de Physique Th\'eorique et Hautes Energies (LPTHE), Sorbonne Universit\'e,\\ CNRS, 4 Place Jussieu, 75005 Paris, France}
\vglue.1in

\end{center}



\vglue.3in

\begin{center}
{\Large\bf Abstract}
\vglue.2in
\end{center}

We compute the bispectrum of primordial density perturbations in CMB to second order in the slow-roll parameters of single field inflation. We correct previous results and found that next-to-leading order corrections can be of the same order as the leading order result in a large class of models, including hilltop inflation.

\newpage

\tableofcontents

\setcounter{footnote}{0}

\section{Introduction}

The next generation of cosmological observations is expected to increase precision in measurements of cosmological observables, providing more accurate experimental tests of theoretical models of inflation. Besides the power spectrum of primordial gravity waves, another important observable is the bispectrum, associated to the three-point function, related to a longstanding open question which is the statistics of primordial perturbations. Although non-gaussianity is expected to be suppressed in slow-roll single field inflation, it may be enhanced in other models and its possible measurement would be a milestone on the search of the underlying theory. 
The computation of the bispectrum in single field inflation was done to lowest order in slow-roll approximation~\cite{Maldacena:2002vr,Acquaviva:2002ud} and is shown to be dominated by the local shape, corresponding to a constant parameter $f_{\text{NL}}\simeq 2\times 10^{-2}$, that can be compared with the PLANCK2018 measurement of $f_{\text{NL}}^{\text{exp}}\sim -0.9\pm 5.1$~\cite{Planck:2019kim}.

Next to leading order (NLO) corrections are expected to be suppressed by two slow-roll parameters to at least $10^{-3}$ level. On the other hand, it is known that perturbation theory of massless minimally coupled fields in de Sitter (dS) space is plagued with infrared divergences due to the logarithmic growth of the propagator at large distances compared to the dS radius $H^{-1}$~\cite{Ford:1977in,Ratra:1984yq,Antoniadis:1985pj}. Such logarithm could typically be as large as the number of e-folds, so that for an inflation period of about 60 e-folds, one can estimate that a typical logarithmic correction could cancel the suppression of one slow-roll parameter.
Thus, if logarithmic divergences survive, NLO corrections to non-gaussianity can be as important as the tree-level result. Even if the question may appear academic at present given the large experimental uncertainty in the measurement, it is still important both theoretically and for future observations, in particular when comparing the predictions of various models of inflation to those of `standard' single field.

In this work, we establish the above claim by performing a complete computation of the bispectrum at the second order in slow-roll parameters. We use the path integral formalism in the usual gauge where the inflaton is fixed to its background value. We find that the dominant contribution is almost local, having a logarithmic dependence on the momenta contributing 
to all shapes of non-gaussianity but predominantly to the squeezed type.

Important properties of the bispectrum that should be verified are the consistency relation between the bispectrum in the squeezed limit and the power spectrum~\cite{Maldacena:2002vr,Creminelli:2004yq} and the time independence of the bispectrum as expected from the adiabaticity of the curvature fluctuation~\cite{Weinberg:2003sw,Lyth:2004gb,Langlois:2005qp}. In a separate companion paper~\cite{Antoniadis:2025jnb}, we check that our bispectrum up to NLO is indeed independent of the time where the three-point function is defined, and also satisfies the consistency relation in the squeezed limit. 
Moreover, after completing the first version of this paper, we noticed the paper~\cite{Burrage:2011hd}, which has computed the bispectrum at NLO in `general single field inflation' by using a different set of cubic vertices. We checked that their cubic action is equivalent to ours up to boundary terms, while their bispectrum in the equilateral limit differs but it depends on time.

The outline of our paper is the following. In Section 2, we expand the action to third order and give the list of all 3-point vertices/operators, as well as the propagators. In Section 3, we compute the  3-point function amplitude as a sum of nine individual contributions. One of them corresponds to an action term which is proportional to the equations of motion but does not vanish when acting on Green functions but produces delta functions (subsection 3.3); we show that this contribution reproduces indeed the contribution one gets from the shift of the gauge invariant perturbation in the Hamiltonian approach~\cite{Maldacena:2002vr}. The total sum is given in subsection 3.4. In Section 4, we use the 3-point amplitude to compute the non-gaussianity of the bispectrum that can be parametrised in terms of a momentum-dependent $f_\text{NL}$. In Section 5, we do a numerical estimate of our result, while Section 6 contains a concluding summary and outlook. Finally, there are five appendices containing technical details on the derivation of the action (Appendix A), the description of the Schwinger-Keldysh path integral formalism (Appendix B), the various expansions in terms of the Hubble flow parameters (Appendix C), the Wightman functions (Appendix D) and the quantitative study of the $f_{\text{NL}}$ parameter (Appendix E).

Before proceeding, for self-containedness, we summarise below the main framework of single field slow-roll inflation and our conventions. Although this paper is intended to be self-contained, a careful reader may also refer to \cite{Collins:2011mz} for a complete review of the non-gaussianity at leading order (LO) in slow-roll parameters, originally computed in \cite{Maldacena:2002vr}.

We study the case of inflation driven by a single scalar field $\phi$ with potential $V(\phi)$. Thus, the action is
\begin{align}\label{inf-action}
    S=\int d^4x\sqrt{-g}\left\{\frac{M_\Pl^2}{2}R(g)-\frac{1}{2}g^{\mu\nu}\partial_\mu\phi\partial_\nu\phi-V(\phi)\right\},
\end{align}
where $M_\Pl=2.2\times10^{18}$\,GeV is the reduced Planck mass, $R(g)$ is the scalar curvature, and the metric $g_{\mu\nu}$ is defined via the line element:
\begin{align}\label{bgmetric}
    ds^{2}=-dt^2+e^{2\rho(t)}\delta_{ij}dx^idx^j\,.
\end{align}
We define the Hubble scale as 
\begin{align}
    H(t)\equiv\dot{\rho}(t)=\frac{d\rho}{dt}.
\end{align}
The time dependent background dynamics is a solution of the equations of motion:
\begin{align}
    3M_\Pl^2H^2&=\frac{1}{2}\dot{\phi}^2+V\,,\label{Friedman1}\\
    M_\Pl^2\dot{H}&=-\frac{1}{2}\dot{\phi}^2\,,\label{Friedman2}\\
    \ddot{\phi}+3H\dot{\phi}&=-\frac{dV}{d\phi}\,.
\end{align}
We will use the Hubble-flow functions defined around the metric of de Sitter space as~\cite{Schwarz:2001vv}
\begin{align}\label{varepsilons}
    \varepsilon_1\equiv&-\frac{\dot{H}}{H^2}=\frac{\dot{\phi}^2}{2H^2},\qquad\text{and}\qquad\varepsilon_{n+1}\equiv\frac{\dot{\varepsilon}_n}{H\varepsilon_n}\,.
\end{align}
They are linked to the more familiar slow-roll parameters defined through the potential:
\begin{align}
    \epsilon_V&\equiv\frac{M_{\mathrm{Pl}}^2}{2V^{2}}\left(\frac{dV}{d\phi}\right)^{2}=\varepsilon_1\left(1+\frac{\varepsilon_2}{6-2\varepsilon_1}\right)^{2}=\varepsilon_1\left(1+{\frac{\varepsilon_2}{3}}\right)+\mathcal{O}\big(\varepsilon^3\big)\,, \label{epsilonV}\\
    \eta_V&\equiv\frac{M_{\mathrm{Pl}}^2}{V}\frac{d^{2}V}{d\phi^{2}}=\frac{12\varepsilon_1-3\varepsilon_2-4\varepsilon_1^2-\frac{\varepsilon_2^2}{2}+5\varepsilon_1\varepsilon_2-\varepsilon_2\varepsilon_3}{6-2\varepsilon_1} \nn\\
    &\qquad\qquad\quad~=2\varepsilon_1-\frac{\varepsilon_2}{2}
 +\varepsilon_2\left(\frac{2}{3}\varepsilon_1-\frac{\varepsilon_2}{12}-\frac{\varepsilon_3}{6}\right)
  +\mathcal{O}\big(\varepsilon^3\big)\,, \label{etaV}
\end{align}
where we used $V=M_\Pl^2H^2(3-\varepsilon_1)$ by combining \eqref{Friedman1}, \eqref{Friedman2}, \eqref{varepsilons}, and 
have expanded to second order in Hubble flow functions $\varepsilon_i$, as needed below for our computation.
We also define the next order slow-roll parameter corresponding to the third derivative of the potential\footnote{Our definition of $\hat\xi_V$ is related to the parameter $\xi_V$ in~\cite{Planck:2018jri} by $\sqrt{2\ep_V}\hat\xi_V \equiv \xi^2_V$.}:
\begin{align} 
\hat\xi_V\equiv\frac{M_{\mathrm{Pl}}^3}{V}\frac{d^{3}V}{d\phi^{3}}
&=\frac{\sqrt{\ep_1}}{3-\vep_1}\left(
-6\sqrt{2}\ep_1+2\sqrt{2}\ep_1^2+\frac{9\ep_2}{\sqrt{2}}-\frac{9\ep_1\ep_2}{\sqrt{2}}+\frac{3\ep_2^2}{\sqrt{2}}\right. \nn\\
&\qquad\qquad\quad
\left.-\frac{3\ep_2\ep_3}{2\sqrt{2}\ep_1}+\frac{7\ep_2\ep_3}{2\sqrt{2}}-\frac{\ep_2^2\ep_3}{2\sqrt{2}\ep_1}-\frac{\ep_2\ep_3^2}{2\sqrt{2}\ep_1}-\frac{\ep_2\ep_3\ep_4}{2\sqrt{2}\ep_1}
\right) \nonumber\\
&=\sqrt{2\varepsilon_1}\left(-2\varepsilon_1+\frac{3}{2}\varepsilon_2- \frac{\varepsilon_2\varepsilon_3}{4\varepsilon_1}
+\mathcal{O}\big(\varepsilon^2\big)\right)\,. \label{hatxiV}
\end{align}
Finally, we define the conformal time:
\begin{align} 
    \tau(t)=-\int_t^\infty dt'e^{-\rho(t')}\,,
\end{align}
which goes from $-\infty$ to $0$, as the cosmic time $t$ is running from $-\infty$ to $\infty$; $\tau$ and $t$ are linked to each other at first order in Hubble flow functions by\footnote{The expansion of all quantities at N3LO in Hubble flow functions is done in Appendix \ref{Appendixexpansion}.}:
\begin{align}
-\tau(t)=\frac{1+\varepsilon_1(t)}{H(t)e^{\rho(t)}}+\mathcal{O}\big(\varepsilon^2\big)\qquad\text{implying} \qquad 
e^{-\rho(t)}=-\tau H(t)(1-\varepsilon_1(t))+\mathcal{O}\big(\varepsilon^2\big)\,.
\end{align}

\section{Non-gaussianity of cosmological perturbations}

In this and next sections, we will compute the bispectrum of the scalar fluctuation around the background inflationary metric \eqref{bgmetric}. The action of the scalar fluctuation up to third order will be presented in this section, together with a review of the computational procedures used to calculate its equal-time three-point function. The next section will be devoted to concrete computations. 

\subsection{Scalar fluctuation and its action}

Around the background configurations of the inflation $\phi(t)$ and the metric \eqref{bgmetric} given in the last section, we parametrise their fluctuations by
\begin{align}
&\phi(t,\vec x)=\phi_0(t)+\psi(t,\vec x),
&ds^2=-N^2dt^2+h_{ij}(N^idt+dx^i)(N^jdt+dx^j),
\end{align}
where $\phi_0(t)$ is the background configuration of the inflaton that was denoted by $\phi(t)$ in the last section, and for the metric we adopted the ADM parametrisation following~\cite{Maldacena:2002vr}. Since the background metric is given by \eqref{bgmetric}, we can further parameterise the metric components $N,N^i$ and $h_{ij}$ as
\begin{align}
    N=1+2\Phi(t,\vec{x}),\qquad N^i=\delta^{ij}\partial_jB(t,\vec{x}),\qquad h_{ij}=e^{2\rho(t)+2\zeta(t,\vec{x})}\delta_{ij},
\end{align}
where we ignored the tensor perturbations in $h_{ij}$ since we are only interested in scalar perturbations in this paper. Under this ADM parameterisation, the constraint Einstein equations are separated as algebraic equations \eqref{eq:constraint1} and \eqref{eq:constraint2} for $N,N^i$. 

While we have two scalar fluctuations $\psi,\zt$ after solving the constraints for $\Phi$ and $B$, the number of the physical fluctuations is just one due to the gauge freedom from time reparametrisation $t \to t+f(x)$. In this paper, we adopt the gauge fixing in which the inflaton fluctuation is set to zero $\psi=0$ and hence $\zt$ is the dynamical scalar fluctuation, which we will call the $\zt$-gauge\footnote{This is also called the comoving gauge.}. This gauge is actually particularly useful since $\zeta$ stays constant outside the horizon. Once we adopt this gauge, we can solve the algebraic equations for $N,N^i$ to express $N,N^i$ in term of $\zeta$. Substituting everything in the $\zt$-gauge into the action \eqref{inf-action} and expanding it in $\zt$, we obtain an action in $\zt$. This contains an infinite number of interaction terms that we decompose as 
\begin{align}
S_\zeta=S^{(\zeta^2)}+S^{(\zeta^3)}+\mathcal{O}\big(\zeta^4\big).
\end{align}
For the first two terms one can find\footnote{A way to obtain this action in $\zt$-gauge is given in~\cite{Collins:2011mz} in full detail. Appendix~\ref{action} gives another derivation of the action, which first adopts another gauge in which $\zt$ is set to zero and hence $\psi$ is the dynamical scalar perturbation ($\psi$-gauge, spatially flat gauge), and then applies a diffeomorphism which converts the $\psi$-gauge to $\zt$-gauge.}
\begin{align}\label{eq:actionz2}
    S^{(\zeta^2)}=M_\Pl^2\int d^4x\ e^{3\rho}\varepsilon_1\left(\dot{\zeta}^2-e^{-2\rho}\partial_k\zeta\partial^k\zeta\right)
\end{align}
and
\begin{align}\label{eq:action}
\begin{split}
S^{(\zeta^3)}=M_\Pl^2\int d^4x\Bigg(&e^{3\rho}\Bigg\{\varepsilon_1^2\left(\zeta\dot{\zeta}^2+e^{-2\rho}\zeta\partial_k\zeta\partial^k\zeta-2\dot{\zeta}\partial_k\zeta\partial^k\partial^{-2}\dot{\zeta}\right)\\
    &\qquad +H^2E\zeta^3+HF\zeta^2\dot{\zeta}+HG\zeta\partial_k\zeta\partial^k\partial^{-2}\dot{\zeta}\\
    &\qquad +\frac{1}{2}\varepsilon_1^3\zeta\left[\partial_i\partial_j\partial^{-2}\dot{\zeta}\partial^i\partial^j\partial^{-2}\dot{\zeta}-\dot{\zeta}^2\right]\Bigg\}\\
    &+\left[\partial_t(2\varepsilon_1e^{3\rho}\dot{\zeta})-2\varepsilon_1e^{\rho}\partial^2\zeta\right]f(\zeta)\Bigg),
    \end{split}
\end{align}
where $f$ is quadratic in $\zt$:
\begin{align}
\begin{split}\label{f(zeta)}
    f(\zeta)&=\frac{\varepsilon_2}{4}\zeta^2+\frac{1}{H}\dot{\zeta}\zeta-\frac{1}{4}\frac{e^{-2\rho}}{H^2}\left[\partial_i\zeta\partial^i\zeta-\partial^{-2}\partial_i\partial_j(\partial^i\zeta\partial^j\zeta)\right]\\
    &\quad+\frac{\varepsilon_1}{2H}\left[\partial_i\zeta\partial^i\partial^{-2}\dot{\zeta}-\partial^{-2}\partial_i\partial_j(\partial^i\zeta\partial^j\partial^{-2}\dot{\zeta})\right],
    \end{split}
\end{align}
and $E,F$ and $G$ are polynomials in Hubble flow functions:
\begin{align}
\begin{split}
    E&=-\frac{1}{6}\Big(6\varepsilon_1^3-2\varepsilon_1^4-9\varepsilon_1^2\varepsilon_2+9\varepsilon_1^3\varepsilon_2-6\varepsilon_1^2\varepsilon_2^2+3\varepsilon_1\varepsilon_2\varepsilon_3\\
    &\qquad\quad\quad -4\varepsilon_1^2\varepsilon_2\varepsilon_3+\varepsilon_1\varepsilon_2^2\varepsilon_3+\varepsilon_1\varepsilon_2\varepsilon_3^2+\varepsilon_1\varepsilon_2\varepsilon_3\varepsilon_4\Big),
    \end{split} \label{E-def} \\
    F&=\varepsilon_1^2\varepsilon_2-\varepsilon_1^3, \label{F-def}\\
    G&=-\varepsilon_1^2\varepsilon_2, \label{G-def}
\end{align}
where total derivatives with respect to time and spacial coordinates are dropped\footnote{We will use Schwinger-Keldysh path integral to compute a three-point correlator of $\zt$. In this method, one can show that interactions in the form of total time derivatives do not contribute to the correlator while the interactions which are proportional to the equation of motion of $\zt$ (the term with $f$ in \eqref{eq:action}) do contribute~\cite{Braglia:2024zsl,Kawaguchi:2024lsw}, as will be shown in the next section and Appendix~\ref{appendix:SK}.}.

\subsection{Equal-time three-point correlator}
A theory that contains more information than what is already included in the two-point function is a non-Gaussian theory. Therefore the three-point function is usually the first and best place to look at the non-gaussianity. This is why we are only interested in the action up to its cubic interactions here. In this paper, we will compute the equal-time expectation value\footnote{This is given in the Heisenberg picture, which is directly related to its path integral expression which we will work with in this paper.}
\begin{align}\label{0ztztzt0}
\langle 0|\zt(t,\vec x)\zt(t,\vec y)\zt(t,\vec z)|0 \rangle,
\end{align}
where $|0\rangle$ is the Bunch-Davies vacuum as the in-state at the past infinity.
We parametrize it as
\begin{align}
    \begin{split}
&\langle0|\zeta(t,\vec{x})\zeta(t,\vec{y})\zeta(t,\vec{z})|0\rangle\\
    &\quad
=\int\frac{d^3\vec{k}_1}{(2\pi)^3}\frac{d^3\vec{k}_2}{(2\pi)^3}\frac{d^3\vec{k}_3}{(2\pi)^3}e^{i\vec{k}_1\cdot\vec{x}}e^{i\vec{k}_2\cdot\vec{y}}e^{i\vec{k}_3\cdot\vec{z}}(2\pi)^3\delta^{(3)}\left(\vec{k}_1+\vec{k}_2+\vec{k}_3\right)\frac{1}{32\varepsilon_1^2}\frac{H^4}{M_\Pl^4}\frac{\mathscr{A}}{k_1^3k_2^3k_3^3},
    \end{split}
\end{align}
where the amplitude $\mathscr{A}$ depends on both time $t$ and the three (comoving) momenta $k_1$, $k_2$ and $k_3$. The computation at tree-level and leading order (LO) in Hubble flow parameters has originally been computed in \cite{Maldacena:2002vr}. Namely we have:
\begin{align}
\begin{split}
    \mathscr{A}_{\text{LO}}&=(\varepsilon_2-2\varepsilon_1)(k_1^3+k_2^3+k_3^3)+\varepsilon_1(k_1^2+k_2^2+k_3^2)(k_1+k_2+k_3)\\
    &\quad +8\varepsilon_1\frac{k_1^2k_2^2+k_1^2k_3^2+k_2^2k_3^2}{k_1+k_2+k_3}+\mathcal{O}\big(\varepsilon^2,\tau\big).
    \end{split}
\end{align}
In practice, we are interested in the fluctuations that have been stretched so much that their wavelength became larger than the size of the horizon by the end of inflation. Consequently, a very large value of $t$ is needed for our purpose so that we only keep the non-vanishing part of the result in the late-time limit, namely $t\rightarrow\infty$ or equivalently $\tau\rightarrow0^-$ in the conformal time. In this paper, we will do the computations up to next-to-leading order (NLO) in Hubble flow parameters at tree-level (no loops). 

To begin with, let us clarify one point: the three-point function \eqref{0ztztzt0} is an expectation value sandwiched by the in-state $|0\rangle$. It can be expressed schematically as 
\begin{align}\label{zt3_UUU}
\langle 0|\hat U(-\infty,t_f)\hat U(t_f,t)\zt(-\infty,\vec x)\zt(-\infty,\vec y)\zt(-\infty,\vec z)\hat U(t,-\infty)|0\rangle
\end{align}
with $\hat U(t_1,t_2)$ the time evolution operator and $t_f$ a time after $t$. This indicates that \eqref{0ztztzt0} involves a sequence of time evolution operators starting from the past infinity, reaching some late time and returning to the past infinity\footnote{This is in contrast to ordinary scattering processes, for which operators are sandwiched by in and out states and therefore time-evolution operators start from the past straight to future infinities.}. Consequently, we will compute \eqref{0ztztzt0} by using the Schwinger-Keldysh formalism. More specifically, we will use its path integral formulation, which is reviewed in Appendix~\ref{appendix:SK}. 
In Schwinger-Keldysh path integrals, integral variable $\zt$ lives along the following time contour $C_+\cup(-C_-)$\footnote{Contour $C_+$ runs from $-\infty(1-i\ep)$ to $t_f$ with $\ep>0$ and $-C_-$ runs from $t_f$ to $-\infty(1+i\ep)$, so that their joint $C_+\cup(-C_-)$ has a well-defined orientation. Thus, $C_-$, the orientation flip of $-C_-$, runs from $-\infty(1+i\ep)$ to $t_f$.}:
\begin{center}
\begin{tikzpicture}
    \draw[thick, ->] (-4.2, 0) -- (7, 0); 
    \draw[thick, ->] (-4, 0.2) -- (3.05, 0.2);
    \draw[thick, -] (2.9, 0.2) -- (6, 0.2);
    \draw[thick] (6, 0.2) arc (90:-90:0.2);
    \draw[thick, ->] (6, -0.2) -- (2.95, -0.2);
    \draw[thick, -] (3.1, -0.2) -- (-4, -0.2);
    \node at (6.2, -0.5) {$t_f$}; 
    \node at (-4, 0.5) {$-\infty(1-i\ep)$}; 
    \node at (-4, -0.5) {$-\infty(1+i\ep)$}; 
    \draw[thin, -] (-4, -0.1) -- (-4, 0.1);
    \draw[thin, -] (1, -0.1) -- (1, 0.1);
    \node at (1, 0.5) {$\zt^3(t)$}; 
    \node at (3.5, -0.5) {$-C_-$};
    \node at (3.5, 0.5) {$C_+$};
    \node at (1, -0.5) {$t$}; 
    \filldraw (1, 0.2) circle (2pt); 
\end{tikzpicture}
\end{center}
in which the time coordinate is complexified with the horizontal axis being the real part, and the small arrows represent the orientation of the time evolution operators as indicated in \eqref{zt3_UUU}. The infinitesimal imaginary parts $\pm i\ep$ at the past infinity are chosen to pick up the Bunch-Davies vacuum. The correlator \eqref{0ztztzt0} can then be expressed as a path integral where the dynamical variable $\zt$ should live along the contour $C_+\cup(-C_-)$:
\begin{align}
\int[d\zt]_{C_+\cup(-C_-)}\,\zt(t,\vec x)\zt(t,\vec y)\zt(t,\vec z)\exp\left(i\int_{C_+\cup(-C_-)}dtd^3\vec x\,\cL\right),  
\end{align}
where $\cL$ is the total Lagrangian density. However, for perturbative computations, it is customary to represent $\zt$ along $C_+\cup(-C_-)$ as two fields $\zt^+$ ($\zt^-$) along $C_+$ ($C_-$), so that the path integral becomes that over two fields $\zt^\pm$
\begin{align}
\int[d\zt^\pm]\zt^+(t,\vec x)\,\zt^+(t,\vec y)\zt^+(t,\vec z)\exp\left(i\int_{C_+}\!\!dt d^3\vec x\,\cL^+-i\int_{C_-}\!\!dt d^3\vec x\,\cL^-\right).
\end{align}
where $\cL^\pm$ is obtained by replacing $\zt$ in $\cL$ by $\zt^\pm$, and $\zt^\pm$ should be equal (glued together) at $t_f$. Note that $\zt^-$ lives along $C_-$ and not along $-C_-$. 

One might wonder why the operators $\zt(t,\vec x)\zt(t,\vec y)\zt(t,\vec z)$ are located on $C_+$. It does not lose any genericity because the path integral is independent of the choice of the contours for these external operators. One might also wonder whether the correlator is independent of the choice of $t_f$. The answer is that it is independent of the choice of $t_f$ as long as $t_f \geq t$. Both statements are proven in Appendix~\ref{appendix:SK}.

Let us proceed to the perturbative expansion in interaction. As derived in Appendix~\ref{appendix:SK}, it is based on the following  master formula:
\begin{align}
\begin{split}
&\langle 0|\zt(t,\vec x)\zt(t,\vec y)\zt(t,\vec z)|0 \rangle \\
&\quad 
=\Bbra \zt(t,\vec x)\zt(t,\vec y)\zt(t,\vec z)
\exp\left(i\int_{C_+}dt'd^3\vec x\,\cL_I^+(t')-i\int_{C_-}dt'd^3\vec x\,\cL_I^-(t')\right)
\Kket, \label{master-zt^3}
\end{split}
\end{align}
where $\cL_I$ is the interaction Lagrangian density and $\cL_I^\pm$ is obtained by replacing $\zt$ in $\cL_I$ by $\zt^\pm$. The double bracket $\bbra\cA\kket$ denotes the Wick contraction of operator $\cA$ with the following rules\footnote{They are justified in Appendix~\ref{appendix:SK}. They are essentially the same as those for ordinary in-out correlators in path integral, except that the fields are doubled.}:
\begin{itemize}
\item Each Wick pair $\zt^a(t_1,\vec x_1)\zt^b(t_2,\vec x_2)$ with $a,b\in\{+,-\}$ is replaced by the Green function $G^{ab}(t_1,\vec x_1;t_2,\vec x_2)$.
\item When $\zt^a,\zt^b$ in the pair are subject to differential operators, we first replace the pair $\zt^a\zt^b$ by $G^{ab}$, and then apply the differential operators to $G^{ab}$ with corresponding spacetime coordinates. 
\item All $\zt^\pm$ should be contracted, which forces correlators with odd numbers of $\zt^\pm$ to vanish. 
\end{itemize} 
Here the four Green functions $G^{ab}$ satisfy the following equations of motion:
\begin{align}
\cD_{t_1\vec x_1}G^{++}(t_1,\vec x_1;t_2,\vec x_2)&=-i\de(t_1-t_2)\de^3(\vec x_1-\vec x_2), \label{sec2:eq-G++}\\
\cD_{t_1\vec x_1}G^{--}(t_1,\vec x_1;t_2,\vec x_2)&=i\de(t_1-t_2)\de^3(\vec x_1-\vec x_2), \label{sec2:eq-G--}\\
\cD_{t_1\vec x_1}G^{\pm\mp}(t_1,\vec x_1;t_2,\vec x_2)&=0, \label{sec2:eq-W}
\end{align}
where $\cD$ is the differential operator in the free equation of motion from the quadratic action \eqref{eq:actionz2}, which reads
\begin{align}\label{freeeom-D}
\cD_{t\vec x}f(t,\vec x)=M_\Pl^2\left[\pd_t\left(2a(t)^3\vep_1(t)\pd_tf(t,\vec x)\right)-2a(t)\vep_1(t)\pd^2f(t,\vec x)\right]. 
\end{align}
$G^{++}$ ($G^{--}$) is the time (anti-time) ordered propagator on $C_+$ ($C_-$). Actually, the equations of motion \eqref{sec2:eq-W} for $G^{\pm\mp}$ are not enough to fix them. In order to determine them, we need continuity conditions, which come from the fact that the path integral was originally defined with a single field $\zt$ and with the time contour $C_+\cup(-C_-)$ and therefore the four Green functions are a single Green function along the contour ``in disguise''. The Green functions can be expressed as
\begin{align}
    G^{++}(t,\vec{x};t',\vec{y})&=\theta(t-t')G^>(t,\vec{x};t',\vec{y})+\theta(t'-t)G^<(t,\vec{x};t',\vec{y}),\\
     G^{+-}(t,\vec{x};t',\vec{y})&=G^<(t,\vec{x};t',\vec{y}),\\
      G^{-+}(t,\vec{x};t',\vec{y})&=G^>(t,\vec{x};t',\vec{y}),\\
       G^{--}(t,\vec{x};t',\vec{y})&=\theta(t'-t)G^>(t,\vec{x};t',\vec{y})+\theta(t-t')G^<(t,\vec{x};t',\vec{y}),
\end{align}
where $\theta(t)$ is the step function defined by 
\begin{align}
\theta(t)=
 \begin{cases}
  1 & (t>0) \\
  0 & (t<0) \\
  1/2 & (t=0).
 \end{cases}
\end{align}
$G^>$ and $G^<$ are the Wightman functions, which in operator language read
\begin{align}
G^>(t,\vec{x};t',\vec{y})&=\langle0|\zeta(t,\vec{x})\zeta(t',\vec{y})|0\rangle_{\text{free}}, \\
G^<(t,\vec{x};t',\vec{y})&=\langle0|\zeta(t',\vec{y})\zeta(t,\vec{x})|0\rangle_{\text{free}},
\end{align}
where the subscript `free' implies that they are two-point functions without interaction. 
They can be constructed through their Fourier modes $G^>_k,G^<_k$ by using a mode function $\zt_k(t)$ and its complex conjugate $\bar\zt_k(t)$ as
\begin{align}\label{eq:Gright}
G^>(t,\vec{x};t',\vec{y})
&=\int\frac{d^3\vec{k}}{(2\pi)^3}e^{i\vec{k}\cdot(\vec{x}-\vec{y})}G^>_k(t,t')
=\int\frac{d^3\vec{k}}{(2\pi)^3}e^{i\vec{k}\cdot(\vec{x}-\vec{y})}\zt_k(t)\bar\zt_k(t'),\\
G^<(t,\vec{x};t',\vec{y})
&=\int\frac{d^3\vec{k}}{(2\pi)^3}e^{i\vec{k}\cdot(\vec{x}-\vec{y})}G^<_k(t,t')
=\int\frac{d^3\vec{k}}{(2\pi)^3}e^{i\vec{k}\cdot(\vec{x}-\vec{y})}\zt_k(t')\bar\zt_k(t),
\label{eq:Gleft}
\end{align}
where the mode functions are fixed by the equations of motion and the normalisation of the Wronskian:
\begin{align}
&\pd_t(2a^3\vep_1\pd_t\zt_k(t))+2a\vep_1k^2\zt_k(t)=0, \label{sec2:ztk-eom}\\
&2a^3\vep_1\pd_t\zt_k(t)\bar\zt_k(t)-2a^3\vep_1\pd_t\bar\zt_k(t)\zt_k(t)=-iM_\Pl^{-2}. \label{sec2:ztk-wronskian}
\end{align}
The mode functions $\zt_k$ are derived perturbatively in Hubble flow parameters in Appendix~\ref{wightman}.

Before proceeding to concrete computations, let us classify the cubic vertices in \eqref{eq:action}.
For this, we write the interaction Lagrangian \eqref{eq:action} as
\begin{align}
L_I=M_\Pl^2\int d^3\vec{x}\left(\sum\limits_{n=1}^9\mathscr{O}_n+\text{higher order terms}\right) \label{LI-O}
\end{align}
with cubic vertices labelled as:
\begin{align}
\mathscr{O}_1&=\varepsilon_1^2e^{3\rho}\zeta\dot{\zeta}^2,\\
    \mathscr{O}_2&=\varepsilon_1^2e^{\rho}\zeta\partial_k\zeta\partial^k\zeta,\\
    \mathscr{O}_3&=-2\varepsilon_1^2e^{3\rho}\dot{\zeta}\partial_k\zeta\partial^k\partial^{-2}\dot{\zeta},\\
    \mathscr{O}_4&=e^{3\rho}H^2E\zeta^3,\\
    \mathscr{O}_5&=e^{3\rho}HF\zeta^2\dot{\zeta},\\
    \mathscr{O}_6&=e^{3\rho}HG\zeta\partial_k\zeta\partial^k\partial^{-2}\dot{\zeta},\\
    \mathscr{O}_7&=\frac{1}{2}\varepsilon_1^3e^{3\rho}\zeta\partial_i\partial_j\partial^{-2}\dot{\zeta}\partial^i\partial^j\partial^{-2}\dot{\zeta},\\
    \mathscr{O}_8&=-\frac{1}{2}\varepsilon_1^3e^{3\rho}\zeta\dot{\zeta}^2, \\
    \mathscr{O}_9&=\left[\partial_t\left(2\varepsilon_1e^{3\rho}\dot{\zeta}\right)-2\varepsilon_1e^{\rho}\partial^2\zeta\right]f(\zeta).
\end{align}
We call three vertices $\mathscr{O}_1,\mathscr{O}_2,\mathscr{O}_3$ ``LO vertices'', five operators $\mathscr{O}_4,\cdots,\mathscr{O}_8$ ``NLO vertices'', and the rest $\mathscr{O}_9$ ``EoM vertex'', since the expression in the square bracket is the free equation of motion.
In these notations, the three-point function reads
\begin{align}\begin{split}
&\langle0|\zeta^+(t,\vec{x})\zeta^+(t,\vec{y})\zeta^+(t,\vec{z})|0\rangle\\
    &\quad=i\int_{-\infty}^{t}dt'\,\Bbra\zeta^+(t,\vec{x})\zeta^+(t,\vec{y})\zeta^+(t,\vec{z})[L_I^+(t')-L_I^-(t')]\Kket+\text{loops}\\
    &\quad=M_\Pl^2\sum_{n=1}^9\langle\mathscr{O}_n\rangle+\text{loops},\end{split}
\end{align}
where we have introduced the notation
\begin{align}\label{<On>}
    \langle\mathscr{O}_n\rangle=i\int_{-\infty}^{t}dt'\int d^3\vec{w}\, \Bbra \zeta^+(t,\vec{x})\zeta^+(t,\vec{y})\zeta^+(t,\vec{z})\left[\mathscr{O}_n^+(t',\vec{w})-\mathscr{O}_n^-(t',\vec{w})\right] \Kket,
\end{align}
where $\mathscr{O}_n^\pm$ is obtained by replacing $\zt$ by $\zt^\pm$.

A comment has to be made here on $M_\Pl$-dependence. The nine vertices are of order $M_\Pl^2$, and the propagators are of order $M_\Pl^{-2}$ because of $M_\Pl^2$ in the definition of $\cD$ for the free equation of motion of $\zt$. Therefore, at tree level, the three-point function is of order $M_\Pl^{-4}$. This can be shown easily for the vertices $\mathscr{O}_1,\cdots,\mathscr{O}_8$, while it is non-trivial for the EoM vertex $\mathscr{O}_9$, which will be demonstrated in the next section. The part with ``loops'' are therefore of higher order in $M_\Pl^{-2}$. 

Another comment is about cubic interactions of the form of time total derivatives. They were suppressed in \eqref{eq:action} and also not included in the vertices $\mathscr{O}_1,\cdots,\mathscr{O}_9$. This is based on the property proven in~\cite{Braglia:2024zsl,Kawaguchi:2024lsw} and also in Appendix~\ref{appendix:SK} that \emph{as long as we use the Schwinger-Keldysh path integral}, such terms do not conbtribute to our three-point function\footnote{If we adopt the operator formalism with canonical quantisation of $\zt$ and the Hamiltonian in the interaction picture, it is the temporal surface terms that give the same result as the EoM vertex $\mathscr{O}_9$ in path integral, and the EoM vertex in the operator formalism do not give any contribution. This is complementary to our path integral computations. See~\cite{Braglia:2024zsl,Kawaguchi:2024lsw} for more details.}.

Related to this comment, we also make a comparison of our action with another set of cubic vertices used in~\cite{Burrage:2011hd}, which reads in our notation:\footnote{When we completed the first version of this paper, we were not aware of~\cite{Burrage:2011hd}.}
\begin{align}
L_I^{\text{BRS}}=M_\Pl^2\int d^3\vec{x}\left(\sum\limits_{n=1}^5\widehat{\mathscr{O}}_n+\text{higher order terms}\right) \label{LI-BRS}
\end{align}
with five cubic vertices labelled as:
\begin{align}
\begin{split}
    \widehat{\mathscr{O}}_1&=(\varepsilon_1^2-\vep_1\vep_2)e^{3\rho}\zeta\dot{\zeta}^2, \qquad
    \widehat{\mathscr{O}}_2=(\varepsilon_1^2+\vep_1\vep_2)e^{\rho}\zeta\partial_k\zeta\partial^k\zeta, \\
    \widehat{\mathscr{O}}_3&=\left(-2\vep_1^2+\frac{1}{2}\vep_1^3\right)e^{3\rho}\dot{\zeta}\partial_k\zeta\partial^k\partial^{-2}\dot{\zeta}, \qquad
    \widehat{\mathscr{O}}_4=\frac{1}{4}\varepsilon_1^3e^{3\rho}\pd^2\zeta\partial_i\partial^{-2}\dot{\zeta}\partial^i\partial^{-2}\dot{\zeta}, \\
    \widehat{\mathscr{O}}_5&=\left[\partial_t\left(2\varepsilon_1e^{3\rho}\dot{\zeta}\right)-2\varepsilon_1e^{\rho}\partial^2\zeta\right]\left[f(\zeta)-\frac{\vep_2}{2}\zt^2\right].
\end{split}
\end{align}
The main difference from ours is the absence of the vertices with at most one time derivative such as our $\mathscr{O}_4,\mathscr{O}_5,\mathscr{O}_6$. Note also that the EoM vertex $\widehat{\mathscr{O}}_5$ does not have the non-derivative term $\zt^2$ which is cancelled in the square bracket of the right hand side [see~\eqref{f(zeta)}]. However, the two actions are equal up to temporal and spatial boundary terms, which can be seen from \eqref{eq:S3Maldacena}; the $\zt^2$ part in our EoM vertex $\mathscr{O}_9$ is combined with $\mathscr{O}_4+\mathscr{O}_5+\mathscr{O}_6$ to form a total derivative and contributes $\vep_2$ in $\widehat{\mathscr{O}}_1,\widehat{\mathscr{O}}_2$. On top of this, our $\mathscr{O}_7+\mathscr{O}_8$ is rewritten through integrations by part into $\widehat{\mathscr{O}}_4$ plus the $\vep_1^3$ part of $\widehat{\mathscr{O}}_3$. Since the difference between the two actions amounts to boundary terms, they should give the same result for the three-point function as long as we  employ the Schwinger-Keldysh path integral formalism.

\section{Computing the three-point function}
This section will be devited to concrete computations of the three-point function \eqref{0ztztzt0}. We first compute it for the NLO vertices $\mathscr{O}_4,\cdots,\mathscr{O}_8$, next for the LO vertices $\mathscr{O}_1,\cdots,\mathscr{O}_3$, and then finally for the EoM vertex $\mathscr{O}_9$. For these computations, we need the propagators expanded in terms of Hubble flow functions which are obtained in Appendix~\ref{wightman}. In this section, we set $M_\Pl=1$ except in Section~\ref{sec:shift}.

\subsection{The NLO vertices}
For clarity, we begin by calculating the contributions of the vertices that do not yield leading-order terms. Those vertices are the ones which are at least of order three in Hubble flow functions:
\begin{align}
    \mathscr{O}_4&=e^{3\rho}H^2\left(E^{(3)}+\mathcal{O}\big(\varepsilon^4\big)\right)\zeta^3,\\
    \mathscr{O}_5&=e^{3\rho}HF\zeta^2\dot{\zeta},\\
    \mathscr{O}_6&=e^{3\rho}HG\zeta\partial_k\zeta\partial^k\partial^{-2}\dot{\zeta},\\
    \mathscr{O}_7&=\frac{1}{2}\varepsilon_1^3e^{3\rho}\zeta\partial_i\partial_j\partial^{-2}\dot{\zeta}\partial^i\partial^j\partial^{-2}\dot{\zeta}, \\
    \mathscr{O}_8&=-\frac{1}{2}\varepsilon_1^3e^{3\rho}\zeta\dot{\zeta}^2,
\end{align}
where $F$ and $G$ are defined in \eqref{F-def} and \eqref{G-def}, and we are only interested in the third order part of $E$ \eqref{E-def}:
\begin{align}
    E^{(3)}=-\varepsilon_1^3+\frac{3}{2}\varepsilon_1^2\varepsilon_2-\frac{1}{2}\varepsilon_1\varepsilon_2\varepsilon_3,
\end{align}
as we perform the computations up to NLO in Hubble flow functions. For $\mathscr{O}_4$, we have\footnote{Here we set the upper boundary of the time integral to $t$ instead of $t_f$ because the three-point function is independent of the choice of $t_f$ as long as $t_f \geq t$, as explained in the last section. We will do this for the other vertices except for $\mathscr{O}_9$ which involves the time integral of a delta function with a peak at $t$.}
\begin{align}
\begin{split}
    \langle\mathscr{O}_4(t)\rangle&=i\int_{-\infty}^{t}dt'\int d^3\vec{w}\ \Bbra \zeta^+(t,\vec{x})\zeta^+(t,\vec{y})\zeta^+(t,\vec{z})\left[\mathscr{O}_4^+(t',\vec{w})-\mathscr{O}_4^-(t',\vec{w})\right] \Kket\\
    &=i\int_{-\infty}^t dt'e^{3\rho(t')}H^2(t')\left(E^{(3)}(t')+\mathcal{O}\big(\varepsilon^4(t')\big)\right)\\
    &\qquad\times\int d^3\vec{w}\,\Bbra \zeta^+(t,\vec{x})\zeta^+(t,\vec{y})\zeta^+(t,\vec{z})\left[\zeta^+(t',\vec{w})^3-\zeta^-(t',\vec{w})^3\right] \Kket \\
    &=\int\frac{d^3\vec{k}_1}{(2\pi)^3}\frac{d^3\vec{k}_2}{(2\pi)^3}\frac{d^3\vec{k}_3}{(2\pi)^3}e^{i\vec{k}_1\cdot\vec{x}}e^{i\vec{k}_2\cdot\vec{y}}e^{i\vec{k}_3\cdot\vec{z}}(2\pi)^3\delta^{(3)}(\vec{k}_1+\vec{k}_2+\vec{k}_3)I(t)
    \end{split}
\end{align}
with
\begin{align}
\begin{split}
    I(t)=&6i\int_{-\infty}^t dt'e^{3\rho(t')}\dot{\rho}^2(t')\left(E^{(3)}(t')+\mathcal{O}\big(\varepsilon^4(t')\big)\right)\\
    &\times\big\{G_{k_1}^>(t,t')G_{k_2}^>(t,t')G_{k_3}^>(t,t')-G_{k_1}^<(t,t')G_{k_2}^<(t,t')G_{k_3}^<(t,t')\big\}.
    \end{split}
\end{align}
The propagators are given at lowest order in Hubble flow functions by\footnote{The computations of the Wightman functions $G^>,G^<$ are given in Appendix~\ref{wightman}.}
\begin{align}
    G_{k}^>(t,t')&=\frac{H(t)H(t')}{4k^3\sqrt{\varepsilon_1(t)\varepsilon_1(t')}}(1+ik\tau)(1-ik\tau')e^{-ik(\tau-\tau')}+\mathcal{O}\big(\varepsilon^0\big),\\
    G_{k}^<(t,t')&=\frac{H(t)H(t')}{4k^3\sqrt{\varepsilon_1(t)\varepsilon_1(t')}}(1-ik\tau)(1+ik\tau')e^{ik(\tau-\tau')}+\mathcal{O}\big(\varepsilon^0\big),
\end{align}
where $\tau=\tau(t)$ and $\tau'=\tau(t')$ are the conformal times. We also introduce
\begin{align}
    1=e^{\rho(t')}\frac{\partial\tau'}{\partial t'}
\end{align}
in the integral to prepare the change of variable $t'\rightarrow\tau'$ and expand $e^{4\rho(t')}$ at lowest order in Hubble flow functions at time $t$ as
\begin{align}
    e^{4\rho(t')}=\frac{1}{(\tau'H(t'))^4}+\mathcal{O}(\varepsilon)
\end{align}
and expand $H(t')$ and $\varepsilon_1(t')$ around their values at $t$ at lowest order, giving
\begin{align}\label{E3-exp}
    \frac{H(t)^3}{H(t')}\frac{E^{(3)}(t')+\mathcal{O}\big(\varepsilon^4(t')\big)}{(\varepsilon_1(t)\varepsilon_1(t'))^{3/2}}=\frac{H(t)^2E^{(3)}(t)}{\varepsilon_1(t)^3}+\mathcal{O}(\varepsilon(t))=\frac{H^2E^{(3)}}{\varepsilon_1^3}+\mathcal{O}(\varepsilon).
\end{align}

One comment has to be made here: the Hubble flow \emph{functions} are functions of time. In order to do a proper perturbative expansion, we therefore have to select a given time (pivot time) $t^*$ at which we evaluate the Hubble flow functions. Therefore, the Hubble flow functions \emph{at this given time} $\varepsilon_i(t^*)$ are \emph{constants}, which we will therefore call Hubble flow \emph{parameters}.  We can then make perturbative expansions of time-dependent quantities (for example, the conformal time, the scale factor, and Hubble flow functions) at an arbitrary time in terms of $\vep_i(t^*)$ with the assumption $\varepsilon_i(t^*)=\mathcal{O}(\varepsilon(t^*))$\footnote{This is demonstrated in full detail up to third order in Hubble flow parameters (N3LO) at an arbitrary pivot time in Appendix~\ref{Appendixexpansion}.}. This expansion can be considered as a time-derivative (gradient) expansion in Hubble flow parameters at the pivot time $t^*$. As we have already done above, we choose the pivot time $t^*$ to be the time at which we evaluate the three-point function $\langle0|\zeta(t,\vec{x})\zeta(t,\vec{y})\zeta(t,\vec{z})|0\rangle$. In what follows, when no explicit time dependence is given, it implicitly means that the quantities are evaluated at $t^*=t$. 

That being said, we now perform the change of variable $t'\rightarrow\tau'$ and find at lowest order\footnote{Note that the $\pm i\ep$ shift in the integration contour at the infinity past makes the oscillating pieces exponentially damping in the infinite past.  
}
\begin{align}
\begin{split}
    I(t)&=\frac{3i}{32k_1^3k_2^3k_3^3}\frac{\dot{\rho}^2E^{(3)}}{\varepsilon_1^3}\int_{-\infty}^\tau\frac{d\tau'}{\tau'^4}\Big\{(1+ik_1\tau)(1-ik_1\tau')(1+ik_2\tau)\\
    &\qquad\times(1-ik_2\tau')(1+ik_3\tau)(1-ik_3\tau')e^{-i(k_1+k_2+k_3)(\tau-\tau')}\\
    &\qquad-(1-ik_1\tau)(1+ik_1\tau')(1-ik_2\tau)\\
    &\qquad\times(1+ik_2\tau')(1-ik_3\tau)(1+ik_3\tau')e^{i(k_1+k_2+k_3)(\tau-\tau')}\Big\}+\mathcal{O}(\varepsilon)
    \end{split}\\
    \begin{split}
    &=-\frac{E^{(3)}}{16\varepsilon_1^3}\frac{H^4}{k_1^3k_2^3k_3^3}\Big\{k_1k_2k_3-k_1^2(k_2+k_3)-k_2^2(k_1+k_3)-k_3^2(k_1+k_2)\\
    &\qquad+(k_1^3+k_2^3+k_3^3)\left[\gamma_E-1+\ln\Big(\!-\tau(k_1+k_2+k_3)\Big)\right]\Big\}+\mathcal{O}(\varepsilon,\tau),
    \end{split}
\end{align}
where we only keep the non-vanishing part of the result in the late-time limit, namely $t\rightarrow\infty$ or equivalently $\tau\rightarrow0^-$. 

We can now compute the contributions of the other operators similarly. By performing a spatial integration by parts on $\mathscr{O}_6$ in $L_I$ of \eqref{LI-O}, it becomes
\begin{align}
    \mathscr{O}_6\longrightarrow\mathscr{O}_{6'}=e^{3\rho}H\left(\frac{-G}{2}\right)\zeta^2\dot{\zeta}.
\end{align}
Similarly, the computations for $\mathscr{O}_5$ and $\mathscr{O}_6$ can be done at the same time. For $\mathscr{O}_5+\mathscr{O}_{6'}\equiv\mathscr{O}_{5'}$ we have
\begin{align}
    \begin{split}\langle\mathscr{O}_{5'}(t)\rangle&=i\int_{-\infty}^{t}dt'\int d^3\vec{w}\ \Bbra\zeta^+(t,\vec{x})\zeta^+(t,\vec{y})\zeta^+(t,\vec{z})\left[\mathscr{O}_{5'}^+(t',\vec{w})-\mathscr{O}_{5'}^-(t',\vec{w})\right]\Kket\\
    &=i\int_{-\infty}^t dt'e^{3\rho(t')}H(t')\left(F(t')-\frac{G(t')}{2}\right)\\
    &\qquad\times\int d^3\vec{w}\,\Bbra\zeta^+(t,\vec{x})\zeta^+(t,\vec{y})\zeta^+(t,\vec{z})\dot{\zeta}^+_n(t',\vec{w})\left[\zeta^+_n(t',\vec{w})^2-\zeta^-(t',\vec{w})^2\right]\Kket\\
    &=\int\frac{d^3\vec{k}_1}{(2\pi)^3}\frac{d^3\vec{k}_2}{(2\pi)^3}\frac{d^3\vec{k}_3}{(2\pi)^3}e^{i\vec{k}_1\cdot\vec{x}}e^{i\vec{k}_2\cdot\vec{y}}e^{i\vec{k}_3\cdot\vec{z}}(2\pi)^3\delta^{(3)}(\vec{k}_1+\vec{k}_2+\vec{k}_3)J(t)
    \end{split}
\end{align}
with
\begin{align}
    \begin{split}
        J(t)&=2i\int_{-\infty}^t dt'e^{3\rho(t')}H(t')\left(F(t')-\frac{G(t')}{2}\right)\\
    &\qquad\times\left\{\dot{G}_{k_1}^>(t,t')G_{k_2}^>(t,t')G_{k_3}^>(t,t')-\dot{G}_{k_1}^<(t,t')G_{k_2}^<(t,t')G_{k_3}^<(t,t')\right.\\
    &\qquad~~\quad+G_{k_1}^>(t,t')\dot{G}_{k_2}^>(t,t')G_{k_3}^>(t,t')-G_{k_1}^<(t,t')\dot{G}_{k_2}^<(t,t')G_{k_3}^<(t,t')\\
    &\qquad~~\quad+\left.G_{k_1}^>(t,t')G_{k_2}^>(t,t')\dot{G}_{k_3}^>(t,t')-G_{k_1}^<(t,t')G_{k_2}^<(t,t')\dot{G}_{k_3}^<(t,t')\right\},
    \end{split}
\end{align}
where the dots on the Wigthman functions denote derivatives with respect to their second arguments, namely $t'$ here. Performing the computations at lowest order in Hubble flow parameters one can find:
\begin{align}
\begin{split}
    J(t)=\frac{F-\frac{G}{2}}{16\varepsilon_1^3}&\frac{H^4}{k_1^3k_2^3k_3^3}\Big\{k_1k_2k_3-k_1^2(k_2+k_3)-k_2^2(k_1+k_3)-k_3^2(k_1+k_2)\\
    &+(k_1^3+k_2^3+k_3^3)\left[\gamma_E-1+\ln\Big(\!-\tau(k_1+k_2+k_3)\Big)\right]\Big\}+\mathcal{O}(\varepsilon,\tau).
    \end{split}
\end{align}
Therefore, the total contribution to the amplitude $\mathscr{A}$ coming from the operators $\mathscr{O}_4$, $\mathscr{O}_{5}$ and $\mathscr{O}_{6}$ is:
\begin{align}
\begin{split}
    \mathscr{A}_{4+5+6}=&\frac{2F-G-2E^{(3)}}{\varepsilon_1}\Big\{k_1k_2k_3-k_1^2(k_2+k_3)-k_2^2(k_1+k_3)-k_3^2(k_1+k_2)\\
    &+(k_1^3+k_2^3+k_3^3)\left[\gamma_E-1+\ln\Big(\!-\tau(k_1+k_2+k_3)\Big)\right]\Big\}+\mathcal{O}(\varepsilon,\tau)
    \end{split}
\end{align}
with
\begin{align}
    \frac{2F-G-2E^{(3)}}{\varepsilon_1}=\varepsilon_2\varepsilon_3.
\end{align}
We now understand why the NLO-vertex contributions could play a crucial role. Indeed, even if they are reduced by (one) more Hubble flow parameters than the LO-vertex part (which is $\mathcal{O}(\varepsilon)$), they are also enhanced by a logarithmically divergent factor in the late-time limit $\tau\rightarrow0^-$ already at tree level. 

Before looking at the other operators, we would like to give a comment on the coherence of the result upon \emph{temporal} integrations by parts of the vertices $\mathscr{O}_4,\mathscr{O}_5,\mathscr{O}_6$ at NLO in Hubble flow functions. If one does the same analysis by using the action \eqref{eq:S3afterIPP}, the equivalent operator to look at is the third order part of
\begin{align}
    \mathscr{O}_{\tilde{4}}=e^{3\rho}H^2\tilde{E}\zeta^3
\end{align}
with
\begin{align}
    \tilde{E}=-\frac{1}{6}\big[3\varepsilon_1\varepsilon_2\varepsilon_3-\varepsilon_1^2\varepsilon_2\varepsilon_3+\varepsilon_1\varepsilon_2^2\varepsilon_3+\varepsilon_1\varepsilon_2\varepsilon_3^2+\varepsilon_1\varepsilon_2\varepsilon_3\varepsilon_4\big].
\end{align}
Performing exactly the same computations as for $\mathscr{O}_4$, one can conclude that at lowest order in Hubble flow parameters, one gets $\mathscr{A}_{\tilde{4}}=\mathscr{A}_{4+5'}$ as $-2\tilde{E}^{(3)}=\varepsilon_1\varepsilon_2\varepsilon_3=2F-G-2E^{(3)}$. Moreover, by using the action  \eqref{eq:S3Maldacena}, the equivalent operator is now
\begin{align}
    \mathscr{O}_{\tilde{5}}=e^{3\rho}H\tilde{F}\zeta^2\dot{\zeta}\qquad\text{with}\qquad\tilde{F}=\frac{1}{2}\varepsilon_1\varepsilon_2\varepsilon_3.
\end{align}
Performing exactly the same computations as for $\mathscr{O}_5'$, one can conclude that at lowest order in Hubble flow parameters, one gets $\mathscr{A}_{\tilde{5}}=\mathscr{A}_{4+5'}$ as $2\tilde{F}=\varepsilon_1\varepsilon_2\varepsilon_3=2F-G-2E^{(3)}$. In the three cases, we indeed recover the same NLO contribution. 

Similar computations for the vertices $\mathscr{O}_7$ and $\mathscr{O}_8$ give:
\begin{align}\begin{split}
    \mathscr{A}_7=&\varepsilon_1^2\left[-\big(k_1^3+k_2^3+k_3^3\big)+3\frac{k_1^4+k_2^4+k_3^4}{k_1+k_2+k_3}\right.\\
    &\qquad\left.+2k_1k_2k_3\frac{k_1^2+k_2^2+k_3^2}{\left(k_1+k_2+k_3\right)^2}-\frac{k_1^5+k_2^5+k_3^5}{(k_1+k_2+k_3)^2}\right]+\mathcal{O}\big(\varepsilon^3,\tau\big),\end{split}\\
    \begin{split}
        \mathscr{A}_8=&-\varepsilon_1^2\left[-\big(k_1^3+k_2^3+k_3^3\big)+\frac{k_1^4+k_2^4+k_3^4}{k_1+k_2+k_3}+\frac{\big(k_1^2+k_2^2+k_3^2\big)^2}{k_1+k_2+k_3}\right.\\
    &\qquad\left.+2k_1k_2k_3\frac{k_1^2+k_2^2+k_3^2}{\left(k_1+k_2+k_3\right)^2}-\frac{k_1^5+k_2^5+k_3^5}{(k_1+k_2+k_3)^2}\right]+\mathcal{O}\big(\varepsilon^3,\tau\big).
    \end{split}
\end{align}

\subsection{The lowest order vertices}
Let us now consider the contributions coming from the LO vertices $\mathscr{O}_1$, $\mathscr{O}_2$ and $\mathscr{O}_3$ up to NLO in Hubble flow parameters. We start with $\mathscr{O}_1$:
\begin{align}
    \langle\mathscr{O}_{1}(t)\rangle=\int\frac{d^3\vec{k}_1}{(2\pi)^3}\frac{d^3\vec{k}_2}{(2\pi)^3}\frac{d^3\vec{k}_3}{(2\pi)^3}e^{i\vec{k}_1\cdot\vec{x}}e^{i\vec{k}_2\cdot\vec{y}}e^{i\vec{k}_3\cdot\vec{z}}(2\pi)^3\delta^{(3)}(\vec{k}_1+\vec{k}_2+\vec{k}_3)K(t)
\end{align}
with
\begin{align}
    \begin{split}
        K(t)&=2i\int_{-\infty}^t \left(\frac{\partial\tau'}{\partial t'}dt'\right)e^{4\rho(t')}\varepsilon_1(t')^2\\
    &\qquad\times\left\{G_{k_1}^>(t,t')\dot{G}_{k_2}^>(t,t')\dot{G}_{k_3}^>(t,t')-G_{k_1}^<(t,t')\dot{G}_{k_2}^<(t,t')\dot{G}_{k_3}^<(t,t')\right.\\
    &\qquad\qquad+\dot{G}_{k_1}^>(t,t')G_{k_2}^>(t,t')\dot{G}_{k_3}^>(t,t')-\dot{G}_{k_1}^<(t,t')G_{k_2}^<(t,t')\dot{G}_{k_3}^<(t,t')\\
    &\qquad\qquad+\left.\dot{G}_{k_1}^>(t,t')\dot{G}_{k_2}^>(t,t')G_{k_3}^>(t,t')-\dot{G}_{k_1}^<(t,t')\dot{G}_{k_2}^<(t,t')G_{k_3}^<(t,t')\right\}.
    \end{split}
\end{align}
The computation is then performed in exactly the same way with the difference that we have to expand all the quantities up to NLO in Hubble flow parameters as the vertex brings less powers (by one) of $\varepsilon_i$. Such expansions are derived in Appendix \ref{Appendixexpansion} and \ref{wightman}\footnote{One can use these expansions with the replacement of $(t,t^*)$ there by $(t',t)$ here.}. The complexity of the computations is mostly increased by the Wightman functions, which are found to be
\begin{align}
\begin{split}
    G^>_k(t,t')&=\frac{H^2}{4k^3\varepsilon_1}e^{-ik(\tau-\tau')}(1+ik\tau)(1-ik\tau')\\
    &\quad\times\Bigg\{1-2\varepsilon_1+\left(\varepsilon_1+\frac{\varepsilon_2}{2}\right)\Bigg[\frac{2}{1+ik\tau}+\frac{2}{1-ik\tau'}\\
    &\qquad\quad-e^{2ik\tau}\frac{1-ik\tau}{1+ik\tau}\Big(\text{Ei}(-2ik\tau)-i\pi\Big)\\
    &\qquad\quad-e^{-2ik\tau'}\frac{1+ik\tau'}{1-ik\tau'}\Big(\text{Ei}(2ik\tau')+i\pi\Big)+\ln\left(\frac{\tau'}{\tau}\right)\Bigg]\Bigg\}+\mathcal{O}(\varepsilon).
\end{split}
\end{align}
Its time derivative with respect to $t'$ is found to be
\begin{align}
\begin{split}
    \dot{G}_k^>(t,t')=&-\frac{H^3}{4k\varepsilon_1}e^{-ik(\tau-\tau')}(1+ik\tau)\tau'^2\\
    &\times\Bigg\{1-3\varepsilon_1+\left(\varepsilon_1+\frac{\varepsilon_2}{2}\right)\Bigg[\frac{2}{1+ik\tau}-e^{-2ik\tau'}\big(\text{Ei}(2ik\tau')+i\pi\big)\\
    &\quad-e^{2ik\tau}\frac{1-ik\tau}{1+ik\tau}\big(\text{Ei}(-2ik\tau)-i\pi\big)\Bigg]+\left(2\varepsilon_1+\frac{\varepsilon_2}{2}\right)\ln\left(\frac{\tau'}{\tau}\right)\Bigg\}+\mathcal{O}(\varepsilon),
\end{split}
\end{align}
The other ordering $G^<$ is simply given by the complex conjugate. Here, the function $\mathrm{Ei}$ is the exponential integral defined by
\begin{align}
\text{Ei}(z)=-\int_{-z}^\infty\frac{e^{-t}}{t}dt.
\end{align} 
Finally one finds:
\begin{align}
\begin{split}
    \mathscr{A}_1=&-2\varepsilon_1\Bigg\{\left(-1+(4\varepsilon_1+\varepsilon_2)\left[\gamma_E-1+\ln\Big(\!-\tau(k_1+k_2+k_3)\Big)\right]\right)\\
    &\qquad\qquad\times\left[\frac{k_1^2k_2^2+k_1^2k_3^2+k_2^2k_3^2}{k_1+k_2+k_3}+k_1k_2k_3\frac{k_1k_2+k_1k_3+k_2k_3}{\left(k_1+k_2+k_3\right)^2}\right]\\
    &\qquad\quad+(2\varepsilon_1+\varepsilon_2)\ln\left(\frac{2k_1}{k_1+k_2+k_3}\right)(k_2+k_3)\\
    &\qquad\qquad\times\left[\frac{k_1^2k_2^2+k_1^2k_3^2+k_2^2k_3^2}{\left(k_1+k_2+k_3\right)\left(-k_1+k_2+k_3\right)}+k_1^2k_2k_3\frac{k_1^2+k_2^2+k_3^2}{\left(k_1^2-(k_2+k_3)^2\right)^2}\right]\\
    &\qquad\quad+(2\varepsilon_1+\varepsilon_2)\ln\left(\frac{2k_2}{k_1+k_2+k_3}\right)(k_1+k_3)\\
    &\qquad\qquad\times\left[\frac{k_1^2k_2^2+k_1^2k_3^2+k_2^2k_3^2}{\left(k_1+k_2+k_3\right)\left(k_1-k_2+k_3\right)}+k_1k_2^2k_3\frac{k_1^2+k_2^2+k_3^2}{\left(k_2^2-(k_1+k_3)^2\right)^2}\right]\\
    &\qquad\quad+(2\varepsilon_1+\varepsilon_2)\ln\left(\frac{2k_3}{k_1+k_2+k_3}\right)(k_1+k_2)\\
    &\qquad\qquad\times\left[\frac{k_1^2k_2^2+k_1^2k_3^2+k_2^2k_3^2}{\left(k_1+k_2+k_3\right)\left(k_1+k_2-k_3\right)}+k_1k_2k_3^2\frac{k_1^2+k_2^2+k_3^2}{\left(k_3^2-(k_1+k_2)^2\right)^2}\right]\\
    &\qquad\quad+2(2\varepsilon_1+\varepsilon_2)\frac{k_1^2k_2^2k_3^2\left(k_1^2+k_2^2+k_3^2\right)}{(-k_1+k_2+k_3)(k_1-k_2+k_3)(k_1+k_2-k_3)(k_1+k_2+k_3)^2}\\
    &\qquad\quad-\varepsilon_2\left[2k_1k_2k_3\frac{k_1k_2+k_1k_3+k_2k_3}{(k_1+k_2+k_3)^2}+3\frac{k_1^2k_2^2+k_1^2k_3^2+k_2^2k_3^2}{k_1+k_2+k_3}\right]\Bigg\}+\mathcal{O}\big(\varepsilon^3,\tau\big).
\end{split}
\end{align}
Similarly, we have for $\mathscr{O}_2$,
\begin{align}
\begin{split}
    \mathscr{A}_2=&-\varepsilon_1\left(k_1^2+k_2^2+k_3^2\right)\Bigg\{\left(-1+(4\varepsilon_1+\varepsilon_2)\left[\gamma_E-1+\ln\Big(\!-\tau(k_1+k_2+k_3)\Big)\right]\right)\\
    &\quad\qquad\times\left[k_1+k_2+k_3-\frac{k_1k_2+k_1k_3+k_2k_3}{k_1+k_2+k_3}-\frac{k_1k_2k_3}{\left(k_1+k_2+k_3\right)^2}\right]\\
    &\quad\quad-(2\varepsilon_1+\varepsilon_2)\ln\left(\frac{2k_1}{k_1+k_2+k_3}\right)(k_2+k_3)\\
    &\quad\qquad\times\left[\frac{k_1^2\big(k_2^2-k_2k_3+k_3^2\big)-(k_2+k_3)^2\big(k_2^2+k_2k_3+k_3^2\big)}{\left(k_1+k_2+k_3\right)^2\left(-k_1+k_2+k_3\right)^2}\right]\\
    &\quad\quad-(2\varepsilon_1+\varepsilon_2)\ln\left(\frac{2k_2}{k_1+k_2+k_3}\right)(k_1+k_3)\\
    &\quad\qquad\times\left[\frac{k_2^2\big(k_1^2-k_1k_3+k_3^2\big)-(k_1+k_3)^2\big(k_1^2+k_1k_3+k_3^2\big)}{\left(k_1+k_2+k_3\right)^2\left(k_1-k_2+k_3\right)^2}\right]\\
    &\quad\quad-(2\varepsilon_1+\varepsilon_2)\ln\left(\frac{2k_3}{k_1+k_2+k_3}\right)(k_1+k_2)\\
    &\quad\qquad\times\left[\frac{k_3^2\big(k_1^2-k_1k_2+k_2^2\big)-(k_1+k_2)^2\big(k_1^2+k_1k_2+k_2^2\big)}{\left(k_1+k_2+k_3\right)^2\left(k_1+k_2-k_3\right)^2}\right]\\
    &\quad\quad+4(2\varepsilon_1+\varepsilon_2)\frac{k_1^2k_2^2k_3^2}{(-k_1+k_2+k_3)(k_1-k_2+k_3)(k_1+k_2-k_3)(k_1+k_2+k_3)^2}\\
    &\quad\quad-\varepsilon_2\left[2(k_1+k_2+k_3)-3\frac{k_1k_2+k_1k_3+k_2k_3}{k_1+k_2+k_3}-2\frac{k_1k_2k_3}{\left(k_1+k_2+k_3\right)^2}\right]\Bigg\}+\mathcal{O}\big(\varepsilon^3,\tau\big),
\end{split}
\end{align}
and for $\mathscr{O}_3$,
\begin{align}
\begin{split}
    \mathscr{A}_3=&-\varepsilon_1\Bigg\{\left(-1+(4\varepsilon_1+\varepsilon_2)\left[\gamma_E-1+\ln\Big(\!-\tau(k_1+k_2+k_3)\Big)\right]\right)\\
    &\qquad\qquad\times\left[-\big(k_1^3+k_2^3+k_3^3\big)+6\frac{k_1^2k_2^2+k_1^2k_3^2+k_2^2k_3^2}{k_1+k_2+k_3}\right.\\
    &\qquad\qquad\qquad\left.-\frac{k_1^4+k_2^4+k_3^4}{k_1+k_2+k_3}+2k_1k_2k_3\frac{k_1^2+k_2^2+k_3^2}{\left(k_1+k_2+k_3\right)^2}\right]\\
    &\qquad\quad+(2\varepsilon_1+\varepsilon_2)\ln\left(\frac{2k_1}{k_1+k_2+k_3}\right)(k_2+k_3)\\
    &\qquad\qquad\times\left[(k_1-k_2+k_3)(k_1+k_2-k_3)+6\frac{k_1^2k_2^2+k_1^2k_3^2+k_2^2k_3^2}{\left(k_1+k_2+k_3\right)\left(-k_1+k_2+k_3\right)}\right.\\
    &\qquad\qquad\left.+\big(k_1^2+k_2^2+k_3^2\big)\frac{k_1^2\big(k_2^2-3k_2k_3+k_3^2\big)-(k_2+k_3)^2\big(k_2^2+k_2k_3+k_3^2\big)}{\left(k_1+k_2+k_3\right)^2\left(-k_1+k_2+k_3\right)^2}\right]\\
    &\qquad\quad+(2\varepsilon_1+\varepsilon_2)\ln\left(\frac{2k_2}{k_1+k_2+k_3}\right)(k_1+k_3)\\
     &\qquad\qquad\times\left[(-k_1+k_2+k_3)(k_1+k_2-k_3)+6\frac{k_1^2k_2^2+k_1^2k_3^2+k_2^2k_3^2}{\left(k_1+k_2+k_3\right)\left(k_1-k_2+k_3\right)}\right.\\
    &\qquad\qquad\left.+\big(k_1^2+k_2^2+k_3^2\big)\frac{k_2^2\big(k_1^2-3k_1k_3+k_3^2\big)-(k_1+k_3)^2\big(k_1^2+k_1k_3+k_3^2\big)}{\left(k_1+k_2+k_3\right)^2\left(k_1-k_2+k_3\right)^2}\right]\\
    &\qquad\quad+(2\varepsilon_1+\varepsilon_2)\ln\left(\frac{2k_3}{k_1+k_2+k_3}\right)(k_1+k_2)\\
     &\qquad\qquad\times\left[(-k_1+k_2+k_3)(k_1-k_2+k_3)+6\frac{k_1^2k_2^2+k_1^2k_3^2+k_2^2k_3^2}{\left(k_1+k_2+k_3\right)\left(k_1+k_2-k_3\right)}\right.\\
     &\qquad\qquad\left.+\big(k_1^2+k_2^2+k_3^2\big)\frac{k_3^2\big(k_1^2-3k_1k_2+k_2^2\big)-(k_1+k_2)^2\big(k_1^2+k_1k_2+k_2^2\big)}{\left(k_1+k_2+k_3\right)^2\left(k_1+k_2-k_3\right)^2}\right]\\
    &\qquad\quad-4(2\varepsilon_1+\varepsilon_2)\left[\frac{2k_1^2k_2^2k_3^2\left(k_1^2+k_2^2+k_3^2\right)}{(-k_1+k_2+k_3)(k_1-k_2+k_3)(k_1+k_2-k_3)(k_1+k_2+k_3)^2}\right.\\
     &\qquad\qquad\qquad\left.-\frac{k_1^2k_2^2+k_1^2k_3^2+k_2^2k_3^2}{k_1+k_2+k_3}\right]\\
    &\qquad\quad-\varepsilon_2\left[-2\big(k_1^3+k_2^3+k_3^3\big)+16\frac{k_1^2k_2^2+k_1^2k_3^2+k_2^2k_3^2}{k_1+k_2+k_3}\right.\\
    &\qquad\qquad\qquad\left.-4\frac{k_1^4+k_2^4+k_3^4}{k_1+k_2+k_3}+4k_1k_2k_3\frac{k_1^2+k_2^2+k_3^2}{\left(k_1+k_2+k_3\right)^2}\right]\Bigg\}+\mathcal{O}\big(\varepsilon^3,\tau\big).
\end{split}
\end{align}
Their sum simplifies nicely to
\begin{align}
\begin{split}
    \mathscr{A}_{1+2+3}&=\varepsilon_1\Bigg\{\left(1-(4\varepsilon_1+\varepsilon_2)\left[\gamma_E-1+\ln\Big(\!-\tau(k_1+k_2+k_3)\Big)\right]\right)\\
    &\qquad\qquad\times\Bigg[8\frac{k_1^2k_2^2+k_1^2k_3^2+k_2^2k_3^2}{k_1+k_2+k_3}-2(k_1^3+k_2^3+k_3^3)\\
    &\qquad\qquad\qquad+(k_1+k_2+k_3)(k_1^2+k_2^2+k_3^2)\Bigg]\\
    &\qquad\quad+(2\varepsilon_1+\varepsilon_2)\frac{k_1^4+k_2^4+k_3^4-10\big(k_1^2k_2^2+k_1^2k_3^2+k_2^2k_3^2\big)}{k_1+k_2+k_3}\\
    &\qquad\qquad\times\Bigg[\ln\left(\frac{2k_1}{k_1+k_2+k_3}\right)\frac{k_2+k_3}{-k_1+k_2+k_3}\\
    &\qquad\qquad\qquad+\ln\left(\frac{2k_2}{k_1+k_2+k_3}\right)\frac{k_1+k_3}{k_1-k_2+k_3}\\
    &\qquad\qquad\qquad+\ln\left(\frac{2k_3}{k_1+k_2+k_3}\right)\frac{k_1+k_2}{k_1+k_2-k_3}\Bigg]\\
    &\qquad\quad-4(2\varepsilon_1+\varepsilon_2)\frac{k_1^2k_2^2+k_1^2k_3^2+k_2^2k_3^2}{k_1+k_2+k_3}\\
    &\qquad\quad-\varepsilon_2\Bigg[7(k_1^3+k_2^3+k_3^3)+3k_1k_2k_3-20\frac{k_1^2k_2^2+k_1^2k_3^2+k_2^2k_3^2}{k_1+k_2+k_3}\\
    &\qquad\qquad\qquad-3(k_1+k_2+k_3)(k_1^2+k_2^2+k_3^2)\Bigg]\Bigg\}+\mathcal{O}\big(\varepsilon^3,\tau\big).
\end{split}
\end{align}

\subsection{EoM vertex $\mathscr{O}_9$\label{sec:shift}}
Let us now consider the EoM vertex $\mathscr{O}_9$:
\begin{align}
    \mathscr{O}_9&=\left[\partial_t(2\varepsilon_1e^{3\rho}\dot{\zeta})-2\varepsilon_1e^{\rho}\partial^2\zeta\right]f(\zeta)=M_\Pl^{-2}f(\zt)\cD\zt
\end{align}
with 
\begin{align}
\begin{split}
    f(\zeta)=&\frac{\varepsilon_2}{4}\zeta^2+\frac{1}{\dot{\rho}}\dot{\zeta}\zeta-\frac{1}{4}\frac{e^{-2\rho}}{\dot{\rho}^2}\left[\partial_i\zeta\partial^i\zeta-\partial^{-2}\partial_i\partial_j(\partial^i\zeta\partial^j\zeta)\right]\\
    &+\frac{\varepsilon_1}{2\dot{\rho}}\left[\partial_i\zeta\partial^i\partial^{-2}\dot{\zeta}-\partial^{-2}\partial_i\partial_j(\partial^i\zeta\partial^j\partial^{-2}\dot{\zeta})\right],
    \end{split}
\end{align}
and $\cD$ defined by \eqref{freeeom-D}. We first note that we are not allowed to set $\cD\zt=0$ because $\zt$ is an \emph{off-shell} integration variable of the path integral. We will see shortly that $\cD$ acts on Green functions after Wick contraction providing non-trivial contributions due to the delta-function source terms in the equations of motion of propagators $G^{\pm\pm}$. It turns out~\cite{Braglia:2024zsl,Kawaguchi:2024lsw} that the result is the same as the result of Maldacena's prescription~\cite{Maldacena:2002vr} of shifting $\zt$ to eliminate the EoM vertex $\mathscr{O}_9$.

The three-point function with this vertex is given by
\begin{align}
    \begin{split}
\Bbra\zeta^+(t,\vec{x})\zeta^+(t,\vec{y})\zeta^+(t,\vec{z})\exp\left[
iM_\Pl^2\int_{-\infty}^{t_f}dt'\int d^3\vec{w}\,\left(\mathscr{O}_9^+(t',\vec{w})-\mathscr{O}_9^-(t',\vec{w})\right)\right]\Kket.
    \end{split}
\label{zt^3-O9}
\end{align}
After Wick contraction of the expansion of the exponential, each propagator involves $M_\Pl^{-2}$. When a propagator $G^{\pm\pm}$ gets acted on by $M_\Pl^{-2}\cD$ in $\mathscr{O}_9$, it becomes a delta function of order $M_\Pl^{-2}$ because of the equations of motion \eqref{sec2:eq-G++} and \eqref{sec2:eq-G--}. Namely, the differential operator $M_\Pl^{-2}\cD$ shrinks propagators $G^{\pm\pm}$ to a point, keeping its order $M_\Pl^{-2}$. Taking this into account, the leading order (tree-level) contribution is given by
\begin{align}
&\quad 
iM_\Pl^2\int_{-\infty}^{t_f}\!\!dt'\int d^3w\, 
M_\Pl^{-2}\cD_{t'\vec w}G^{++}(t,\vec z;t',\vec w)\bbra \zt^+(t,\vec x)\zt^+(t,\vec y)f(\zt^+(t',\vec w)) \kket_{M_\Pl^{-4}} \nn\\
&=\bbra \zt^+(t,\vec x)\zt^+(t,\vec y)f(\zt^+(t,\vec z)) \kket_{M_\Pl^{-4}} \nn\\
&=\bbra \zt^+(t,\vec x)\zt^+(t,\vec y)f(\zt^+(t,\vec z)) \kket_{M_\Pl^{-4}}^{\text{no vertex}} \label{ztztf(zt)}
\end{align}
plus permutations in $(\vec x,\vec y,\vec z)$, where we used \eqref{sec2:eq-W}: $\cD G^{\pm\mp}=0$, and in the second equality we used the property that if the four-point correlator involves interaction vertices, it becomes of order $M_\Pl^{-6}$ or smaller. The first line of \eqref{ztztf(zt)} is nothing but $M_\Pl^2\langle\mathscr{O}_9\rangle$.

The last expression in \eqref{ztztf(zt)} does not involve any interaction vertex. It can therefore be regarded as a Gaussian four-point correlator of $\zt^2f(\zt)$. This is a path-integral derivation~\cite{Braglia:2024zsl,Kawaguchi:2024lsw} of the Maldacena's prescription~\cite{Maldacena:2002vr} of shifting $\zt\to\zt+f(\zt)$ in order to eliminate the EoM vertex $\mathscr{O}_9$. The above argument of $M_\Pl$-counting also justifies neglecting contributions with multiple $f(\zt)$ giving rise to Gaussian higher-point correlators: they are of higher order in $M_\Pl^{-2}$.

Let us next compute the last expression of \eqref{ztztf(zt)} explicitly:
\begin{align}
    \begin{split}\label{eq:quartic}
        &\bbra \zt^+(t,\vec x)\zt^+(t,\vec y)f(\zt^+(t,\vec z)) \kket_{M_\Pl^{-4}}^{\text{no vertex}}\\
        &=\int\frac{d^3\vec{k}_1}{(2\pi)^3}\frac{d^3\vec{k}_2}{(2\pi)^3}e^{i\vec{k}_1\cdot(\vec{x}-\vec{z})}e^{i\vec{k}_2\cdot(\vec{y}-\vec{z})}\\
        &\quad\times\Bigg\{\frac{\varepsilon_2}{2}G_{k_1}^{>}(t,t)G_{k_2}^{>}(t,t)+\frac{1}{H}\left[\dot{G}_{k_1}^{>}(t,t)G_{k_2}^{>}(t,t)+G_{k_1}^{>}(t,t)\dot{G}_{k_2}^{>}(t,t)\right]\\
        &\qquad+\frac{e^{-2\rho}}{2H^2}\big(\vec{k}_1\cdot\vec{k}_2\big)G_{k_1}^{>}(t,t)G_{k_2}^{>}(t,t)-\frac{e^{-2\rho}}{2H^2}\frac{\big(k_1^2+\vec{k}_1\cdot\vec{k}_2\big)\big(k_2^2+\vec{k}_1\cdot\vec{k}_2\big)}{|\vec{k}_1+\vec{k}_2|^2}G_{k_1}^{>}(t,t)G_{k_2}^{>}(t,t)\\
        &\qquad+\frac{\varepsilon_1}{2H}\big(\vec{k}_1\cdot\vec{k}_2\big)\left[\frac{1}{k_2^2}G_{k_1}^{>}(t,t)\dot{G}_{k_2}^{>}(t,t)+\frac{1}{k_1^2}\dot{G}_{k_1}^{>}(t,t)G_{k_2}^{>}(t,t)\right]\\
        &\qquad-\frac{\varepsilon_1}{2H}\frac{\big(k_1^2+\vec{k}_1\cdot\vec{k}_2\big)\big(k_2^2+\vec{k}_1\cdot\vec{k}_2\big)}{|\vec{k}_1+\vec{k}_2|^2}\left[\frac{1}{k_2^2}G_{k_1}^{>}(t,t)\dot{G}_{k_2}^{>}(t,t)+\frac{1}{k_1^2}\dot{G}_{k_1}^{>}(t,t)G_{k_2}^{>}(t,t)\right]\Bigg\}.
    \end{split}
\end{align}
Actually, at NLO in Hubble flow parameters and in the late-time limit, only the first term in $f(\zt)$ (i.e. the first term in the curly bracket in \eqref{eq:quartic}) gives a non-vanishing contribution\footnote{A careful reader can be convinced by looking at Appendix \ref{N3LO} where all the ingredients are computed up to N3LO.}. We thus have
\begin{align}
    \begin{split}
        &\quad\bbra \zt^+(t,\vec x)\zt^+(t,\vec y)f(\zt^+(t,\vec z)) \kket_{M_\Pl^{-4}}^{\text{no vertex}}\\
        &=\int\frac{d^3\vec{k}_1}{(2\pi)^3}\frac{d^3\vec{k}_2}{(2\pi)^3}\frac{d^3\vec{k}_3}{(2\pi)^3}e^{i\vec{k}_1\cdot\vec{x}}e^{i\vec{k}_2\cdot\vec{y}}e^{i\vec{k}_3\cdot\vec{z}}(2\pi)^3\delta^{(3)}(\vec{k}_1+\vec{k}_2+\vec{k}_3)\frac{H^4\varepsilon_2}{32\varepsilon_1^2k_1^3k_2^3}\\
        &\quad\times\Big\{1-(4\varepsilon_1+2\varepsilon_2)\big[\gamma_E-1+\frac{1}{2}\ln(-2\tau k_1)+\frac{1}{2}\ln(-2\tau k_2)\big]+2\varepsilon_2+\mathcal{O}\big(\varepsilon^2,\tau)\Big\}.\end{split}
        \end{align}
After adding the other terms obtained by permutations, the contribution of the EoM vertex becomes:
       \begin{align}
    \begin{split}
        \mathscr{A}_{9}&=\big(k_1^3+k_2^3+k_3^3\big)\varepsilon_2\big(1-(4\varepsilon_1+2\varepsilon_2)(\gamma_E-1)+2\varepsilon_2\big)\\
        &\quad-(2\varepsilon_1+\varepsilon_2)\varepsilon_2\Big[\big(k_2^3+k_3^3)\ln(-2\tau k_1)+\big(k_1^3+k_3^3)\ln(-2\tau k_2)\\
        &\qquad\qquad\qquad\qquad+\big(k_1^3+k_2^3)\ln(-2\tau k_3)\Big]+\mathcal{O}\big(\varepsilon^3,\tau\big) \\
        &=\varepsilon_2\Bigg\{\big(k_1^3+k_2^3+k_3^3\big)\Bigg(1+2\varepsilon_2-2(2\varepsilon_1+\varepsilon_2)\left[\gamma_E-1+\ln\Big(-\tau(k_1+k_2+k_3)\Big)\right]\Bigg)\\
        &\qquad~~-(2\varepsilon_1+\varepsilon_2)\Bigg[\big(k_2^3+k_3^3)\ln\left(\frac{2k_1}{k_1+k_2+k_3}\right)+\big(k_1^3+k_3^3)\ln\left(\frac{2k_2}{k_1+k_2+k_3}\right)\\
        &\qquad~~+\big(k_1^3+k_2^3)\ln\left(\frac{2k_3}{k_1+k_2+k_3}\right)\Bigg]\Bigg\}+\mathcal{O}\big(\varepsilon^3,\tau\big).
        \end{split}
        \end{align}

An important remark is in order. In \eqref{zt^3-O9}, we chose the three external $\zt$'s to lie on the contour $C_+$. Let us show that the correlator is independent of the choice of the time contour for the external $\zt$'s. Indeed, one can generalise \eqref{zt^3-O9} to
\begin{align}
    \begin{split}
\Bbra\zeta^c(t,\vec{x})\zeta^a(t,\vec{y})\zeta^b(t,\vec{z})\exp\left[
iM_\Pl^2\int_{-\infty}^{t_f}dt'\int d^3\vec{w}\,\left(\mathscr{O}_9^+(t',\vec{w})-\mathscr{O}_9^-(t',\vec{w})\right)\right]\Kket,
    \end{split}
\label{zt^3-O9-ab}
\end{align}
and consider the case where $\zeta^c$ is contracted with $\cD\zt$ in $\mathscr{O}_9$. When $c=+$, nothing changes from the computation above. When $c=-$, this gives a non-trivial result only when it is contracted with $\cD\zt^-$ in $\mathscr{O}_9^-$. While $\mathscr{O}_9^-$ has the minus sign in contrast to that in front of $\mathscr{O}_9^+$, the delta function source in the equation of motion for $G^{--}$ also has the opposite sign to that for $G^{++}$. Combining the two sign flips gives the same result as in the case $c=+$. We therefore find that \eqref{zt^3-O9-ab} is reduced to
\begin{align}
\bbra \zt^a(t,\vec x)\zt^b(t,\vec y)f(\zt^c(t,\vec z)) \kket_{M_\Pl^{-4}}^{\text{no vertex}}. \label{ztztf(zt)-abc}
\end{align}
We can then conclude that it is independent of the choice of of $a,b,c$ because this involves only equal-time propagators $G^{ab}_k(t,t)$ and therefore we can use 
\begin{align}
G^{+\pm}_k(t,t)=G^{-\pm}_k(t,t)=G^>_k(t,t)=G^<_k(t,t).
\end{align}

\subsection{Total result}
Finally, we sum up the all the contributions and decompose $\mathscr{A}$ order by order as
\begin{align}
    \mathscr{A}=\mathscr{A}_{\text{LO}}+\mathscr{A}_{\text{NLO}}+\mathcal{O}\big(\varepsilon^3\big)
\end{align}
with
\begin{align}
    \mathscr{A}_{\text{LO}}=\varepsilon_1K_1\big(k_1,k_2,k_3\big)+\varepsilon_2K_2\big(k_1,k_2,k_3\big)+\mathcal{O}(\tau)
\end{align}
and
\begin{align}\begin{split}
    \mathscr{A}_{\text{NLO}}=&\left[\gamma_E-1+\ln\Big(\!-\tau(k_1+k_2+k_3)\Big)\right]\Big(\varepsilon_1^2K_3\big(k_1,k_2,k_3\big)\\
    &\qquad+\varepsilon_1\varepsilon_2K_4\big(k_1,k_2,k_3\big)+\varepsilon_2^2K_5\big(k_1,k_2,k_3\big)+\varepsilon_2\varepsilon_3K_6\big(k_1,k_2,k_3\big)\Big)\\
    &~+\big(2\varepsilon_1+\varepsilon_2\big)\ln\left(\frac{2k_1}{k_1+k_2+k_3}\right)\Big(\varepsilon_1K_7\big(k_1,k_2,k_3\big)+\varepsilon_2K_8\big(k_1,k_2,k_3\big)\Big)\\
    &~+\big(2\varepsilon_1+\varepsilon_2\big)\ln\left(\frac{2k_2}{k_1+k_2+k_3}\right)\Big(\varepsilon_1K_7\big(k_2,k_1,k_3\big)+\varepsilon_2K_8\big(k_2,k_1,k_3\big)\Big)\\
    &~+\big(2\varepsilon_1+\varepsilon_2\big)\ln\left(\frac{2k_3}{k_1+k_2+k_3}\right)\Big(\varepsilon_1K_7\big(k_3,k_2,k_1\big)+\varepsilon_2K_8\big(k_3,k_2,k_1\big)\Big)\\
    &~+\varepsilon_1^2K_9\big(k_1,k_2,k_3\big)+\varepsilon_1\varepsilon_2K_{10}\big(k_1,k_2,k_3\big)\\
    &~+\varepsilon_2^2K_{11}\big(k_1,k_2,k_3\big)+\varepsilon_2\varepsilon_3K_{12}\big(k_1,k_2,k_3\big)+\mathcal{O}(\tau),\end{split}
\end{align}
where the $K_i$ functions are given by:
\begin{align}
    \begin{split}
    K_1\big(k_1,k_2,k_3\big)&=8\frac{k_1^2k_2^2+k_1^2k_3^2+k_2^2k_3^2}{k_1+k_2+k_3}-2(k_1^3+k_2^3+k_3^3)\\
    &\qquad+(k_1+k_2+k_3)(k_1^2+k_2^2+k_3^2),
    \end{split} \label{K1}\\
    K_2\big(k_1,k_2,k_3\big)&=k_1^3+k_2^3+k_3^3, \label{K2}\\
    K_3\big(k_1,k_2,k_3\big)&=-4K_1\big(k_1,k_2,k_3\big), \label{K3}\\
    K_4\big(k_1,k_2,k_3\big)&=-K_1\big(k_1,k_2,k_3\big)-4K_2\big(k_1,k_2,k_3\big), \label{K4}\\
    K_5\big(k_1,k_2,k_3\big)&=-2K_2\big(k_1,k_2,k_3\big), \label{K5}\\
    K_6\big(k_1,k_2,k_3\big)&=K_2\big(k_1,k_2,k_3\big), \label{K6}\\
    K_7\big(k_1,k_2,k_3\big)&=\frac{k_2+k_3}{-k_1+k_2+k_3}\Big(2k_1k_2k_3-K_1\big(k_1,k_2,k_3\big)\Big), \label{K7}\\
    K_8\big(k_1,k_2,k_3\big)&=k_1^3-K_2\big(k_1,k_2,k_3\big), \label{K8}\\
    K_9\big(k_1,k_2,k_3\big)&=2k_1k_2k_3-K_1\big(k_1,k_2,k_3\big),\\
    \begin{split}K_{10}\big(k_1,k_2,k_3\big)&=2K_1\big(k_1,k_2,k_3\big)-3K_2\big(k_1,k_2,k_3\big)-3k_1k_2k_3\\
    &\qquad+(k_1+k_2+k_3)(k_1^2+k_2^2+k_3^2),\end{split}\\
    K_{11}\big(k_1,k_2,k_3\big)&=2K_2\big(k_1,k_2,k_3\big),\\
    \begin{split}K_{12}\big(k_1,k_2,k_3\big)&=k_1k_2k_3+K_2\big(k_1,k_2,k_3\big) 
    -(k_1+k_2+k_3)(k_1^2+k_2^2+k_3^2).\end{split}
\end{align}

We now compare our expression for the bispectrum with that in~\cite{Burrage:2011hd}. We verified that the squeezed limit of the bispectra coincide. However, in the equilateral shape $k\equiv k_1=k_2=k_3$, the two expressions do not agree. Indeed, our amplitude reads
\begin{align}\label{Aeq-our}
\begin{split}
k^{-3}\mathscr{A}^\text{eq}_{\text{our}}
&=11\vep_1+3\vep_2+[\ga_\text{E}-1+\ln(-3k\tau)](-44\vep_1^2-23\vep_1\vep_2-6\vep_2^2+3\vep_2\vep_3) \\
&\quad-9\vep_1^2+19\vep_1\vep_2+6\vep_2^2-5\vep_2\vep_3
+6(2\vep_1+\vep_2)(9\vep_1+\vep_2)\ln\frac{3}{2},
\end{split}
\end{align}
while the amplitude of \cite{Burrage:2011hd} reads\footnote{For the comparison, we used the correspondence of the parameters of \cite{Burrage:2011hd} to ours: $c_s=1, s=\lambda=0$, $\varepsilon\to\varepsilon_1, \eta\to\varepsilon_2, \xi\to\varepsilon_3$ and $k_*\to -1/\tau$. Recall that the pivot time $\tau^*$ has been chosen to be $\tau$, as explained below \eqref{E3-exp}.}
\begin{align}\label{Aeq-BRS}
\begin{split}
k^{-3}\mathscr{A}_{\text{BRS}}^\text{eq}
=k^{-3}\mathscr{A}^\text{eq}_{\text{our}}
+\frac{1}{36}&\Big[-\vep_2(11\vep_1+3\vep_3)(\ga_\text{E}-1+\ln(-3k\tau)) \\
&\quad+9\vep_1^2+3\vep_1\vep_2+5\vep_2\vep_3-32(2\vep_1+\vep_2)\ln\frac{3}{2}
\Big]\,.
\end{split}
\end{align}

Let us check the time independence of the bispectrum in the equilateral shape $\frac{H^4}{32k^6\vep_1^2}\mathscr{A}^\text{eq}$, which is expected to hold from the adiabaticity of the curvature fluctuation $\zeta$~\cite{Weinberg:2003sw,Lyth:2004gb,Langlois:2005qp}. The $\tau$-dependence arises not only from the explicit logarithm $\log(-k\tau)$ but also from the implicit time dependence of the parameters $H,\vep_i$ defined at the pivot time $\tau$. In order to manifest their $\tau$-dependence up to NLO, we can use the slow-roll expansions of $H,\vep_i$ in Appendix~\ref{Appendixexpansion} around another pivot time $\tau^*$, offering more logarithms $\log(\tau/\tau^*)$. We can then show that the logarithm $\log(-k\tau)$ in our $\mathscr{A}^\text{eq}_{\text{our}}$ cancels those from the parameters $H,\vep_i$ up to NLO, so that our three-point function is indeed independent of $\tau$ up to NLO. However, the amplitude $\mathscr{A}^\text{eq}_{\text{BRS}}$ of~\cite{Burrage:2011hd} depends on $\tau$ due to the $\log(-k\tau)$ at NLO in the difference between the two results, given in the first line of \eqref{Aeq-BRS}.

\section{Amplitude of the non-gaussianity}
The non-gaussianity was initially parameterised through a non-linear correction to a Gaussian field $\zeta_g$:
\begin{align}\label{fNL}
    \zeta=\zeta_g+\frac{3}{5}f_{\text{NL}}\zeta_g^2.
\end{align}
By defining the power spectrum\footnote{Note that the power spectrum is usually defined as $\frac{k^3}{2\pi^2}\mathcal{P}$.} and the bispectrum by
\begin{align}
    \langle\zeta(\vec{k}_1)\zeta(\vec{k}_2)\rangle&=(2\pi)^3\delta^{(3)}(\vec{k}_1+\vec{k}_2)\mathcal{P}(\vec{k}_1),\\
    \langle\zeta(\vec{k}_1)\zeta(\vec{k}_2)\zeta(\vec{k}_3)\rangle&=(2\pi)^3\delta^{(3)}(\vec{k}_1+\vec{k}_2+\vec{k}_3)\mathcal{B}(\vec{k}_1,\vec{k}_2,\vec{k}_3),
\end{align}
the shift \eqref{fNL} would give the relation
\begin{align}
    \mathcal{B}(\vec{k}_1,\vec{k}_2,\vec{k}_3)=\frac{6}{5}\left(\mathcal{P}(\vec{k}_1)\mathcal{P}(\vec{k}_2)+\mathcal{P}(\vec{k}_1)\mathcal{P}(\vec{k}_3)+\mathcal{P}(\vec{k}_2)\mathcal{P}(\vec{k}_3)\right)f_{\text{NL}}.
\end{align}
In the slow-roll limit, we have
\begin{align}
    \mathcal{P}(\vec{k})=\frac{H^2}{4\varepsilon_1 k^3}
\end{align}
and we would therefore find
\begin{align}
    \mathcal{B}(\vec{k}_1,\vec{k}_2,\vec{k}_3)=\frac{1}{32\varepsilon_1^2}\frac{H^4}{k_1^3k_2^3k_3^3}\cdot\frac{12}{5}\big(k_1^3+k_2^3+k_3^3\big)f_{\text{NL}}.
\end{align}
From our results, we see that the naive shift \eqref{fNL} does not provide the correct shape for the bispectrum. We can however encapsulate the amplitude of the non-gaussianity in the same way by defining a $k$-dependent $f_{\text{NL}}$ as
\begin{align}
    f_{\text{NL}}\equiv\frac{5}{12}\frac{\mathscr{A}}{k_1^3+k_2^3+k_3^3}.
\end{align}
We can then define a $f_{\text{NL}}$ at each order as
\begin{align}
    f_{\text{NL}}=f_{\text{NL,LO}}+f_{\text{NL,NLO}}+\mathcal{O}\big(\varepsilon^3\big)
\end{align}
and get
\begin{align}\label{fNLLO}
    f_{\text{NL,LO}}=&\varepsilon_1\tilde{K}_1\big(k_1,k_2,k_3\big)+\varepsilon_2\tilde{K}_2\big(k_1,k_2,k_3\big),\\
    \begin{split}f_{\text{NL,NLO}}=&\left[\gamma_E-1+\ln\Big(\!-\tau(k_1+k_2+k_3)\Big)\right]\Big(\varepsilon_1^2\tilde{K}_3\big(k_1,k_2,k_3\big)\\
    &+\varepsilon_1\varepsilon_2\tilde{K}_4\big(k_1,k_2,k_3\big)+\varepsilon_2^2\tilde{K}_5\big(k_1,k_2,k_3\big)+\varepsilon_2\varepsilon_3\tilde{K}_6\big(k_1,k_2,k_3\big)\Big)\\
    &+\big(2\varepsilon_1+\varepsilon_2\big)\Big(\varepsilon_1\tilde{K}_{7'}\big(k_1,k_2,k_3\big)+\varepsilon_2\tilde{K}_{8'}\big(k_1,k_2,k_3\big)\Big)\\
    &+\varepsilon_1^2\tilde{K}_9\big(k_1,k_2,k_3\big)+\varepsilon_1\varepsilon_2\tilde{K}_{10}\big(k_1,k_2,k_3\big)\\
    &+\varepsilon_2^2\tilde{K}_{11}\big(k_1,k_2,k_3\big)+\varepsilon_2\varepsilon_3\tilde{K}_{12}\big(k_1,k_2,k_3\big),\end{split}
    \label{fNLNLO}
\end{align}
where the functions $\tilde{K}_i$ are defined as rescaled $K_i$:
\begin{align}\label{Ktilde}
    \tilde{K}_i\big(k_1,k_2,k_3\big)\equiv\frac{5}{12}\frac{K_i\big(k_1,k_2,k_3\big)}{k_1^3+k_2^3+k_3^3}.
\end{align}
We have also introduced the function $\tilde{K}_{7'}$
\begin{align}\label{Ktilde7}
    \begin{split}\tilde{K}_{7'}\big(k_1,k_2,k_3\big)=&\ln\left(\frac{2k_1}{k_1+k_2+k_3}\right)\tilde{K}_7\big(k_1,k_2,k_3\big)\\
    &+\ln\left(\frac{2k_2}{k_1+k_2+k_3}\right)\tilde{K}_7\big(k_2,k_1,k_3\big)\\
    &+\ln\left(\frac{2k_3}{k_1+k_2+k_3}\right)\tilde{K}_7\big(k_3,k_2,k_1\big)
    \end{split}
\end{align}
as well as $\tilde{K}_{8'}$ defined similarly. 

All the above functions are actually $\mathcal{O}(1)$ quantities for any values of the momenta, except $\tilde{K}_{7'}$ and $\tilde{K}_{8'}$. Their values are found to be
\begin{equation}
    \begin{array}{cclccclcccl}\label{ranges}
       \tilde{K}_1&\in&\left[\frac{5}{6},\frac{55}{36}\right],&\qquad&\tilde{K}_2&=&\frac{5}{12},&\qquad&\tilde{K}_3&\in&\left[-\frac{55}{9},-\frac{10}{3}\right],\\
       \tilde{K}_4&\in&\left[-\frac{115}{36},-\frac{5}{2}\right],&\qquad&\tilde{K}_5&=&-\frac{5}{6},&\qquad&\tilde{K}_6&=&\frac{5}{12},\\
       \tilde{K}_{7'}&\in&\left[\frac{3}{8}(1+\ln(64)),\infty\right),&\qquad&\tilde{K}_{8'}&\in&\left[\frac{5}{6}\ln\right(\frac{3}{2}\left),\infty\right),&\qquad&\tilde{K}_9&\in&\left[-\frac{5}{4},-\frac{3}{4}\right],\\
       \tilde{K}_{10}&\in&\left[\frac{5}{54}\big(4\sqrt{13}-1\big),\frac{95}{36}\right],&\qquad&\tilde{K}_{11}&=&\frac{5}{6},&\qquad&\tilde{K}_{12}&\in&\left[-\frac{25}{36},-\frac{5}{12}\right],
    \end{array}
\end{equation}
and the non-constant ones are plotted in Fig. \ref{fig:Ki}. The divergence of $\tilde{K}_{7'}$ and $\tilde{K}_{8'}$ occurs in the squeezed limit where  one of the momenta goes to zero (corresponding to the upper left corner of the plots of Fig. \ref{fig:Ki}).
\begin{figure}[t!]
    \begin{subfigure}[t]{0.5\textwidth}
    \centering\includegraphics[width=0.8\linewidth]{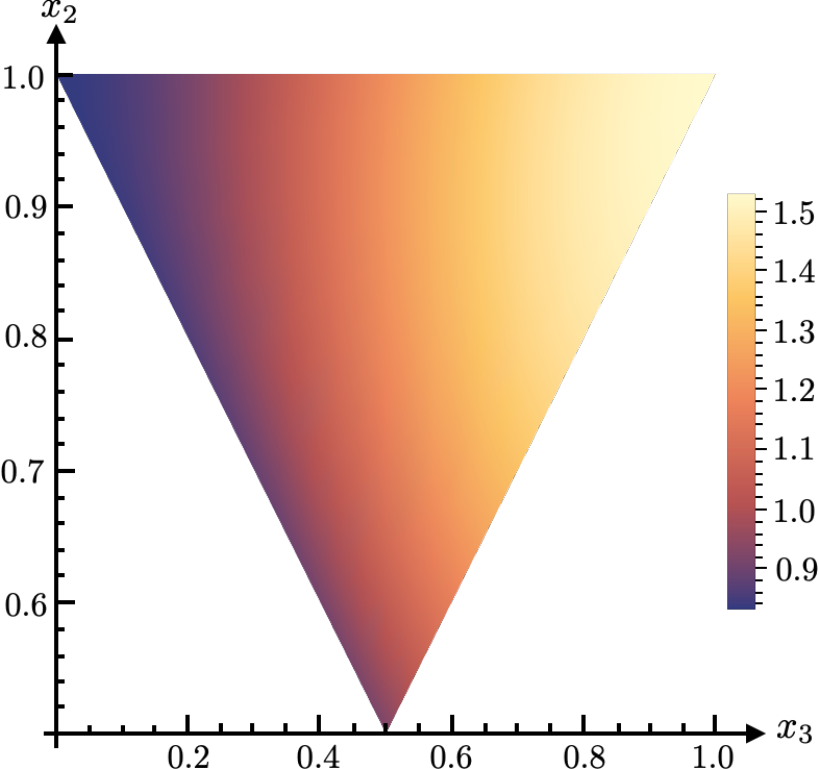}
    \subcaption{$\tilde{K}_1(1,x_2,x_3)$}
    \label{fig:K1}
    \end{subfigure}
    \begin{subfigure}[t]{0.5\textwidth}
    \centering\includegraphics[width=0.8\linewidth]{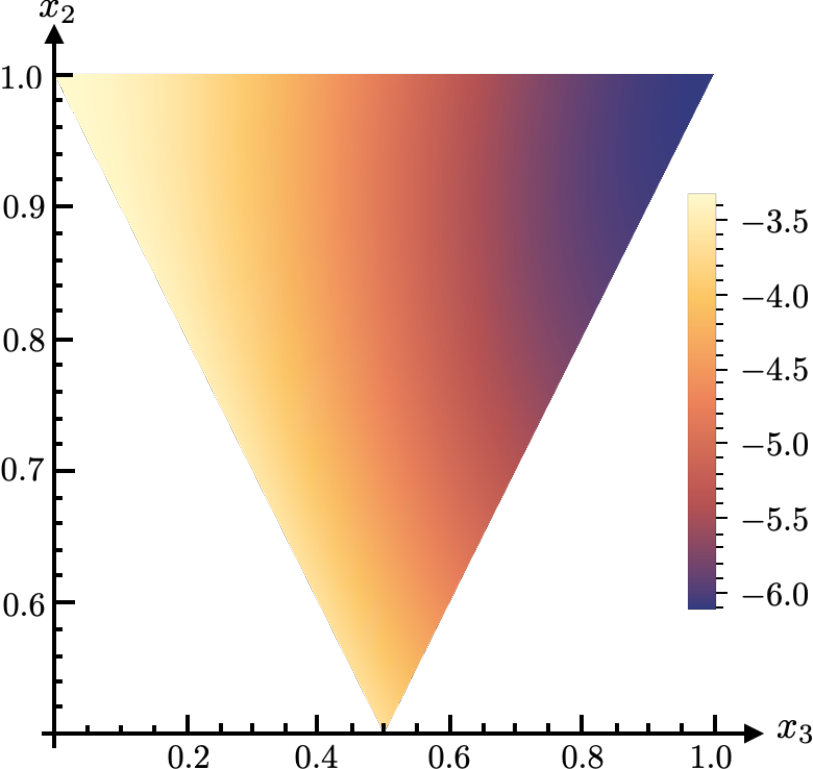}
    \subcaption{$\tilde{K}_3(1,x_2,x_3)$}
    \label{fig:K3}
    \end{subfigure}
    \begin{subfigure}[t]{0.5\textwidth}
    \centering\includegraphics[width=0.8\linewidth]{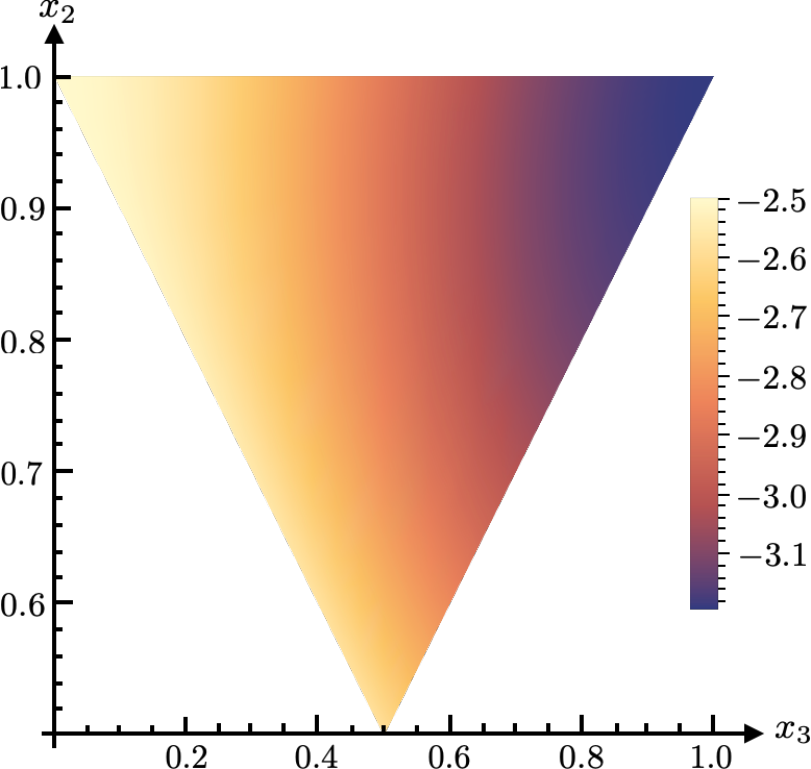}
    \subcaption{$\tilde{K}_4(1,x_2,x_3)$}
    \label{fig:K4}
    \end{subfigure}
    \begin{subfigure}[t]{0.5\textwidth}
    \centering\includegraphics[width=0.8\linewidth]{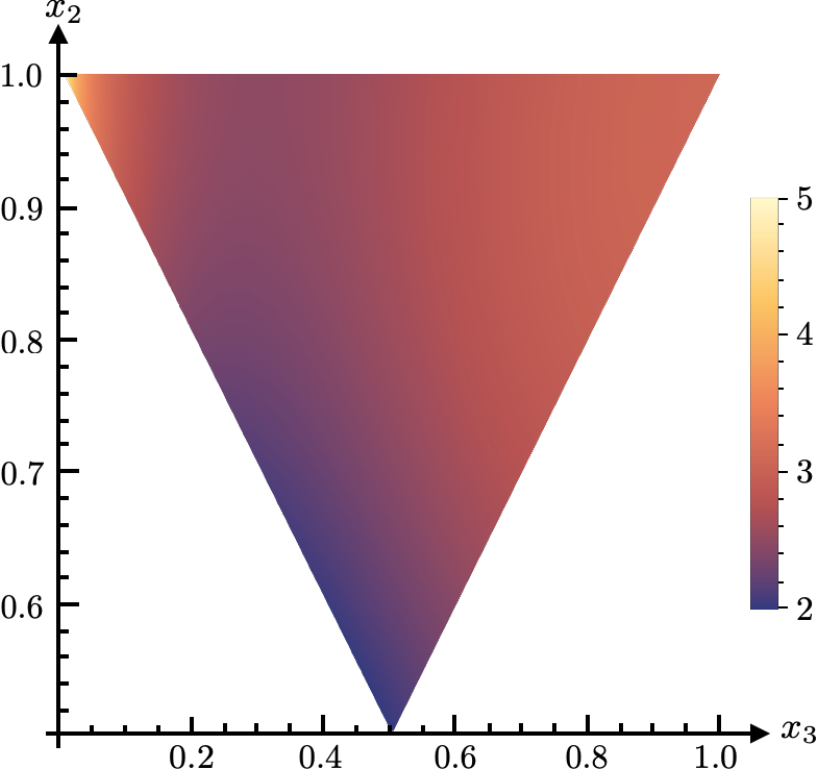}
    \subcaption{$\tilde{K}_{7'}(1,x_2,x_3)$}
    \label{fig:K7}
    \end{subfigure}
 \end{figure}

\begin{figure}[t!]\ContinuedFloat
     \begin{subfigure}[t]{0.5\textwidth}
    \centering\includegraphics[width=0.8\linewidth]{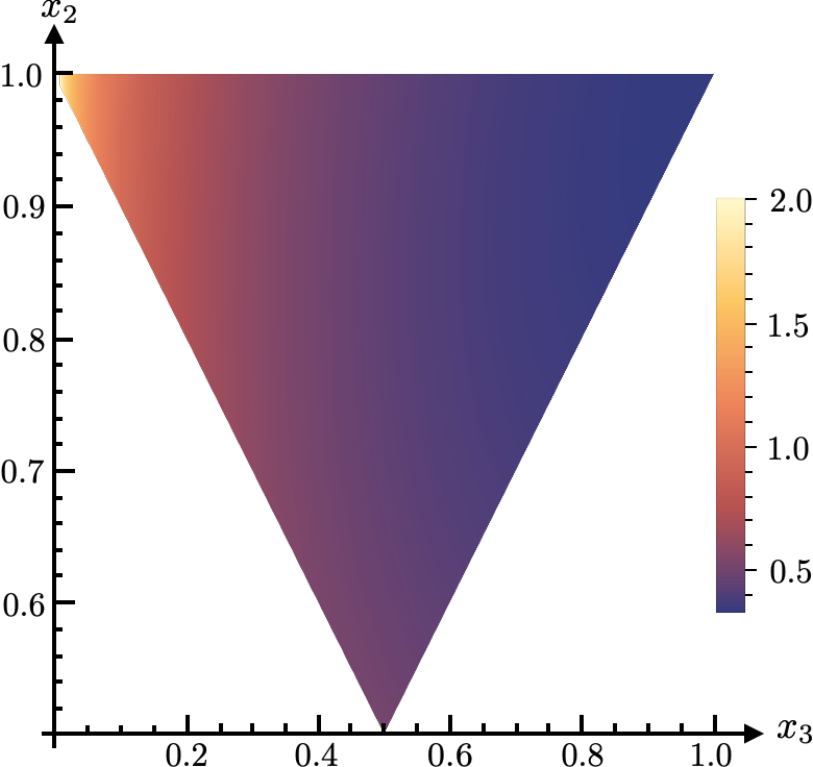}
    \subcaption{$\tilde{K}_{8'}(1,x_2,x_3)$}
    \label{fig:K8}
    \end{subfigure}
    \begin{subfigure}[t]{0.5\textwidth}
    \centering\includegraphics[width=0.8\linewidth]{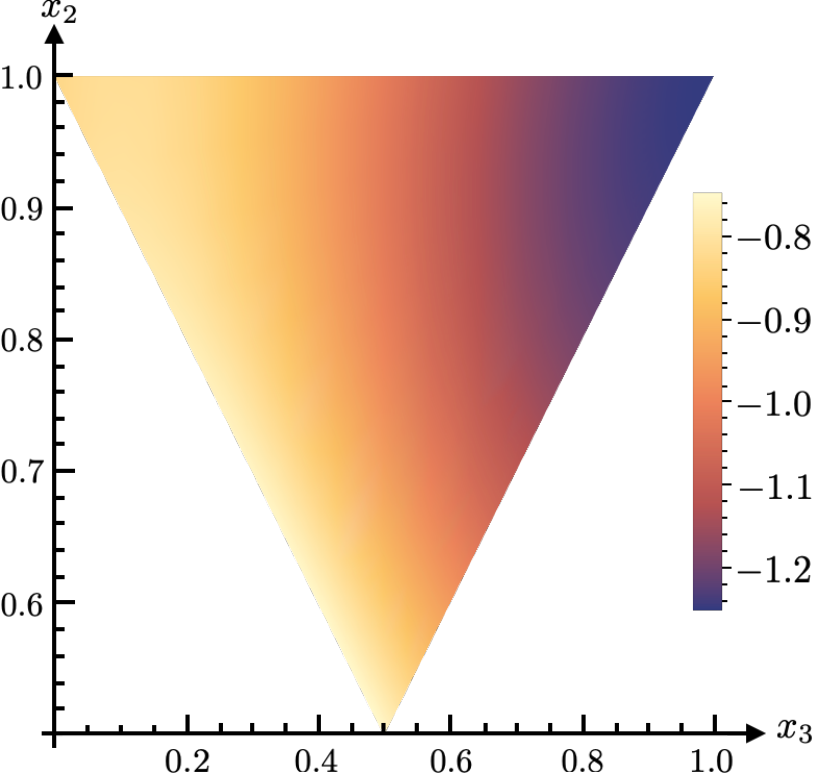}
    \subcaption{$\tilde{K}_9(1,x_2,x_3)$}
    \label{fig:K9}
    \end{subfigure}
  \begin{subfigure}[t]{0.5\textwidth}
    \centering\includegraphics[width=0.8\linewidth]{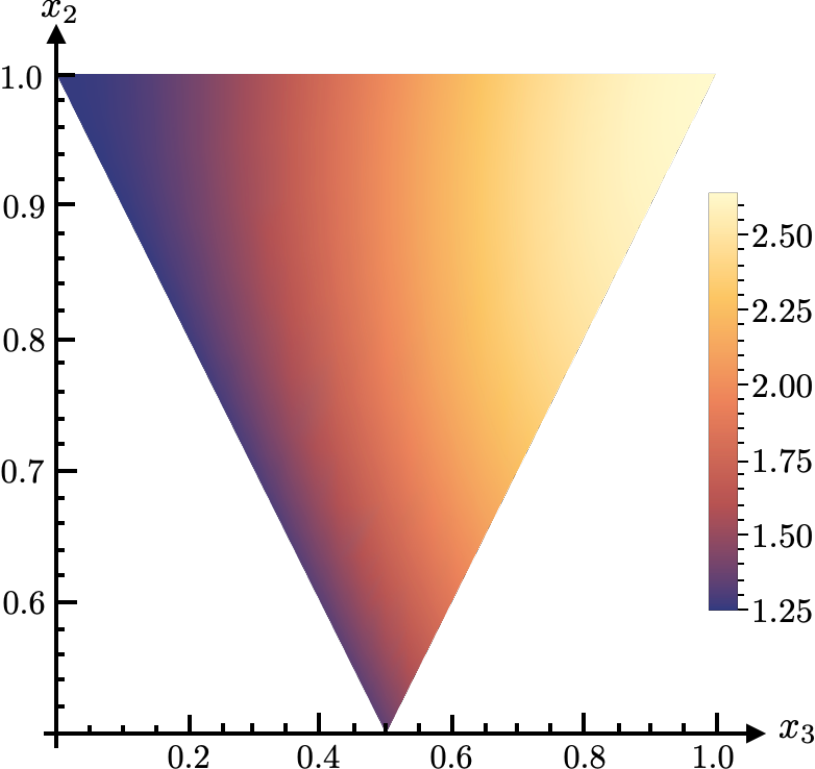}
    \subcaption{$\tilde{K}_{10}(1,x_2,x_3)$}
    \label{fig:K10}
    \end{subfigure}
    \begin{subfigure}[t]{0.5\textwidth}
    \centering\includegraphics[width=0.8\linewidth]{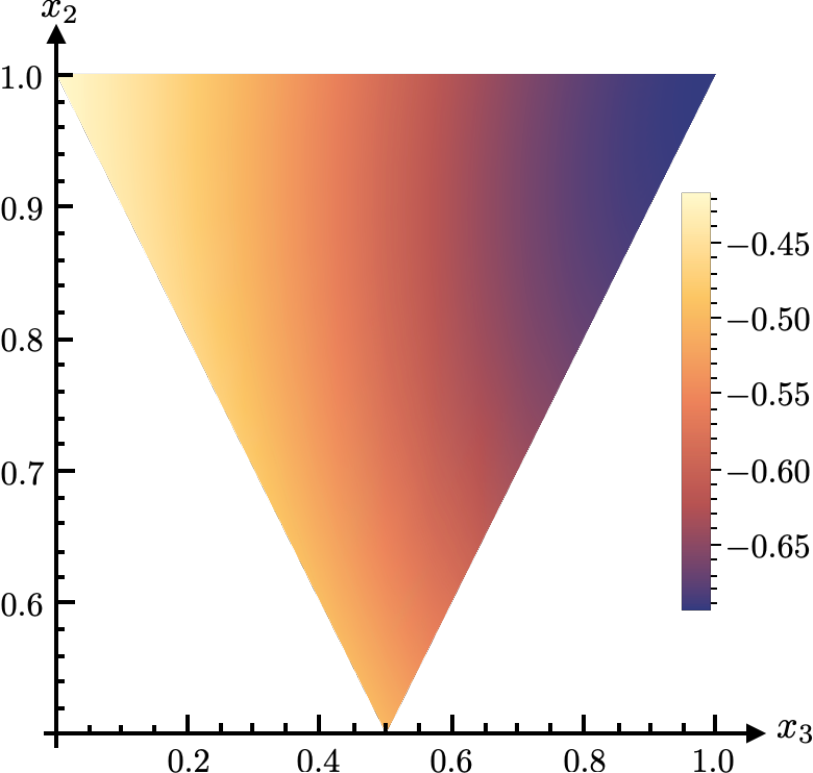}
    \subcaption{$\tilde{K}_{12}(1,x_2,x_3)$}
    \label{fig:K12}
    \end{subfigure}
  \caption{Values of the functions $\tilde{K}_i(k_1,k_2,k_3)$ plotted for the rescaled momenta $x_1=1,\ x_2=k_2/k_1$ and $x_3=k_3/k_1$. The momenta are chosen to be ordered as $x_3\leq x_2\leq x_1$ and obey the triangle inequality $x_2+x_3\geq 1$.}
  \label{fig:Ki}
\end{figure}

\section{Numerical estimate\label{sec:numest}}

To make a numerical estimate of our results summarized in the previous section, we first use observational constraints on the slow-roll parameters. Indeed $\varepsilon_1$ and $\varepsilon_2$ determine the spectral index of scalar perturbations $n_s$ and the tensor-to-scalar ratio $r$ by
\begin{align}
1-n_s=\varepsilon_2+2\varepsilon_1\simeq 6\epsilon_V-2\eta_V; \quad r=16\varepsilon_1\,,
\end{align}
constrained by observations to be~\cite{ParticleDataGroup:2024cfk}\footnote{Strictly speaking, these values are defined at the pivot scale $k^* \simeq 0.05\,\mathrm{Mpc}^{-1}$. Since their variations with scale is negligible for our purposes, we adopt these values to fix the relevant orders of magnitude, which is sufficient for the present discussion.}
\begin{align}
n_s=0.965\pm 0.004;\quad r<0.036\ \text{at}\ 95\%\ \text{CL}\,.
\end{align}
It follows that $\varepsilon_1\lesssim 2\times 10^{-3}$ and can be neglected, while $\varepsilon_2\simeq 0.04$ \cite{Pajer:2016ieg}. From \eqref{fNLLO}, one then obtains
\begin{align}\label{fNLLO2}
f_\text{NL,LO}\simeq \frac{5}{12}\varepsilon_2\simeq 0.02\,,
\end{align}
where we used \eqref{Ktilde} and \eqref{K2}.

At next to leading order, dominant contributions come from the logarithm in \eqref{fNLNLO}, from ${\tilde K}_{7'}, {\tilde K}_{8'}$ due to \eqref{Ktilde7} and the denominator in \eqref{K7} and similarly for ${\tilde K}_{8'}$, as well as from $\tilde{K}_{11}$ and $\tilde{K}_{12}$ (see also the ranges of ${\tilde K}$'s in \eqref{ranges}). One then finds\footnote{A more detailed analysis of the $f_{\text{NL}}$  is provided in Appendix~\ref{sec:fNLs}.}:
\begin{align}
&f_\text{NL,NLO}\simeq\varepsilon_2 (\varepsilon_3-2\varepsilon_2)\tilde{K}_2\left[\gamma_E-1+\ln\Big(\!-\tau(k_1+k_2+k_3)\Big)\right] \nn\\
&~+ \varepsilon_2\left[ \ln\left(\frac{2k_1}{k_1+k_2+k_3}\right) \left(\varepsilon_2 \tilde{K}_8\big(k_1,k_2,k_3\big)
+\varepsilon_1\tilde{K}_7\big(k_1,k_2,k_3\big)\right)+(k_1\leftrightarrow k_2)+(k_1\leftrightarrow k_3)\right] \nn\\
&~+\varepsilon_2\big(\varepsilon_2\tilde{K}_{11}+\varepsilon_3\tilde{K}_{12}\big),
\end{align}
where we neglected $\varepsilon_1$ compared to $\varepsilon_{2,3}$. The logarithm in the second line is large when $k_1$ is small, in which case ${\tilde K}_{8}$ can be replaced by $-{\tilde K}_{2}$ (see \eqref{K8}). Moreover from \eqref{K7}, in the limit $k_1\to 0$,
${\tilde K}_{7}$ can be replaced by $-2{\tilde K}_{2}$, since $k_2\simeq k_3$ and $K_1\simeq 2K_2$ (see \eqref{K1}). This contribution is however suppressed by the coefficient $\varepsilon_1$ and can be neglected. On the other hand, ${\tilde K}_{7}$ is enhanced when the denominator in \eqref{K7} vanishes for, say, $k_1\simeq k_2+k_3$. However, in this limit the logarithm vanishes, canceling the pole. It follows that
\begin{align}\begin{split}\label{fNLNLO2}
f_\text{NL,NLO}\simeq& f_\text{NL,LO}\Bigg\{ (\varepsilon_3-2\varepsilon_2)\left[\gamma_E-1+\ln\Big(\!-\tau(k_1+k_2+k_3)\Big)\right]\\
&\qquad\qquad-\varepsilon_2 \ln\left(\frac{2k_1}{k_1+k_2+k_3}\right)_\text{sq}+(2\varepsilon_2-\alpha\varepsilon_3) \Bigg\} \,,\end{split}
\end{align}
where we used \eqref{fNLLO2} to factorize the leading order contribution and we introduced the order one parameter $\alpha=-\tilde{K}_{12}/\tilde{K}_2\in[1,5/3]$, the minimum and maximum being reached for the squeezed limit  and for the equilateral case respectively. The subscript `sq' stands for the squeezed limit $k_1\to 0$ and $k_2\simeq k_3$, while the other terms contribute to all shapes of the 3-point function.

We now turn to the evaluation of the two logarithmic terms, $\ln(-\tau k_t)$ and $\ln(2k_i/k_t)$, with $k_t=k_1+k_2+k_3$. The first logarithm would naively diverge in the late-time limit $-\tau\rightarrow0$, while the second becomes large in the squeezed limit where one of the $k_i$ tends to zero. Although these terms could in principle induce large corrections to the amplitude of the non-Gaussianity, they do not produce dominant modifications to its overall magnitude. Nevertheless, they may still play a significant role in shaping the detailed momentum dependence of the bispectrum, as we discuss below.

To examine the first logarithm, we note that cosmological correlation functions are typically evaluated at the horizon exit, when the fluctuations freeze. In the case of the bispectrum, the three modes cross the horizon at slightly different times, and it is therefore appropriate to evaluate the result at the last horizon exit, when all momenta satisfy $k_i \leq aH$. This corresponds to
\begin{align}\label{kimax-aH}
    k_{i,\max}=aH=-\frac{1}{\tau}+\mathcal{O}(\varepsilon),
\end{align}
where $k_{i,\max}$ is the maximum out of $k_1,k_2,k_3$.
Evaluated at that time, the first logarithm then becomes
\begin{align}
    \ln\left(\frac{k_1+k_2+k_3}{k_{i,\max}}\right)+\mathcal{O}(\varepsilon)\in\big[\ln(2),\ln(3)\big],
\end{align}
where its minimal and maximal values are obtained for the folded and the equilateral cases, respectively. 
We can further evaluate the full bracket to be in the range\footnote{
    Alternative evaluation times have been proposed in the literature. 
    For example, \cite{Pajer:2016ieg} chooses $\tau$ such that this bracket vanishes.}
\begin{equation}\label{eq:betabracket}
    \begin{array}{ccccc}
    0.27&\lesssim&\left[\gamma_E-1+\ln\Big(\!-\tau(k_1+k_2+k_3)\Big)\right]&\lesssim&0.68.
    \end{array}
\end{equation}

Regarding the second logarithm, only a finite range of momenta is experimentally accessible. 
Because the bispectrum can be measured only over this range, there exists a minimum wavenumber even in the squeezed limit, set by the size of the observable universe. 
In practice, the comoving momenta probed by CMB scalar fluctuations span roughly three orders of magnitude, 
such that in the squeezed limit we can have at most\footnote{
    For a discussion of the experimental sensitivity in the squeezed limit, see, for example, \cite{Kenton:2015lxa}. 
    Moreover, we consider here the full range of scales probed by Planck, from multipoles $l=2$ to $l=2500$. 
    It is important to note, however, that the CMB exhibits anomalies at large scales, 
    and that the uncertainties are particularly large at low multipoles, 
    where scale invariance cannot be assumed \cite{Schwarz:2015cma}.} 
$\ln\big(k_\text{min}/k_\text{max}\big)\simeq -7$.

We can now make an estimate of the third Hubble flow parameter $\varepsilon_3$ related to the third derivative of the potential $\hat\xi_V$ by \eqref{hatxiV}, leading to $\varepsilon_3\simeq -2(\sqrt{2\varepsilon_1}/\varepsilon_2)\hat\xi_V=-2\xi_V^2/\varepsilon_2$. It follows that $\varepsilon_3$ can be as large as $0.3-0.5$~\cite{Ballardini:2024irx} keeping the slow-roll parameter $\hat\xi_V$ small (of order 0.1), within the validity of the slow-roll expansion. By re-writting \eqref{fNLNLO2} as
\begin{align}\label{fNLNLO3}
f_\text{NL,NLO}\simeq& f_\text{NL,LO}\Bigg\{ \varepsilon_2\left(2(1-\beta)-\ln\left(\frac{2k_1}{k_1+k_2+k_3}\right)_\text{sq}\right)-\varepsilon_3(\alpha-\beta) \Bigg\} \,
\end{align}
with $\beta$ the bracket \eqref{eq:betabracket} and notincing that $2(1-\beta)\sim\mathcal{O}(1)$ and $(\alpha-\beta)\sim\mathcal{O}(1)$, we see that the NLO contribution to the $f_\text{NL}$ parameter could be as large as the LO contribution. Note that the  sign of $\varepsilon_3$ is also model dependent.

As a concrete example, below we investigate the range of values of $\varepsilon_3$ in a class of hilltop models called `inflation by supersymmetry breaking' where the inflaton is identified with the supersymmetric partner of the goldstino, in the presence of a $U(1)$ that gauges the R-symmetry~\cite{Antoniadis:2017gjr, Antoniadis:2018cpq}. For the sake of computational simplicity, we choose the D-term realisation alone, where the scalar potential has a supersymmetric minimum with vanishing energy and a flat region around the maximum at the origin where $U(1)$ is restored and inflation takes place~\cite{Antoniadis:2018cpq}.
The effective supergravity has vanishing superpotential and a K\"ahler potential
\begin{align}\label{Kahler}
K(X,{\bar X})=X{\bar X}+A(X{\bar X})^2+B(X{\bar X})^3\,,
\end{align}
where $X$ is the complexified inflaton and $A,B$ are parameters representing higher order corrections to canonical kinetic terms $(A,B < 1)$ and we have set $M_\text{Pl}=1$. Since $X$ is charged under the $U(1)$ with charge $q$ the scalar potential has only D-term contribution:
\begin{align}\label{ScalarPotential}
V=\frac{q^2}{2}D^2\quad;\quad D=b+\rho^2+2A\rho^4+3B\rho^6+b'\rho^{\frac{2}{3}b}e^{K(\rho,\rho)/3}\,,
\end{align}
where $\rho=|X|$ and $b, b'$ are two Fayet-Iliopoulos parameters~\cite{Antoniadis:2018cpq}. The phase of $X$ is absorbed by the $U(1)$ which becomes massive away from the origin $(\langle\rho\rangle\ne 0)$. The requirement that the potential and all its derivatives are regular at the origin fixes $b=3$, while the requirement that the origin is maximum implies $b'<-1$. Thus, the model has one scale $q$ (that rescales the potential) and three parameters $A,B,b'$. Inflation takes place around the origin which has a plateau when $b'+1$ is nearly vanishing. Imposing the spectral index to have the observed value, leads to a constraint in the parameter space which is reduced to two dimensional.

\begin{figure}[t!]
    \centering\includegraphics[width=0.8\linewidth]{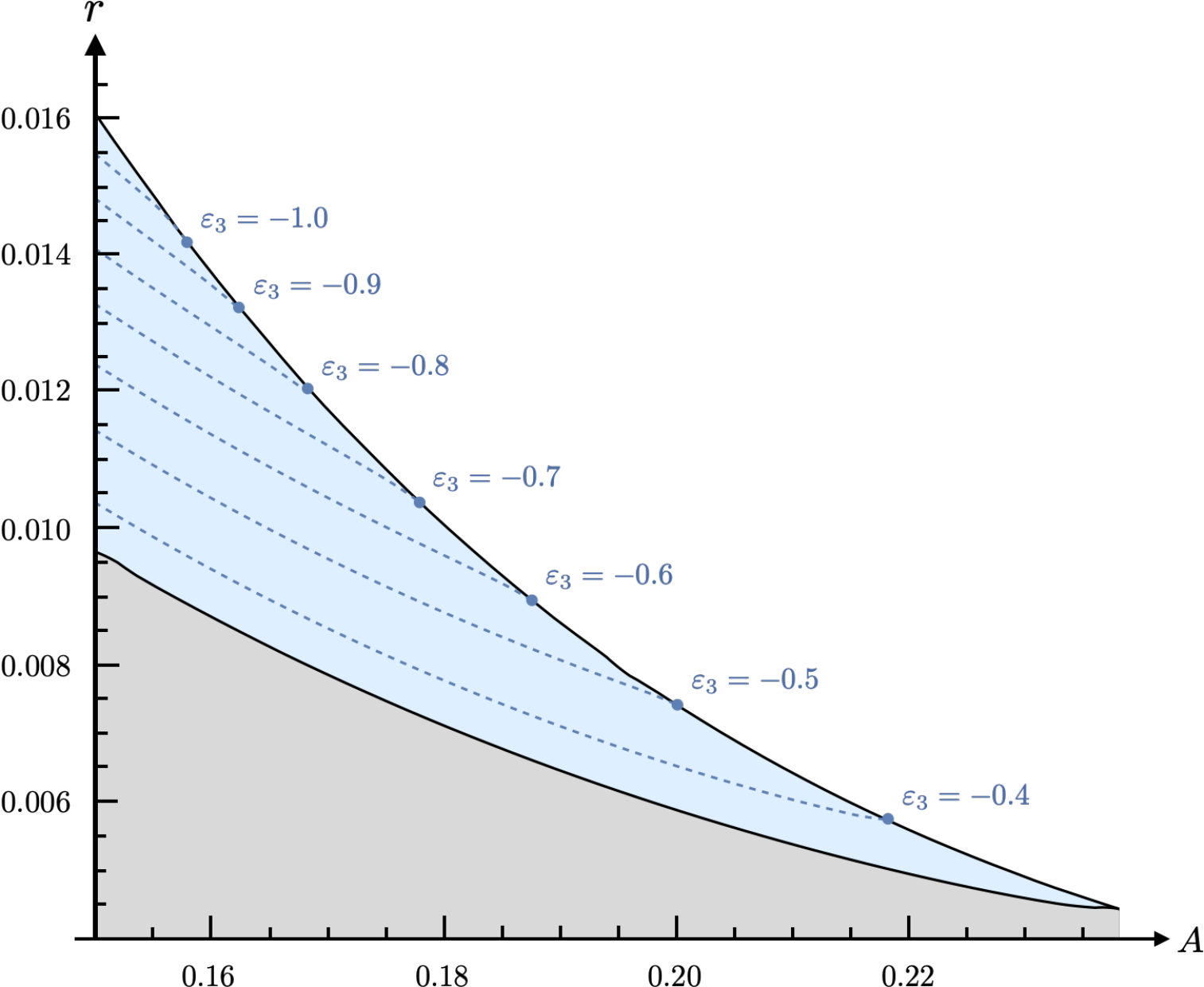}
        \caption{Contour plot of the third Hubble flow parameter $\varepsilon_3$ as a function of the two free parameters of the model \eqref{Kahler}-\eqref{ScalarPotential} $A$ and $r$ (tensor-to-scalar ratio). The blue region corresponds to $|\varepsilon_3|\ge 10\,\varepsilon_2$ and the dashed curves to fixed values of $\varepsilon_3$ shown in the figure.}
    \label{fig:varepsilon3}
 \end{figure}

We now invert relations~\eqref{epsilonV}-\eqref{hatxiV} by using the leading order terms on their right hand sides, and express the Hubble-flow parameters in terms of the potential slow roll parameters~\cite{Ballardini:2024irx}:
\begin{align}
\varepsilon_1 &\simeq \epsilon_V \label{vep1-pSR} \\
\varepsilon_2 &\simeq 4\epsilon_V-2\eta_V \label{vep2-pSR}\\
\varepsilon_3 &\simeq \frac{2}{\varepsilon_2}\left(8\epsilon_V^2-6\epsilon_V\eta_V -\sqrt{2\epsilon_V}\hat\xi_V\right)\,. \label{vep3-pSR}
\end{align}
In Fig.\ref{fig:varepsilon3}, we plot the region of $\varepsilon_3$ (contour plot) such that $|\varepsilon_3|$ takes values between $10|\varepsilon_2|$ and $1$, as a function of the two parameters of the model that we choose to be $A$ and the tensor-to-scalar ratio $r$.

\section{Conclusion}

In summary, in this work we computed the NLO corrections to the bispectrum of primordial scalar perturbations in single field inflation, parametrised in terms of a shape-dependent function $f_\text{NL}$ expressing the non gaussianities. We found that the next-to-leading order corrections can be parametrically as large as the leading-order result. This occurs for two reasons: first, one factor of $\varepsilon_2$ is partially compensated by a logarithmic enhancement in the squeezed limit; and second, the third slow-roll parameter $\varepsilon_3$ can be much larger in magnitude than the second $\varepsilon_2$, which is essentially fixed by the spectral index, as we show in explicit hilltop inflation models within the framework of inflation by supersymmetry breaking. 
This is not exceptional since $\varepsilon_2$ is also much larger than $\varepsilon_1$ due to the upper experimental bound on the primordial tensor modes.
It is also worth noting that upcoming experiments aim to extend the maximum value of the co-moving momenta $k_{\max}$ by one or two orders of magnitude, which would slightly enhance the possible impact of the logarithmic divergence. 
These results may play an important role in distinguishing between next-to-leading order contributions and the effects of heavy particles, 
which have been shown to produce distinctive signatures in the bispectrum by the cosmological collider physics program \cite{Arkani-Hamed:2015bza,Meerburg:2016zdz}.
The exact comparison requires a dedicated analysis which goes beyond the scope of this paper.

It should also be noted that we evaluated the bispectrum at the time of the last horizon exit, 
which is the standard choice in studies of cosmological correlators. 
At that moment, all fluctuations have frozen, fixing the amplitude of the bispectrum. 
In principle, one could instead evaluate it at a later time, which would not significantly affect the result, since the fluctuations remain frozen. 
One may therefore choose a time near the end of inflation, corresponding to about 50--60 e-folds later for the scales observed in the CMB. 
In this case, the time independence of the bispectrum simply means that it retains the same functional form at any time, as shown in \cite{Antoniadis:2025jnb}. 
However, the logarithmic term $\ln(-\tau k_t)$ would then be of order 50--60, as suggested in the introduction, seemingly spoiling the conclusions of that paper. 
Yet in this situation, all quantities, including the slow-roll parameters, should consistently be evaluated at that later time, and their values would differ from those relevant for the scales probed in the CMB. 
Interestingly, the slow-roll parameters would also receive substantial corrections in their expansions, as suggested by \eqref{eq:eistar}, 
where the logarithmic factors are themselves of order 50--60. It would therefore be worthwhile to study these expansions in detail, 
to determine whether they could impose constraints on the slow-roll parameters over long time intervals and to examine how the amplitude of the primordial correlators behaves under this choice of evaluation time. 
Such an analysis, however, lies beyond the scope of the present work.


An interesting extension of our work would be to compute the non-gaussianties at a different time from the initial horizon exit where some Hubble flow parameters get larger than one, for instance during a possible short period of ultra-slow roll inflation where $\varepsilon_2\simeq-6$ (for a review, see~\cite{Dimopoulos:2017ged} and references therein). In this case, perturbative treatment is not valid and one should use the exact expressions of the vertices \eqref{eq:action} with \eqref{E-def}-\eqref{G-def}. Moreover, the expansions of the Green functions in Appendix~\ref{wightman} have to be replaced by exact expressions appropriate for ultra-slow roll.

\section*{Acknowledgements}

I.A. is supported by the Second Century Fund (C2F), Chulalongkorn University. A.C. and H.I. have been supported by Thailand NSRF via PMU-B, grant number B37G660014 and B13F670063. We are grateful to Spyros Sypsas, Shinji Tsujikawa and Vicharit Yingcharoenrat for valuable discussions.

\newpage
\appendix
\section{Action}\label{action}
In order to derive the action up to third order in perturbation, we will use the ADM formalism. Namely, we write the metric as
\begin{align}
\begin{split}
        ds^2&=-N^2dt^2+h_{ij}(N^idt+dx^i)(N^jdt+dx^j)\\
        &=-(N^2-N_iN^i)dt^2+2N_idtdx^i+h_{ij}dx^idx^j.
        \end{split}
\end{align}
Note that the function $N_i$ is defined as $N_i\equiv h_{ij}N^j$. We therefore distinguish the quantities computed according to the full metric $g_{\mu\nu}$ and those related to the metric $h_{ij}$: the later are denoted with a hat. Then, the action becomes
\begin{align}
\begin{split}
    S&=\int d^4x\sqrt{-g}\left\{\frac{1}{2}R-\frac{1}{2}g^{\mu\nu}\partial_\mu\phi\partial_\nu\phi-V(\phi)\right\}\\
    &=\frac{1}{2}\int d^4x\sqrt{h}\left\{N\hat{R}+\frac{1}{N}(E_{ij}E^{ij}-E^2)+\frac{1}{N}(\dot{\phi}-N^i\partial_i\phi)^2-Nh^{ij}\partial_i\phi\partial_j\phi-2NV\right\},
    \end{split}
\label{S-ADM}
\end{align}
where we have defined
\begin{align}
    \begin{split}
        E_{ij}&=\frac{1}{2}\left[\dot{h}_{ij}-\hat{\nabla}_iN_j-\hat{\nabla}_jN_i\right]\\
        &=\frac{1}{2}\left[\dot{h}_{ij}-h_{ik}\partial_jN^k-h_{jk}\partial_iN^k-N^k\partial_kh_{ij}\right].
    \end{split}
\end{align}
This formulation is particularly useful since $N$ and $N^i$ are just Lagrange multipliers giving respectively the two constraints
($E_i^j=E_{ik}h^{kj}$, $E^{ij}=h^{ik}h^{j\ell}E_{k\ell}$, $E=h^{ij}E_{ij}$),
\begin{align}
    0&=\hat{R}-N^{-2}(E_{ij}E^{ij}-E^{2})-N^{-2}(\dot{\phi}-N^i\partial_i\phi)^2-h^{ij}\partial_i\phi\partial_j\phi-2V,\label{eq:constraint1}\\
    0&=\hat{\nabla}_j\left[N^{-1}(E_i^j-\delta_i^jE)\right]-N^{-1}(\dot{\phi}-N^j\partial_j\phi)\partial_i\phi.\label{eq:constraint2}
\end{align}
A general parametrisation of the scalar fluctuations around the background is given by\footnote{In this appendix, the scale factor is parameterised as $a=e^\rho$, thereby $H=\dot a/a=\dot\rho$.}
\begin{align}
    N=1+2\Phi(t,\vec{x}),\qquad N^i=\delta^{ij}\partial_j\chi(t,\vec{x}),\qquad h_{ij}=e^{2\rho(t)+2\zeta(t,\vec{x})}\delta_{ij}+e^{2\rho(t)}\partial_i\partial_j\xi
\end{align}
for the metric together with 
\begin{align}
    \phi=\phi_0(t)+\psi(t,\vec{x})
\end{align}
for the inflaton. In the following, we will drop the subscript $0$ and simply denote the background by $\phi$ as it should not bring any confusion.

We can now use the gauge freedom to fix $\xi=0$ and $\zeta=0$ (the spatially flat gauge, or $\psi$-gauge in short) and the two constraint equations \eqref{eq:constraint1} and \eqref{eq:constraint2} to express $\Phi$ and $\chi$ in terms of $\psi$. Actually, we only need to express them at first order in $\psi$. Indeed, for the action at second order, the second order corrections to $\Phi$ and $\chi$ will respectively multiply the constraint equations $\delta S/\delta N=0$ and $\delta S/\delta N^i=0$ at order zero which vanish trivially. Similarly, for the action at third order, the third order corrections of $\Phi$ and $\chi$ will multiply the constraints at order zero and the second order corrections will multiply the constraints at order 1 which vanish thanks to the action at second order. In that gauge and at first order one gets
\begin{align}
    \Phi=\frac{\dot{\phi}}{4\dot{\rho}}\psi+\cdots\qquad\text{and}\qquad\partial^2\chi=-\frac{1}{2}\frac{\dot{\phi}^2}{\dot{\rho}^2}\frac{d}{dt}\left(\frac{\dot{\rho}}{\dot{\phi}}\psi\right)+\cdots,
\end{align}
where $\partial^2=\partial_k\partial^k$ and where the indices of the derivatives are moved up/down with the metric $\delta_{ij}$. A direct expansion of the action gives at second order
\begin{align}\label{eq:2nd}
\begin{split}
    S^{(2)}=\frac{1}{2}\int d^4x\ e^{3\rho}\Bigg\{&\dot{\psi}^2-\frac{\dot{\phi}^2}{\dot{\rho}}\psi\dot{\psi}-e^{-2\rho}\partial_k\psi\partial^k\psi-\left[\frac{d^2V}{d\phi^2}-\frac{3}{2}\dot{\phi}^2-\frac{1}{4}\frac{\dot{\phi}^4}{\dot{\rho}^2}-\frac{\dot{\phi}\ddot{\phi}}{\dot{\rho}}\right]\psi^2\\
    &+\partial_k\left[\partial_j\chi\partial^j\partial^k\chi-\partial^k\chi\partial^2\chi-2\dot{\phi}\psi\partial^k\chi\right]\Bigg\}
    \end{split}
\end{align}
and at third order
\begin{align}\label{eq:3rd}
\begin{split}
    S^{(3)}=\int d^4x\ e^{3\rho}\Bigg\{&-\frac{1}{4}\frac{\dot{\phi}}{\dot{\rho}}\psi\dot{\psi}^2-\frac{1}{4}e^{-2\rho}\frac{\dot{\phi}}{\dot{\rho}}\psi\partial_k\psi\partial^k\psi-\dot{\psi}\partial_k\psi\partial^k\chi\\
    &+\frac{3}{8}\frac{\dot{\phi}^3}{\dot{\rho}}\psi^3-\frac{1}{16}\frac{\dot{\phi}^5}{\dot{\rho}^3}\psi^3-\frac{1}{4}\frac{\dot{\phi}}{\dot{\rho}}\frac{d^2V}{d\phi^2}\psi^3-\frac{1}{6}\frac{d^3V}{d\phi^3}\psi^3+\frac{1}{4}\frac{\dot{\phi}^3}{\dot{\rho}^2}\psi^2\dot{\psi}\\
    &+\frac{1}{4}\frac{\dot{\phi}^2}{\dot{\rho}}\psi^2\partial^2\chi-\frac{1}{4}\frac{\dot{\phi}}{\dot{\rho}}\psi\left[\partial_i\partial_j\chi\partial^i\partial^j\chi-(\partial^2\chi)^2\right]\\
    &+\partial_k\left[\frac{1}{4}\frac{\dot{\phi^2}}{\dot{\rho}}\psi^2\partial^k\chi\right]\Bigg\},
    \end{split}
\end{align}
where the derivatives of the potential are evaluated at the background values. We have used this gauge so far because it is particularly simple to compute the second and third order with it. It is mostly important to notice that no integration by part has been used to obtain \eqref{eq:2nd} and \eqref{eq:3rd}. However, we are interested in $\zeta$ rather than $\psi$ as $\zeta$ is constant outside the horizon. 

In order to get the action at third order in $\zeta$, we will now see, following~\cite{Maldacena:2002vr}, how to go from the $\psi$-gauge used so far in this appendix, namely from
\begin{align}\label{psi-gauge}
    N=1+2\Phi(t,\vec{x}),\qquad N^i=\delta^{ij}\partial_j\chi(t,\vec{x}),\qquad h_{ij}=e^{2\rho(t)}\delta_{ij},\qquad\phi=\phi_0(t)+\psi(t,\vec{x}),
\end{align}
to the $\zeta$-gauge 
\begin{align} \label{zeta-gauge}
    N=1+2\Phi(t,\vec{x}),\qquad N^i=\delta^{ij}\partial_jB(t,\vec{x}),\qquad h_{ij}=e^{2\rho(t)+2\zeta(t,\vec{x})}\delta_{ij},\qquad\phi=\phi_0(t),
\end{align}
where again the subscript `0' will be dropped. We then perform a shift in spacetime coordinates as
\begin{align}
    t&\rightarrow\Tilde{t}=t+T(t,\vec{x}), \\
    x^i&\rightarrow\Tilde{x^i}=x^i+\epsilon^i(t,\vec{x}),
\end{align}
where we expand all quantities in series of $\zeta$, namely
\begin{align}
    T=T_1+T_2+\cdots\qquad\text{and}\qquad\epsilon^{i}=\epsilon^i_1+\epsilon^i_2+\cdots.
\end{align}
Under this shift, the scalar field changes according to $\phi(t,\vec{x})\rightarrow\Tilde{\phi}(\tilde{t},\tilde{\vec{x}})=\phi(t,\vec{x})$ with
\begin{align}
   \Tilde{\phi}(\tilde{t},\tilde{\vec{x}})=\tilde{\phi}(t,\vec{x})+T\frac{\partial\tilde{\phi}}{\partial\tilde{t}}+\epsilon^i\frac{\partial\tilde{\phi}}{\partial\tilde{x}^i}+T\epsilon^i\frac{\partial^2\tilde{\phi}}{\partial\tilde{t}\partial\tilde{x}^i}+\frac{1}{2}T^2\frac{\partial\tilde{\phi}}{\partial\tilde{t}^2}+\frac{1}{2}\epsilon^i\epsilon^j\frac{\partial^2\tilde{\phi}}{\partial\tilde{x}^i\partial\tilde{x}^j}+\cdots,
\end{align}
where the derivatives are evaluated in $(t,\vec{x})$. In these two gauges we have
\begin{align}
    \Tilde{\phi}(\tilde{t},\tilde{\vec{x}})&=\phi_0(\tilde{t}),\\
    \phi(t,\vec{x})&=\phi_0(t)+\psi(t,\vec{x}),
\end{align}
where the subscript $0$ have been re-introduced here for clarity. One can then conclude that $\psi$ up to second order in $\zeta$ is given by
\begin{align}\label{eq:psi}
    \psi=T_1\dot{\phi}+T_2\dot{\phi}+\frac{1}{2}T_1^2\ddot{\phi}+\cdots.
\end{align}
In order to see how the metric transforms, let's look at
\begin{align}
\begin{split}\label{eq:metric}
    ds^2&=-(N^2-N_iN^i)dt^2+2N_idtdx^i+h_{ij}dx^idx^j\\
    &=-(\tilde{N}^2-\tilde{N}_i\tilde{N}^i)d\tilde{t}^2+2\tilde{N}_id\tilde{t}d\tilde{x}^i+\tilde{h}_{ij}d\tilde{x}^id\tilde{x}^j.
    \end{split}
\end{align}
Applying the change of coordinates to the differentials as
\begin{align}
    d\tilde{t}&=dt+\frac{\partial T}{\partial t}dt+\frac{\partial T}{\partial x^i}dx^i,\\
    d\tilde{x}^i&=dx^i+\frac{\partial\epsilon^i}{\partial t}dt+\frac{\partial\epsilon^i}{\partial x^j}dx^j,
\end{align}
we deduce that
\begin{align}\label{eq:h}
\begin{split}
    h_{ij}(t,\vec{x})&=(\tilde{h}_{kl}\tilde{N}^k\tilde{N}^l-\tilde{N}^2)\partial_iT\partial_jT+\tilde{N}_i\partial_jT+\tilde{N}_j\partial_iT+\tilde{N}_k\partial_iT\partial_j\epsilon^k+\tilde{N}_k\partial_jT\partial_i\epsilon^k\\
    &\quad+\tilde{h}_{ij}+\tilde{h}_{ik}\partial_j\epsilon^k+\tilde{h}_{kj}\partial_i\epsilon^k+\tilde{h}_{kl}\partial_i\epsilon^k\partial_j\epsilon^l
    \end{split}
\end{align}
where the tilded quantities are functions of $\tilde{t}$ and $\tilde{x}^i$ here. In these two gauges we have
\begin{align}
    \tilde{h}_{ij}(\tilde{t},\tilde{\vec{x}})&=e^{2\rho(\tilde{t})+2\zeta(\tilde{t},\tilde{\vec{x}})}\delta_{ij},\\
    h_{ij}(t,\vec{x})&=e^{2\rho(t)}\delta_{ij},
\end{align}
and we can expand $\tilde{h}$ as
\begin{align}
\begin{split}
    \tilde{h}_{ij}(\tilde{t},\tilde{\vec{x}})&=e^{2\rho+2\zeta}\delta_{ij}\left[1+2(\dot{\rho}+\dot{\zeta})T+2\epsilon^i\partial_i\zeta+T^2(\ddot{\rho}+2\dot{\rho}^2)+\cdots\right]\\
    &=e^{2\rho}\delta_{ij}\left[1+2\dot{\rho}T+2\zeta+2\dot{\zeta}T+2\epsilon^i\partial_i\zeta+T^2(\ddot{\rho}+2\dot{\rho}^2)+4\dot{\rho}\zeta T+2\zeta^2+\cdots\right],
    \end{split}
\end{align}
where all the right-hand side quantities are functions of $t$ and $\vec{x}$. Equation \eqref{eq:h} is then trivially satisfied at zeroth order and gives at order one
\begin{align}
\delta_{ij}\left[2\dot{\rho}T_1+2\zeta\right]+\delta_{ik}\partial_j\epsilon_1^k+\delta_{jk}\partial_i\epsilon_1^k=0
\end{align}
leading to
\begin{align}
    T_1=-\frac{1}{\dot{\rho}}\zeta\qquad\text{and}\qquad\epsilon_1^i=0.
\end{align}
Using \eqref{eq:metric}, we can compute $\tilde{N}_i$ at first  order as
\begin{align}
    \begin{split}\tilde{N}_i(\tilde{t},\tilde{\vec{x}})=&e^{2\rho}\partial_i\chi+\partial_iT_1+\cdots\\
    =&\frac{1}{2}e^{2\rho}\frac{\dot{\phi}^2}{\dot{\rho}^2}\partial_i\partial^{-2}\dot{\zeta}-\frac{1}{\dot{\rho}}\partial_i\zeta+\cdots
    \end{split}
\end{align}
where all the right-hand side quantities are again functions of $t$ and $\vec{x}$
and where $\partial^{-2}$ is defined by $\partial^{-2}\partial^2=1$. Finally, \eqref{eq:h} at order 2 is
\begin{align}\begin{split}
    0&=\delta_{ij}\left[2\dot{\rho}T_2+2\dot{\zeta}T_1+T_1^2(\ddot{\rho}+2\dot{\rho}^2)+4\dot{\rho}\zeta T_1+2\zeta^2\right]\\
    &\quad+\left[\frac{1}{2}e^{2\rho}\frac{\dot{\phi}^2}{\dot{\rho}^2}\partial_i\partial^{-2}\dot{\zeta}-\frac{1}{\dot{\rho}}\partial_i\zeta\right]\partial_jT_1+\left[\frac{1}{2}e^{2\rho}\frac{\dot{\phi}^2}{\dot{\rho}^2}\partial_j\partial^{-2}\dot{\zeta}-\frac{1}{\dot{\rho}}\partial_j\zeta\right]\partial_iT_1\\
    &\quad+\delta_{ik}\partial_j\epsilon_2^k+\delta_{jk}\partial_i\epsilon_2^k-e^{-2\rho}\partial_iT_1\partial_jT_1.\\
    \end{split}
\end{align}
In order to simplify the calculations, we consider the trace of this equation as well as the result once we apply $\partial^{-2}\partial^i\partial^j$ on it:
\begin{align}
    0&=6\dot{\rho}T_2-\frac{6}{\dot{\rho}}\dot{\zeta}\zeta+3\frac{\ddot{\rho}}{\dot{\rho}^2}\zeta^2+2\partial_k\epsilon_2^k-\frac{\dot{\phi}^2}{\dot{\rho}^3}\partial^{-2}\partial_i\dot{\zeta}\partial^i\zeta+\frac{1}{\dot{\rho}^2}e^{-2\rho}\partial_i\zeta\partial^i\zeta,\\
    0&=2\dot{\rho}T_2-\frac{2}{\dot{\rho}}\dot{\zeta}\zeta+\frac{\ddot{\rho}}{\dot{\rho}^2}\zeta^2+2\partial_k\epsilon_2^k-\frac{1}{\dot{\rho}}\partial^{-2}\partial^i\partial^j\left\{\frac{\dot{\phi}^2}{\dot{\rho}^2}\partial^{-2}\partial_i\dot{\zeta}\partial_j\zeta-\frac{1}{\dot{\rho}}e^{-2\rho}\partial_i\zeta\partial_j\zeta\right\}.
\end{align}
We can now find $T_2$ by summing those two equations, and in the end, \eqref{eq:psi} becomes~\cite{Maldacena:2002vr}
\begin{align}
    \psi=-\frac{\dot{\phi}}{\dot{\rho}}(\zeta-f(\zeta))+\cdots
\end{align}
with 
\begin{align}
\begin{split}
    f(\zeta)&=\frac{1}{2}\left[\frac{\ddot{\phi}}{\dot{\phi}\dot{\rho}}+\frac{1}{2}\frac{\dot{\phi}^2}{\dot{\rho}^2}\right]\zeta^2+\frac{1}{\dot{\rho}}\dot{\zeta}\zeta-\frac{1}{4}\frac{e^{-2\rho}}{\dot{\rho}^2}\left[\partial_i\zeta\partial^i\zeta-\partial^{-2}\partial_i\partial_j(\partial^i\zeta\partial^j\zeta)\right]\\
    &\quad+\frac{1}{4}\frac{\dot{\phi}^2}
    {\dot{\rho}^3}\left[\partial_i\zeta\partial^i\partial^{-2}\dot{\zeta}-\partial^{-2}\partial_i\partial_j(\partial^i\zeta\partial^j\partial^{-2}\dot{\zeta})\right].
    \end{split}
\end{align}

We can now obtain the action expressed in terms of $\zeta$ instead of $\psi$. Let's start with the second order action. In order to simplify the expressions, we will do an integration by parts to change the $\psi\dot{\psi}$ term into $\psi^2$. We also write the total derivative as $\mathscr{D}$ and then we re-write \eqref{eq:2nd} as
\begin{align}\label{eq:2nd2}
\begin{split}
    S^{(2)}=\frac{1}{2}\int d^4x\ e^{3\rho}\Bigg\{&\dot{\psi}^2-e^{-2\rho}\partial_k\psi\partial^k\psi-\left[\frac{d^2V}{d\phi^2}-3\dot{\phi}^2-\frac{1}{2}\frac{\dot{\phi}^4}{\dot{\rho}^2}-2\frac{\dot{\phi}\ddot{\phi}}{\dot{\rho}}\right]\psi^2+\mathscr{D}\Bigg\}.
    \end{split}
\end{align}
When we expand this expression at third order in $\zeta$, the $\dot{\psi}^2$ term will give a term proportional to $\partial_t f(\zeta)$. We then perform a second integration by parts to get rid of this derivative on $f$. We also perform a third integration by parts to get rid of $\partial_k f(\zeta)$. Otherwise, nothing else is necessary than a simple expansion to get the third order contribution:
\begin{align}
    \int d^4x\left\{\left[\partial_t\left(e^{3\rho}\frac{\dot{\phi}^2}{\dot{\rho}^2}\dot{\zeta}\right)-e^{\rho}\frac{\dot{\phi}^2}{\dot{\rho}^2}\partial^2\zeta\right]f(\zeta)+\mathscr{D}_2\right\},
\end{align}
where $\mathscr{D}_2$ is the total derivative of \eqref{eq:2nd} expanded at third order in $\zeta$. This form is actually particularly useful since it only gives a contribution proportional to the linear equation of motion of $\zeta$ as one can find the second order action to be
\begin{align}\label{eq:actionzeta2}
    S^{(\zeta^2)}=&\int d^4x\ e^{3\rho}\frac{1}{2}\frac{\dot{\phi}^2}{\dot{\rho}^2}\left(\dot{\zeta}^2-e^{-2\rho}\partial_k\zeta\partial^k\zeta+\mathscr{D}_1\right),
\end{align}
where $\mathscr{D}_1$ is the total derivative of \eqref{eq:2nd} expanded at second order in $\zeta$. A direct expansion of \eqref{eq:3rd} then leads to the third order action:
\begin{align}\label{eq:S3noIPP}
\begin{split}
    S^{(\zeta^3)}=\int d^4x\Bigg(&\left[\partial_t\left(e^{3\rho}\frac{\dot{\phi}^2}{\dot{\rho}^2}\dot{\zeta}\right)-e^{\rho}\frac{\dot{\phi}^2}{\dot{\rho}^2}\partial^2\zeta\right]f(\zeta)\\
    &+e^{3\rho}\frac{\dot{\phi}^4}{\dot{\rho}^4}\Bigg\{\frac{1}{4}\zeta\dot{\zeta}^2+\frac{1}{2}\left(\frac{\ddot{\phi}}{\dot{\phi}}+\frac{1}{4}\frac{\dot{\phi}^2}{\dot{\rho}}\right)\zeta^2\dot{\zeta}+\frac{1}{4}e^{-2\rho}\zeta\partial_k\zeta\partial^k\zeta\\
    &-\frac{1}{2}\left(\frac{\ddot{\phi}}{\dot{\phi}}+\frac{1}{2}\frac{\dot{\phi}^2}{\dot{\rho}}\right)\zeta\partial_k\zeta\partial^k\partial^{-2}\dot{\zeta}-\frac{1}{2}\dot{\zeta}\partial_k\zeta\partial^k\partial^{-2}\dot{\zeta}\\
    &+\left[-\frac{3}{8}\dot{\phi}^2+\frac{1}{4}\frac{\ddot{\phi}^2}{\dot{\phi}^2}+\frac{1}{4}\frac{d^2V}{d\phi^2}+\frac{1}{6}\frac{\dot{\rho}}{\dot{\phi}}\frac{d^3V}{d\phi^3}\right]\zeta^3\\
    &+\frac{1}{16}\frac{\dot{\phi}^2}{\dot{\rho}^2}\zeta\left[\partial_i\partial_j\partial^{-2}\dot{\zeta}\partial^i\partial^j\partial^{-2}\dot{\zeta}-\dot{\zeta}^2\right]\Bigg\}+\mathscr{D}_2+\mathscr{D}_3\Bigg),
    \end{split}
\end{align}
where $\mathscr{D}_3$ is the total derivative of \eqref{eq:3rd} expanded at third order in $\zeta$. This action can be written in terms of the Hubble flow functions as
\begin{align}
\begin{split}
S^{(\zeta^3)}=\int d^4x\Bigg(&\left[\partial_t\left(2\varepsilon_1e^{3\rho}\dot{\zeta}\right)-2\varepsilon_1 e^{\rho}\partial^2\zeta\right]f(\zeta)\\
    &+e^{3\rho}\Bigg\{\varepsilon_1^2\left(\zeta\dot{\zeta}^2+e^{-2\rho}\zeta\partial_k\zeta\partial^k\zeta-2\dot{\zeta}\partial_k\zeta\partial^k\partial^{-2}\dot{\zeta}\right)\\
    &+\dot{\rho}^2E\zeta^3+\dot{\rho}F\zeta^2\dot{\zeta}+\dot{\rho}G\zeta\partial_k\zeta\partial^k\partial^{-2}\dot{\zeta}\\
    &+\frac{1}{2}\varepsilon_1^3\zeta\left[\partial_i\partial_j\partial^{-2}\dot{\zeta}\partial^i\partial^j\partial^{-2}\dot{\zeta}-\dot{\zeta}^2\right]\Bigg\}+\mathscr{D}_2+\mathscr{D}_3\Bigg)
    \end{split}
\end{align}
with 
\begin{align}
\begin{split}
    E=&-\frac{1}{6}\Big[6\varepsilon_1^3-2\varepsilon_1^4-9\varepsilon_1^2\varepsilon_2+9\varepsilon_1^3\varepsilon_2-6\varepsilon_1^2\varepsilon_2^2+3\varepsilon_1\varepsilon_2\varepsilon_3\\
    &\qquad\ -4\varepsilon_1^2\varepsilon_2\varepsilon_3+\varepsilon_1\varepsilon_2^2\varepsilon_3+\varepsilon_1\varepsilon_2\varepsilon_3^2+\varepsilon_1\varepsilon_2\varepsilon_3\varepsilon_4\Big],
    \end{split}\\
    F=&\varepsilon_1^2\varepsilon_2-\varepsilon_1^3,\\
    G=&-\varepsilon_1^2\varepsilon_2.
\end{align}
One can also perform two more integrations by part to regroup the terms in $\zeta^3$, and replacing the derivatives of the potential we get
\begin{align}
\begin{split}
    S^{(\zeta^3)}=\int d^4x\Bigg(&\left[\partial_t\left(e^{3\rho}\frac{\dot{\phi}^2}{\dot{\rho}^2}\dot{\zeta}\right)-e^{\rho}\frac{\dot{\phi}^2}{\dot{\rho}^2}\partial^2\zeta\right]f(\zeta)\\
    &+e^{3\rho}\frac{\dot{\phi}^4}{\dot{\rho}^4}\Bigg\{\frac{1}{4}\zeta\dot{\zeta}^2+\frac{1}{4}e^{-2\rho}\zeta\partial_k\zeta\partial^k\zeta-\frac{1}{2}\dot{\zeta}\partial_k\zeta\partial^k\partial^{-2}\dot{\zeta}\\
    &+\Bigg[-\frac{1}{4}\dot{\phi}^2-\frac{5}{24}\frac{\dot{\phi}^4}{\dot{\rho}^2}-\frac{3}{4}\frac{\dot{\rho}\ddot{\phi}}{\dot{\phi}}-\frac{\dot{\phi}\ddot{\phi}}{\dot{\rho}}+\frac{1}{2}\frac{\dot{\rho}^2\ddot{\phi}^2}{\dot{\phi}^4}\\
    &-\frac{1}{2}\frac{\ddot{\phi}^2}{\dot{\phi}^2}-\frac{1}{2}\frac{\dot{\rho}^2\phi^{(3)}}{\dot{\phi}^3}-\frac{1}{2}\frac{\phi^{(3)}}{\dot{\phi}}+ \frac{1}{6}\frac{\dot{\rho}\ddot{\phi}\phi^{(3)}}{\dot{\phi}^4}-\frac{1}{6}\frac{\dot{\rho}\phi^{(4)}}{\dot{\phi}^3}\Bigg]\zeta^3\\
    &+\frac{1}{16}\frac{\dot{\phi}^2}{\dot{\rho}^2}\zeta\left[\partial_i\partial_j\partial^{-2}\dot{\zeta}\partial^i\partial^j\partial^{-2}\dot{\zeta}-\dot{\zeta}^2\right]\Bigg\}\Bigg),
    \end{split}
\end{align}
where we dropped the total derivatives. In terms of the Hubble flow functions it becomes
\begin{align}\label{eq:S3afterIPP}
\begin{split}
    S^{(\zeta^3)}=\int d^4x\Bigg(&\left[\partial_t\left(2\varepsilon_1 e^{3\rho}\dot{\zeta}\right)-2\varepsilon_1 e^{\rho}\partial^2\zeta\right]f(\zeta)\\
    &+e^{3\rho}\Bigg\{\varepsilon_1^2\left(\zeta\dot{\zeta}^2+e^{-2\rho}\zeta\partial_k\zeta\partial^k\zeta-2\dot{\zeta}\partial_k\zeta\partial^k\partial^{-2}\dot{\zeta}\right)\\
    &-\frac{\dot{\rho}^2}{6}\big[3\varepsilon_1\varepsilon_2\varepsilon_3-\varepsilon_1^2\varepsilon_2\varepsilon_3+\varepsilon_1\varepsilon_2^2\varepsilon_3+\varepsilon_1\varepsilon_2\varepsilon_3^2+\varepsilon_1\varepsilon_2\varepsilon_3\varepsilon_4\big]\zeta^3\\
    &+\frac{1}{2}\varepsilon_1^3\zeta\left[\partial_i\partial_j\partial^{-2}\dot{\zeta}\partial^i\partial^j\partial^{-2}\dot{\zeta}-\dot{\zeta}^2\right]\Bigg\}\Bigg).
    \end{split}
\end{align}
One can also perform its integrations by part differently to get:
\begin{align}\label{eq:S3Maldacena}
\begin{split}
    S^{(\zeta^3)}=\int d^4x\Bigg(&\left[\partial_t\left(2\varepsilon_1 e^{3\rho}\dot{\zeta}\right)-2\varepsilon_1 e^{\rho}\partial^2\zeta\right]f(\zeta)\\
    &+e^{3\rho}\Bigg\{\varepsilon_1^2\left(\zeta\dot{\zeta}^2+e^{-2\rho}\zeta\partial_k\zeta\partial^k\zeta-2\dot{\zeta}\partial_k\zeta\partial^k\partial^{-2}\dot{\zeta}\right)\\
    &+\frac{1}{2}\dot{\rho}\varepsilon_1\varepsilon_2\varepsilon_3\zeta^2\dot{\zeta}+\frac{1}{2}\varepsilon_1^3\zeta\left[\partial_i\partial_j\partial^{-2}\dot{\zeta}\partial^i\partial^j\partial^{-2}\dot{\zeta}-\dot{\zeta}^2\right]\Bigg\}\Bigg).
    \end{split}
\end{align}

\paragraph{Comment on the comoving gauge.}
Here we make a comment on the comoving gauge \eqref{zeta-gauge}. In this gauge, the lapse and shift are expanded as
\begin{align}
N=\frac{\dot\zt}{\dot\rho}+\cdots, \quad N^i=\pd_i\left[-\frac{e^{-2\rho}}{\dot\rho}\zt+\frac{\dot\phi^2}{2\dot\rho^2}\pd^{-2}\dot\zt\right]+\cdots.
\end{align}
Note that we only need fluctuations up to linear order by the same argument for the $\psi$-gauge. Substituting them into the action \eqref{S-ADM}, we can observe that the resulting action contains only $\zt$ and $\dot\zt$, without $\zt$ with higher order time derivatives, at least up to the third order in $\zt$. This is because first time derivatives in the action \eqref{S-ADM} act only on $\phi$ and $h_{ij}$. Therefore, we are free to move to the path integral formulation, as explained in a bit more detail in Section~\ref{subsec:deriv-coupling}.

\section{The Schwinger-Keldysh path integral}\label{appendix:SK}

In this appendix, we briefly review the Schwinger-Keldysh path integral to compute expectation values. See, for example, \cite{Weinberg:2005vy,Chen:2017ryl,Abolhasani:2022twf,Palma:2023idj,Braglia:2024zsl,Kawaguchi:2024lsw,Kawaguchi:2024rsv} 
for more detailed discussions.

Let us work with the physical time $t$, though most of the reviews use the conformal time, and consider a system with a Lagrangian of the form
\begin{align}
\cL(\zt,\dot\zt)=\cL_2(\zt,\dot\zt)+\cL_{\mathrm{int}}(\zt,\dot\zt), 
\end{align}
where $\cL_2$ is quadratic in $\zt$ and $\dot\zt$ specified and $\cL_{\mathrm{int}}$ cubic or of higher order. We are interested in expectation values of equal-time operators of the form\footnote{This definition of $\langle\mathscr{O}(t)\rangle$ is valid only in this Appendix. In the main text, $\langle\mathscr{O}(t)\rangle$ is a different quantity defined by \eqref{<On>}.}
\begin{align}
\langle\mathscr{O}(t)\rangle\equiv\langle 0|\hat{\mathscr O}(t)|0 \rangle, \label{<O>}
\end{align}
where $\hat{\mathscr O}(t)$ is an operator at time $t$ being a polynomial of Heisenberg-picture operator $\hat\zt(t)$ (with different positions or momenta), and $|0\rangle$ is a vacuum state of the system given as an in-state at past infinity. Since the expectation value \eqref{<O>} can be written schematically as $\langle 0|\hat U(-\infty,t_f)\hat U(t_f,t)\hat{\mathscr O}\hat U(t,-\infty)|0\rangle$ with $\hat U(t_1,t_2)$ the time-evolution operator and $t_f$ a time after $t$, one can see from the standard time-slice argument that its corresponding path integral involves a time contour that starts from $-\infty$, reaches time $t_f$ after $t$, and then goes back to $-\infty$. Furthermore, to select the vacuum state, we make the $i\ep$-prescription with the complexified time. Therefore, the path integral for $\langle\mathscr{O}(t)\rangle$ is given by
\begin{align}
\langle\mathscr{O}(t)\rangle=\int[d\zt]_C\,\mathscr{O}(t)\exp\left(i\int_Cdtd^3x\,[\cL(\zt,\zt')+\zt J]\right),
\end{align}
where the subscript $C$ on the measure $[d\zt]$ dictates that $\zt$ should lie along a time contour $C$, which is defined as follows: $C$ is a curve $t(\lm)$ in complex plane parameterised by a real parameter $\lm\in(-\infty,\infty)$; It starts from $t(-\infty)=-\infty(1-i\ep)$, reaches a time $t(\lm_f)$, and goes back to $t(\infty)=-\infty(1+i\ep)$, with an infinitesimal parameter $\ep>0$. The operator $\mathscr{O}(t)$ is located at some point $\lm$ on $C$ such that $t=t(\lm)$. The time contour $C$ with the characteristic times is depicted as follows:
\begin{center}
\begin{tikzpicture}
    \draw[thick, ->] (-4.2, 0) -- (7, 0); 
    \draw[thick, ->] (-4, 0.2) -- (3.05, 0.2);
    \draw[thick, -] (2.9, 0.2) -- (6, 0.2);
    \draw[thick] (6, 0.2) arc (90:-90:0.2);
    \draw[thick, ->] (6, -0.2) -- (2.95, -0.2);
    \draw[thick, -] (3.1, -0.2) -- (-4, -0.2);
    \node at (6.2, -0.5) {$t(\lm_f)$}; 
    \node at (-4, 0.5) {$t(-\infty)$}; 
    \node at (-4, -0.5) {$t(\infty)$}; 
    \draw[thin, -] (-4, -0.1) -- (-4, 0.1);
    \draw[thin, -] (1, -0.1) -- (1, 0.1);
    \node at (1, 0.5) {$\mathscr{O}(t)$}; 
    \node at (3.5, -0.5) {$-C_-$};
    \node at (3.5, 0.5) {$C_+$};
    \node at (1, -0.5) {$t=t(\lm)$}; 
    \filldraw (1, 0.2) circle (2pt); 
\end{tikzpicture}
\end{center}
In this picture, $\mathscr{O}$ is inserted at $t=t(\lm)$ with $\lm<\lm_f$. We also denoted the contour $C$ as the joint of the forward part $C_+$ and the backward part $-C_-$, where the minus sign means that contour $C_-$ is defined by flipping the orientation of $-C_-$, thereby running from $t(\infty)$ to $t(\lm_f)$.

Let us look at a general path integral of the form
\begin{align}
\int[d\zt]_C\,\mathscr{O}_1(t_1)\cdots\mathscr{O}_n(t_n)\exp\left(i\int_Cdtd^3x\,[\cL(\zt,\zt')+\zt J]\right),  
\end{align}
where $t_i$ satisfies $t_i=\mathrm{Re}(t(\lm_i))$ with some $\lm_i$ $(1\leq i\leq n)$. This corresponds to 
\begin{align}
\langle 0|\, \mathrm{T}_C[\,\hat{\mathscr O}_1(t(\lm_1))\cdots\hat{\mathscr O}_n(t(\lm_n))\,] \,|0 \rangle,
\end{align}
where the time-ordering operator $\mathrm{T}_C$ along $C$ is defined with respect to the ordering of the real numbers $\lm_1,\cdots,\lm_n$. For example, 
\begin{align}
\mathrm{T}_C[\,\hat{\mathscr O}_1(t(\lm_1))\hat{\mathscr O}_2(t(\lm_2))\,]&=\hat{\mathscr O}_1(t(\lm_1))\hat{\mathscr O}_2(t(\lm_2)) \quad \text{ if } \lm_1>\lm_2, \\
\mathrm{T}_C[\,\hat{\mathscr O}_1(t(\lm_1))\hat{\mathscr O}_2(t(\lm_2))\,]&=\hat{\mathscr O}_2(t(\lm_2))\hat{\mathscr O}_1(t(\lm_1)) \quad \text{ if } \lm_1<\lm_2.
\end{align}
Note that $\mathrm{T}_C$ \emph{has nothing to do with the ordering of} $\mathrm{Re}(t(\lm_1)),\cdots,\mathrm{Re}(t(\lm_n))$.

While path integral expressions with $C$ are convenient to find their operator version, it is customary to decompose $C$ into $C_\pm$ for pertubative computations. Recall that both $C_+$ and $C_-$ are defined to be oriented from past to future; in particular, $C_-$ has the opposite orientation to the backward part of $C$. 

Let us focus on the case $n=2$ and consider two operators $\hat{\mathscr O}_1(t(\lm_1)),\hat{\mathscr O}_2(t(\lm_2))$ on $C$. We use the shorthands $t_i:=t(\lm_i)$. When the times $t_1,t_2$ are on $C_+$, the ordering with respect to $\lm_1,\lm_2$ coincides with that with respect to $\mathrm{Re}(t_1),\mathrm{Re}(t_2)$. However, when $t_1,t_2$ are on $C_-$, the ordering with respect to $\lm_1,\lm_2$ is \emph{opposite} to that with respect to $\mathrm{Re}(t_1),\mathrm{Re}(t_2)$ due to the choice of the orientation of $C_-$; namely $\mathrm{Re}(t_1)<\mathrm{Re}(t_2)$ for $\lm_1>\lm_2$. We therefore find the following relations:
\begin{align}
\mathrm{T}_C[\,\hat{\mathscr O}_1(t(\lm_1))\hat{\mathscr O}_2(t(\lm_2))\,]=\mathrm{T}[\,\hat{\mathscr O}_1(t(\lm_1))\hat{\mathscr O}_2(t(\lm_2))\,], \quad \text{ when } ~ t(\lm_1), t(\lm_2) \in C_+, \\ 
\mathrm{T}_C[\,\hat{\mathscr O}_1(t(\lm_1))\hat{\mathscr O}_2(t(\lm_2))\,]=\ol{\mathrm{T}}[\,\hat{\mathscr O}_1(t(\lm_1))\hat{\mathscr O}_2(t(\lm_2))\,], \quad \text{ when } ~ t(\lm_1), t(\lm_2) \in C_-,
\end{align}
where $\mathrm{T}$ ($\ol{\mathrm{T}}$) is the time (anti-time) ordering operator with respect to $\mathrm{Re}(t(\lm_1)),\mathrm{Re}(t(\lm_2))$. When the two operators are inserted along different contours, they have no chance of collide and hence we do not need time ordering operators. Generalisations to cases with insertions of more operators are straightforward.

\subsection{Master formula for perturbative expansions}\label{subsec:master}
The master formula for in-in correlators is the generating functional $Z[J]$ with source $J$,
\begin{align} \label{ZJ}
Z[J]=\int[d\zt]_C\exp\left(i\int_Cdtd^3x\,[\cL(\zt,\zt')+\zt J]\right).
\end{align}
Following the decomposition of $C$ into $C_\pm$, let us introduce the following fields\footnote{Here we suppressed the other arguments in $\zt$ for notational simplicity.}:
\begin{align}
\zt^+(t), ~ J^+(t) \quad &\text{ if } t \text{ is on } C_+, \nn\\
\zt^-(t), ~ J^-(t) \quad &\text{ if } t \text{ is on } C_-, \nn
\end{align}
with the constraint that $\zt^\pm$ and $J^\pm$ should be glued at the turn $t(\lm_f)$ for any $\vec x$:
\begin{align} \label{ztglue}
\begin{split}
\zt^+(t(\lm_f),\vec x)&=\zt^-(t(\lm_f),\vec x), \\
J^+(t(\lm_f),\vec x)&=J^-(t(\lm_f),\vec x),
\end{split}
\end{align}
and rewrite the action in the exponent of \eqref{ZJ} as
\begin{align}\label{S+-}
\int_Cdtd^3x\,[\cL(\zt,\dot\zt)+\zt J]
&=\int_{C_\pm}\!\!dtd^3x\,[\cL(\zt^+,\dot\zt^+)-\cL(\zt^-,\dot\zt^-)+\zt^+J^+-\zt^-J^-],
\end{align}
where the time integral $\int_{C_\pm}$ should be understood as along $C_+$ for $(\zt^+,J^+)$ and along $C_-$ for $(\zt^-,J^-)$\footnote{Note that the paths are not $C_+$ and $-C_-$; both are oriented from past to future.}. The minus signs in front of the terms with $\zt^-,J^-$ in \eqref{S+-} are due to the flip of $-C_-$ in $C$ to $C_-$. As a result, the generating functional \eqref{ZJ} is rewritten as an integral over two fields $\zt^\pm$ with sources $J^\pm$ along time contours $C_\pm$:
\begin{align} \label{ZJ+-}
Z[J^\pm]=\int[d\zt^\pm]\exp\left(i\int_{C_\pm}\!\!dt d^3x\,[\cL(\zt^+,\dot\zt^+)-\cL(\zt^-,\dot\zt^-)+\zt^+J^+-\zt^-J^-]\right),
\end{align}
where recall that the integral variables $\zt^\pm$ satisfy the gluing condition \eqref{ztglue} since they are originally the \emph{single} field $\zt$ along the \emph{single} contour $C$.

Let us rewrite this into a form that is convenient for perturbative expansion following the standard steps for path integrals for in-out correlators and scattering amplitudes. First, we rewrite the interaction part $\cL_{\mathrm{int}}$ in terms of functional derivatives with respect to $J^\pm$:
\begin{align} \label{ZZ0}
Z[J^\pm]=\exp\left(iS_{\mathrm{int}}^+\left[\frac{\de}{i\de J^+}\right]-iS_{\mathrm{int}}^-\left[\frac{\de}{-i\de J^-}\right]\right) Z_0[J^\pm],
\end{align}
where $S_{\mathrm{int}}^\pm[\zt^\pm]$ is defined as
\begin{align}
S_{\mathrm{int}}^\pm[\zt^\pm]\equiv\int_{C_\pm}\!\!dtd^3x\,\cL_{\mathrm{int}}(\zt^\pm,\dot\zt^\pm),
\end{align}
and the free part of the generating functional $Z_0$ is defined as
\begin{align} \label{Z0J+-}
Z_0[J^\pm]=\int[d\zt^\pm]\exp\left(i\int_{C_\pm}\!\!dtd^3x\,[\cL_2(\zt^+,\dot\zt^+)-\cL_2(\zt^-,\dot\zt^-)+\zt^+J^+-\zt^-J^-]\right).
\end{align}
To proceed further, let us assume that $\cL_2$ has the following form:
\begin{align}
\cL_2=-\frac{1}{2}\zt\cD\zt, 
\end{align}
where $\cD$ is a second-order differential operator with respect to $t$ and $\vec x$. We then introduce four propagators $G^{ab}$ with $a,b\in\{+,-\}$ that satisfy the following equations of motion\footnote{Exactly speaking, the argument of the temporal delta-function sources is the real part of $t_1-t_2$, but we suppressed a symbol for taking the real part for notational simplicity.}:
\begin{align}
\cD_{t_1\vec x_1}G^{++}(t_1,\vec x_1;t_2,\vec x_2)&=-i\de(t_1-t_2)\de^3(\vec x_1-\vec x_2), \label{eq-G++}\\
\cD_{t_1\vec x_1}G^{--}(t_1,\vec x_1;t_2,\vec x_2)&=i\de(t_1-t_2)\de^3(\vec x_1-\vec x_2), \label{eq-G--}\\
\cD_{t_1\vec x_1}G^{\pm\mp}(t_1,\vec x_1;t_2,\vec x_2)&=0, \label{eq-W}
\end{align}
where the subscript $t\vec x$ on $\cD$ means that $\cD$ is taken with respect to $(t,\vec x)$.
Note that the equations \eqref{eq-W} are not enough to fix $G^{\pm\mp}$; we need continuity conditions from $G^{\pm\pm}$, which will be discussed later.
Using them, we can integrate out $Z_0$ to obtain
\begin{align}
&Z_0[J^\pm]=\exp(W_0[J^\pm]),
\end{align}
where the exponent is given by
\begin{align}
W_0[J^\pm]:=\int_{C_\pm}dt_1dt_2d^3x_1d^3x_2 \!\!
\sum_{a,b\,\in\,\{+,-\}} \!\! i^{a+b}
J^a(t_1,\vec x_1)G^{ab}(t_1,\vec x_1;t_2,\vec x_2)J^b(t_2,\vec x_2),
\end{align}
where $i^{a+b}$ is $i^{\pm 2}$ when $(a,b)=(\pm,\pm)$ and $i^0=1$ when $(a,b)=(\pm,\mp)$. Recall again that a time variable in $J^a$ is integrated along $C_a$. Second-order derivatives of $Z_0$ with respect to the sources $J^\pm$ give propagators:
\begin{align}
G^{ab}(t_1,\vec x_1;t_2,\vec x_2)
&= \frac{\de^2Z_0[J^\pm]}{i^{a+b}\de J^a(t_1,\vec x_1)\de J^b(t_2,\vec x_2)}\bigg|_{J^\pm=0},
\end{align}
and they correspond to in-in two-point functions \emph{without interaction} as follows:
\begin{align}
G^{++}(t_1,\vec x_1;t_2,\vec x_2)
&=\langle 0|\mathrm{T}[\hat\zt(t_1,\vec x_1)\hat\zt(t_2,\vec x_2)]|0\rangle_{\mathrm{free}}, \label{G++}\\ 
G^{+-}(t_1,\vec x_1;t_2,\vec x_2)
&=\langle 0|\hat\zt(t_2,\vec x_2)\hat\zt(t_1,\vec x_1)|0 \rangle_{\mathrm{free}} , \label{G+-}\\ 
G^{-+}(t_1,\vec x_1;t_2,\vec x_2)
&=\langle 0|\hat\zt(t_1,\vec x_1)\hat\zt(t_2,\vec x_2)|0 \rangle_{\mathrm{free}} , \label{G-+}\\ 
G^{--}(t_1,\vec x_1;t_2,\vec x_2)
&=\langle 0|\ol{\mathrm T}[\hat\zt(t_1,\vec x_1)\hat\zt(t_2,\vec x_2)]|0 \rangle_{\mathrm{free}}, \label{G--}
\end{align} 
where the subscript `free' implies that they are two-point functions without interaction.

Coming back to $Z[J^\pm]$, we can further rewrite \eqref{ZZ0} as follows: 
\begin{align}
&Z[J^\pm]=e^{iS_{\mathrm{int}}^+\left[\frac{\de}{i\de J^+}\right]-iS_{\mathrm{int}}^-\left[\frac{\de}{-i\de J^-}\right]}e^{W_0[J^\pm]+i\int_{C_\pm}dtd^3x(J^+\zt^+-J^-\zt^-)}\big|_{\zt^\pm=0} \nn\\
&\quad=e^{iS_{\mathrm{int}}^+\left[\frac{\de}{i\de J^+}\right]-iS_{\mathrm{int}}^-\left[\frac{\de}{-i\de J^-}\right]}e^{W_0\left[\frac{\de}{\pm i\de\zt^\pm}\right]}e^{i\int_{C_\pm}dtd^3x(J^+\zt^+-J^-\zt^-)}\big|_{\zt^\pm=0} \nn\\
&\quad=e^{W_0\left[\frac{\de}{\pm i\de\zt^\pm}\right]}\exp\left[iS_{\mathrm{int}}^+\left[\zt^+\right]-iS_{\mathrm{int}}^-\left[\zt^-\right]+i\int_{C_\pm}dtd^3x(J^+\zt^+-J^-\zt^-)\right]_{\zt^\pm=0}\,.
\end{align}
Here an important observation is that the differential operator $e^{W_0\left[{\de}/{(\pm i\de\zt^\pm)}\right]}$ yields the Wick contraction of its operand with respect to $\zt^\pm$, in which the Wick pair $\zt^a(t_1,\vec x_1)\zt^b(t_2,\vec x_2)$ is replaced by the propagator $G^{ab}(t_1,\vec x_1;t_2,\vec x_2)$. Setting $\zt^\pm=0$ dictates that all $\zt^\pm$ should be contracted, forcing terms with odd numbers of $\zt^\pm$ to vanish. Hence, we finally obtain a master formula for perturbative expansion\footnote{As is clear from the derivation above, $\zt^\pm$ here are just dummy variables introduced in order to define the Wick contraction.}:
\begin{align}
Z[J^\pm]&=\bigg\langle\kern-5.5pt\bigg\langle
\exp\left[iS_{\mathrm{int}}^+\left[\zt^+\right]-iS_{\mathrm{int}}^-\left[\zt^-\right]+i\int_{C_\pm}dtd^3x(J^+\zt^+-J^-\zt^-)\right]
\bigg\rangle\kern-5.5pt\bigg\rangle,
\end{align}
where the double bracket $\langle\kern-2.5pt\langle\cdots\rangle\kern-2.5pt\rangle$ means the Wick contraction. For example, equal-time correlator $\langle\mathscr{O}(t)\rangle$ is given by
\begin{align}
\langle\mathscr{O}(t)\rangle=
\Bbra 
\mathscr{O}(t)\exp\left(iS_{\mathrm{int}}^+\left[\zt^+\right]-iS_{\mathrm{int}}^-\left[\zt^-\right]\right)
\Kket,
\end{align}
together with expanding the exponential in terms of the vertices in $S_{\mathrm{int}}^\pm$. Diagrammatic rules are derived from this master formula. For more details about the rules, see, for example,~\cite{Chen:2017ryl,Palma:2023idj,Braglia:2024zsl}.

\subsection{More on propagators and correlators}

Let us look at propagators in more detail. As already mentioned, the equations of motion \eqref{eq-W} are not enough to fix $G^{\pm\mp}$. In order to determine them, we can use their relations with (anti-)time ordered propagators $G^{\pm\pm}$. The point is that the four types of propagators are originally a single propagator along the contour $C$. Therefore, $G^{\pm\mp}$ should be defined by continuation from $G^{\pm\pm}$. Let us denote $t_i=t(\lm_i)$ and suppress spatial coordinates in propagators. For example, increasing $\lm_1$ in $G^{++}(t_1;t_2)$, passing through the turn $\lm_f$ and reaching the backward part $-C_-$ of $C$, we obtain $G^{-+}(t_1;t_2)$. This indicates that $G^{-+}(t_1;t_2)$ is a continuation with respect to $t_1$ of the $t_1>t_2$ part of $G^{++}(t_1;t_2)$. By such arguments, we obtain
\begin{align}
&G^{-+}(t_1;t_2) \text{ is a continuation with respect to } \lm_1 \text{ of } 
t_1>t_2 \text{ part of } G^{++}(t_1;t_2), \nn\\
&G^{-+}(t_1;t_2) \text{ is a continuation with respect to } \lm_2 \text{ of } 
t_2>t_1 \text{ part of } G^{--}(t_1;t_2), \nn
\end{align}
and also
\begin{align}
&G^{+-}(t_1;t_2) \text{ is a continuation with respect to } \lm_2 \text{ of } 
t_2>t_1 \text{ part of } G^{++}(t_1;t_2), \nn\\
&G^{+-}(t_1;t_2) \text{ is a continuation with respect to } \lm_1 \text{ of } 
t_1>t_2 \text{ part of } G^{--}(t_1;t_2). \nn
\end{align}

Let us represent the propagators in terms of mode functions. For concreteness, consider the following free equation of motion:
\begin{align}
\cD\zt(t,\vec x)=2\pd_t(a^3\vep_1\pd_t\zt(t,\vec x))-2a\vep_1\pd^2\zt(t,\vec x)=0,
\end{align}
and suppose our Lagrangian $\cL$ is invariant under spatial translation. It is then natural to introduce Fourier mode
function $\zt_k(t)$ as a solution of the equation of motion for $\zt$ in Fourier space:
\begin{align}
\pd_t(2a^3\vep_1\pd_t\zt_k(t))+2a\vep_1k^2\zt_k(t)=0. \label{ztk-eom}
\end{align}
Note that $\zt_k(t)$ depends on $\vec k$ only through its magnitude $k=|\vec k|$.
To fix its normalisation, we impose that the Wronskian should satisfy
\begin{align}
2a^3\vep_1\pd_t\zt_k(t)\bar\zt_k(t)-2a^3\vep_1\pd_t\bar\zt_k(t)\zt_k(t)=-i. \label{ztk-wronskian}
\end{align}
Taking the equations of motion and the continuity conditions of the propagators, we can then express the propagators in terms of the mode functions as\footnote{They can be obtained also from the canonical quantisation of $\zt$ with the mode functions $\zt_k$ and the Green functions expressed in terms of operators given in \eqref{G++}-\eqref{G--}.}
\begin{align}
G^{++}_k(t_1,t_2)&=\te(t_1-t_2)\zt_k(t_1)\bar\zt_k(t_2)+\te(t_2-t_1)\zt_k(t_2)\bar\zt_k(t_1), \\ 
G^{+-}_k(t_1,t_2)&=\zt_k(t_2)\bar\zt_k(t_1), \\ 
G^{-+}_k(t_1,t_2)&=\zt_k(t_1)\bar\zt_k(t_2), \\ 
G^{--}_k(t_1,t_2)&=\te(t_2-t_1)\zt_k(t_1)\bar\zt_k(t_2)+\te(t_1-t_2)\zt_k(t_2)\bar\zt_k(t_1),
\end{align} 
where we introduced $G_k^{ab}(t_1;t_2)$ by
\begin{align}
G^{ab}(t_1,\vec x_1;t_2,\vec x_2)=\int\frac{d^3k}{(2\pi)^3}e^{i\vec k\cdot(\vec x_1-\vec x_2)}G_k^{ab}(t_1,t_2),
\end{align}
with spatial coordinates recovered on the left hand side.
It is also convenient to introduce $G_k^>,G_k^<$ by
\begin{align}
G_k^>(t_1,t_2)&\equiv\zt_k(t_1)\bar\zt_k(t_2), \\
G_k^<(t_1,t_2)&\equiv\zt_k(t_2)\bar\zt_k(t_1).
\end{align}

Let us show some properties of equal-time correlators. We only focus on three-point functions of $\zt$ in tree level 
\begin{align}
\langle \zt^a(t,\vec x_1)\zt^b(t,\vec x_2)\zt^c(t,\vec x_3) \rangle,
\end{align}
and consider a three-point vertex of the form 
\begin{align}\label{3ptvertex-SK}
\cL_{\mathrm{int}}=\cD_1\zt(t,\vec x)\cD_2\zt(t,\vec x)\cD_3\zt(t,\vec x),
\end{align}
where each $\cD_i$ is 1, the temporal first derivative $\pd_t$, or a differential operator with respect to spatial coordinates. This assumption excludes vertices proportional to the free equation of motion of $\zt$ as they contain second time derivatives. After perturbative expansion, this correlator at tree level reads
\begin{align}
\begin{split}
&\quad i\int_{C_+}dt'd^3x'\,\cD'_1G^{a+}(t,\vec x_1;t',\vec x')\cD'_2G^{b+}(t,\vec x_2;t',\vec x')\cD'_3G^{c+}(t,\vec x_3;t',\vec x') \\
&-i\int_{C_-}dt'd^3x'\,\cD'_1G^{a-}(t,\vec x_1;t',\vec x')\cD'_2G^{b-}(t,\vec x_2;t',\vec x')\cD'_3G^{c-}(t,\vec x_3;t',\vec x') \\
&+\text{permutations},
\end{split}
\label{ztaztbztc-1}
\end{align}
where the prime on each $\cD_i$ means that it acts with respect to $t',\vec x'$. Note that the differential operators act on the Green functions. In other words, we first preform the Wick contraction leaving the differential operators aside and then act with them on the resulting Green functions with corresponding coordinates. This is the consequence of the definition of the Wick contraction operator $e^{W_0\left[{\de}/{(\pm i\de\zt^\pm)}\right]}$ that is directly with respect to the fields $\zt^\pm$.

We show that \eqref{ztaztbztc-1} is independent of the choice of $t_f$ and of the choice of $a,b,c$.
To show this, let us look at the part of $C_\pm$ which is later than $t$; namely the part from $t$ to $t_f$ of $C_\pm$:
\begin{align}
\begin{split}
&\quad i\int_{t,C_+}^{t_f}dt'd^3x'\,\cD'_1G^{a+}(t,\vec x_1;t',\vec x')\cD'_2G^{b+}(t,\vec x_2;t',\vec x')\cD'_3G^{c+}(t,\vec x_3;t',\vec x') \\
&-i\int_{t,C_-}^{t_f}dt'd^3x'\,\cD'_1G^{a-}(t,\vec x_1;t',\vec x')\cD'_2G^{b-}(t,\vec x_2;t',\vec x')\cD'_3G^{c-}(t,\vec x_3;t',\vec x')\\
&+\text{permutations}.
\end{split}
\end{align}
However, this vanishes because of $G^{a+}(t,\vec x;t',\vec x')=G^{a-}(t,\vec x;t',\vec x')$ along the paths by construction. Therefore, for vertices of type \eqref{3ptvertex-SK}, the time integration region of the vertices can be just up to the time $t$ of equal-time operators from the past infinity. It is therefore independent of the choice of $t_f$.
Next, it is independent of the choice of $a,b,c$ because of $G^{+a}(t,\vec x;t',\vec x')=G^{-a}(t,\vec x;t',\vec x')$ for $t'$ from infinite past up to $t$ along $C_\pm$.

Let us consider a vertex which is a total derivative with respect to time of the form
\begin{align}
\cL_{\mathrm{int}}=\frac{\pd}{\pd t}\left[\hat\cD_1\zt(t,\vec x)\hat\cD_2\zt(t,\vec x)\hat\cD_3\zt(t,\vec x)\right].
\end{align}
After perturbative expansion, this correlator at tree level reads
\begin{align}
\begin{split}
&\quad i\int d^3x'\,\hat\cD'_1G^{a+}(t,\vec x_1;t_f,\vec x')\hat\cD'_2G^{b+}(t,\vec x_2;t_f,\vec x')\hat\cD'_3G^{c+}(t,\vec x_3;t_f,\vec x') \\
&-i\int d^3x'\,\hat\cD'_1G^{a-}(t,\vec x_1;t_f,\vec x')\hat\cD'_2G^{b-}(t,\vec x_2;t_f,\vec x')\hat\cD'_3G^{c-}(t,\vec x_3;t_f,\vec x') \\
&+\text{permutations}.
\end{split}
\end{align}
However, this vanishes because of $G^{a+}(t,\vec x;t_f,\vec x')=G^{a-}(t,\vec x;t_f,\vec x')$ for $t<t_f$. Therefore, temporal total derivative vertices do not contribute~\cite{Braglia:2024zsl,Kawaguchi:2024lsw}.

In contrast, \cite{Braglia:2024zsl,Kawaguchi:2024lsw} showed that vertices proportional the free equation of motion \emph{do} contribute and give the same result as the contributions from a shift of $\zt$ introduced first in~\cite{Maldacena:2002vr}. This is demonstrated in Section~\ref{sec:shift}.

\subsection{Comment on derivative couplings}
\label{subsec:deriv-coupling}
Expanding the total action in ADM form in $\zt$ directly in the comoving gauge without doing anything else such as integration by part, we obtain an action $\cL_{\mathrm{int}}^0(\zt,\dot\zt)$ that contains only $\zt$ and $\dot\zt$, as explained at the end of Appendix~\ref{action}. Since $\cL_{\mathrm{int}}^0$ contains vertices like $\zt\dot\zt^2$, the canonical momentum $\pi_\zt$ contains higher order terms in $\zt,\pi_\zt$ in addition to $\dot\zt$, and therefore $\dot\zt$ in terms of $\zt,\pi_\zt$ is not only equal to $\pi_\zt$ but also gets higher order corrections in $\zt,\pi_\zt$. But, such corrections give higher order vertices in $\cH_{\mathrm{int}}$ and hence are irrelevant to our computations. We therefore have
\begin{align}
\cH_{\mathrm{int}}^0(\zt,\pi_\zt)|_3=-\cL_{\mathrm{int}}^0(\zt,\dot\zt=\pi_\zt),
\end{align}
where $|_3$ means the truncation up to third order.
Therefore, the Lagrangian in the corresponding path integrals is the original one, $\cL_{\mathrm{free}}+\cL_{\mathrm{int}}^0(\zt,\dot\zt)$. Once we obtain this, we are free to do integrations by part in the path integral in order to obtain \eqref{eq:actionz2} and \eqref{eq:action}. Detailed analysis about the relation between Lagnrangian and Hamiltonian with derivative couplings (aiming at applications to computations at one-loop) and its effect on path integrals is given in~\cite{Abolhasani:2022twf,Kawaguchi:2024rsv}.

\section{Expansion in Hubble flow parameters\label{Appendixexpansion}}
In this appendix, we expand all the relevant quantities at N3LO (up to third order in Hubble flow parameters), following the approach of~\cite{Auclair:2022yxs}. Let us start by relating the conformal time to the scale factor $a(t)=e^{\rho(t)}$. We can write the conformal time as
\begin{align}
    \begin{split}
        -\tau(t)&=\int_t^\infty\frac{dt'}{a(t')}=\int_{a(t)}^\infty\frac{da}{a^2H}\\
        &=\frac{1}{aH}+\int_{a(t)}^\infty\frac{1}{a}\frac{d\big(H^{-1}\big)}{da}da\\
        &=\frac{1}{aH}+\int_{a(t)}^\infty\frac{\varepsilon_1}{a^2H}da,
    \end{split}
\end{align}
where we have performed an integration by parts to get the second line. We therefore see that, by doing successively some integrations by parts, we can perform an expansion of $\tau$ in Hubble flow functions. Up to third order, we get
\begin{align}
    -\tau=\frac{e^{-\rho}}{H}\Big[1+\varepsilon_1\big(1+\varepsilon_1+\varepsilon_2+\varepsilon_1^2+3\varepsilon_1\varepsilon_2+\varepsilon_2^2+\varepsilon_2\varepsilon_3\big)\Big]+\mathcal{O}(\varepsilon^4), \label{eq:etaorder3}
\end{align}
and then
\begin{align}
    e^{-\rho}=-\tau H\Big[1-\varepsilon_1\big(1+\varepsilon_2+\varepsilon_1\varepsilon_2+\varepsilon_2^2+\varepsilon_2\varepsilon_3\big)\Big]+\mathcal{O}(\varepsilon^4).\label{eq:exp}
\end{align}
Next, let us expand the Hubble flow functions $\varepsilon_i(t)$ and Hubble function $H(t)$. To do so, we Taylor-expand these functions around a given time $t^*$. The Hubble flow functions evaluated at this time are now real $constants$, namely, the Hubble flow \emph{parameters}. We then have
\begin{align}
\begin{split}
\varepsilon_i&=\varepsilon_i^*+(t-t^*)H^*\varepsilon_i^*\varepsilon_{i+1}^*+\frac{1}{2}(t-t^*)^2H^{*2}\varepsilon_i^*\varepsilon_{i+1}^*\left(-\varepsilon_1^*+\varepsilon_{i+1}^*+\varepsilon_{i+2}^*\right) \\
&\quad+\frac{1}{6}(t-t^*)^3H^{*3}\varepsilon_i^*\varepsilon_{i+1}^*\big(2\varepsilon_1^{*2}-\varepsilon_1^*\varepsilon_2^*-3\varepsilon_1^*\varepsilon_{i+1}^*+\varepsilon_{i+1}^{*2}-3\varepsilon_1^*\varepsilon_{i+2}^* \\
&\quad+3\varepsilon_{i+1}^*\varepsilon_{i+2}^*+\varepsilon_{i+2}^{*2}+\varepsilon_{i+2}^*\varepsilon_{i+3}^*\big)+\mathcal{O}\big(\varepsilon^{*5}\big),\label{eq:ei}
\end{split} \\
\begin{split}
H&=H^*-(t-t^*)H^{*2}\varepsilon_1^*+\frac{1}{2}(t-t^*)^2H^{*3}\varepsilon_1^*\big(2\varepsilon_1^*-\varepsilon_2^*\big)\\
&\quad
-\frac{1}{6}(t-t^*)^3H^{*4}\varepsilon_1^*\left(6\varepsilon_1^{*2}-7\varepsilon_1^*\varepsilon_2^*+\varepsilon_2^{*2}+\varepsilon_2^*\varepsilon_3^*\right)+\mathcal{O}\left(\varepsilon^{*4}\right),\label{eq:rhodot}\end{split}
\end{align}
where the quantities are functions of $t$ when no explicit time dependence is indicated and of $t^*$ when they carry a star. In the following, we will mainly be interested in those quantities expressed in terms of the conformal time. As a consequence, we have to express $(t-t^*)$ in terms of the conformal time up to second order in Hubble flow parameters. To do so, let us look at $\tau^*-\tau$ and expand it at second order:
\begin{align}
    \begin{split}
        \tau^*-\tau&=\int_{t}^{t^{*}} e^{-\rho(t')}dt'\\
        &=\int_{t}^{t^{*}}\sum\limits_{n=0}^{\infty}\frac{(t'-t^{*})^n}{n!}\frac{\partial^n(e^{-\rho^*})}{\partial t^{*n}}dt'\\
        &=\sum\limits_{n=0}^{\infty}\Bigg[1+\varepsilon_1^*\frac{n(n-1)}{2}\bigg(1+\frac{(n-2)(3n-1)}{12}\varepsilon_1^*-\frac{(n-2)}{3}\varepsilon_2^*\\
        &\qquad\qquad+\frac{n(n-1)(n-2)(n-3)}{24}\varepsilon_1^{*2}-\frac{(n-2)(n-3)(2n-1)}{12}\varepsilon_1^*\varepsilon_2^*\\
        &\qquad\qquad+\frac{(n-2)(n-3)}{12}\big(\varepsilon_2^{*2}+\varepsilon_2^*\varepsilon_3^*\big)\bigg)+\mathcal{O}\big(\varepsilon^{*4}\big)\Bigg]\\
        &\quad\times\frac{(-H^*)^n}{n!} e^{-\rho^*}\int_{t}^{t^*}(t'-t^*)^ndt'\\
        &=-\tau^*\sum\limits_{n=0}^{\infty}\frac{\big((t^*-t)H^*\big)^{n+1}}{(n+1)n!}\Bigg[1+\varepsilon_1^*\frac{(n+1)(n-2)}{2}\\
        &\qquad\qquad+\frac{n+1}{24}\varepsilon_1^*\bigg(n(n-1)(3n-10)\varepsilon_1^*-4(6+n(n-4))\varepsilon_2^*\bigg)\\
        &\qquad\qquad+\frac{n+1}{48}\varepsilon_1^*\bigg(n(n-1)(n-2)(2+n(n-5))\varepsilon_1^{*2}\\
        &\qquad\qquad-\big(48-2n(38-n(39+n(2n-15)))\big)\varepsilon_1^*\varepsilon_2^*\\
        &\qquad\qquad+2(n-4)(6+n(n-3))\big(\varepsilon_2^{*2}+\varepsilon_2^*\varepsilon_3^*\big)\bigg)\Bigg]+\mathcal{O}\big(\varepsilon^{*4}\big)\\
        &=\tau^*\Bigg[1-e^\De\Bigg(1+\Delta\varepsilon_1^*\bigg(\frac{\Delta-2}{2}+\frac{3\Delta-4}{24}\Delta^2\varepsilon_1^{*}-\frac{\Delta^2-3\Delta+6}{6}\varepsilon_2^*\\
        &\qquad\quad+\frac{\Delta^2+2\Delta-4}{48}\Delta^3\varepsilon_1^{*2}-\frac{2\Delta^4-3\Delta^3+8\Delta^2-12\Delta+24}{24}\varepsilon_1^*\varepsilon_2^*\\
        &\qquad\quad+\frac{\Delta^3-4\Delta^2+12\Delta-24}{24}\big(\varepsilon_2^{*2}+\varepsilon_2^*\varepsilon_3^*\big)\bigg)\Bigg)\Bigg]+\mathcal{O}\big(\varepsilon^{*4}\big),
    \end{split}
\end{align}
with $\Delta=(t^*-t)H^*$. From this result, we can now expand $\ln(\tau/\tau^*)$ up to third order as
\begin{align}
\begin{split}
    \ln\left(\frac{\tau}{\tau^*}\right)&=\Delta\Bigg[1+\left(\frac{\Delta}{2}-1\right)\varepsilon_1^*+\left(\frac{\Delta^2}{6}\big(2\varepsilon_1^*-\varepsilon_2^*\big)-\frac{\Delta}{2}\big(\varepsilon_1^*-\varepsilon_2^*\big)-\varepsilon_2^*\right)\varepsilon_1^*\\
    &\qquad~~ +\bigg(\frac{\Delta^3}{24}\big(6\varepsilon_1^{*2}-7\varepsilon_1^*\varepsilon_2^*+\varepsilon_2^{*2}+\varepsilon_2^*\varepsilon_3^*\big)-\frac{\Delta^2}{6}\big(2\varepsilon_1^{*2}-4\varepsilon_1^*\varepsilon_2^*+\varepsilon_2^{*2}+\varepsilon_2^*\varepsilon_3^*\big)\\
    &\qquad~~ -\frac{\Delta}{2}\big(\varepsilon_1^*\varepsilon_2^*-\varepsilon_2^{*2}-\varepsilon_2^*\varepsilon_3^*\big)-\varepsilon_2^*\big(\varepsilon_1^*+\varepsilon_2^*+\varepsilon_3^*\big)\bigg)\varepsilon_1^*\Bigg]+\mathcal{O}\big(\varepsilon^{*4}\big).\end{split}
\end{align}
Solving this equation order by order then gives
\begin{align}
\begin{split}
    \Delta&= \ln\left(\frac{\tau}{\tau^*}\right)\Bigg[1+\varepsilon_1^*\Bigg(\big(1+\varepsilon_1^*+\varepsilon_2^*+\varepsilon_1^{*2}+3\varepsilon_1^*\varepsilon_2^*+\varepsilon_2^{*2}+\varepsilon_2^*\varepsilon_3^*\big)\\
    &\qquad\qquad\quad~
-\frac{1}{2}\ln\left(\frac{\tau}{\tau^*}\right)\big(1+2\varepsilon_1^*+\varepsilon_2^*+3\varepsilon_1^{*2}+5\varepsilon_1^*\varepsilon_2^*+\varepsilon_2^{*2}+\varepsilon_2^*\varepsilon_3^*\big)\\
    &\qquad\qquad\quad~
+\frac{1}{6}\left(\ln\left(\frac{\tau}{\tau^*}\right)\right)^2\big(\varepsilon_1^*+\varepsilon_2^*+3\varepsilon_1^{*2}+6\varepsilon_1^*\varepsilon_2^*+\varepsilon_2^{*2}+\varepsilon_2^*\varepsilon_3^*\big)\\
    &\qquad\qquad\quad~
-\frac{1}{24}\left(\ln\left(\frac{\tau}{\tau^*}\right)\right)^3\big(\varepsilon_1^{*2}+3\varepsilon_1^*\varepsilon_2^*+\varepsilon_2^{*2}+\varepsilon_2^*\varepsilon_3^*\big)\Bigg)\Bigg]+\mathcal{O}\big(\varepsilon^{*4}\big).
    \end{split}
    \end{align}
Inserting this result into \eqref{eq:ei} and \eqref{eq:rhodot}, we finally get:
\begin{align}\begin{split}
    \varepsilon_i&=\varepsilon_i^*\Bigg[1-\ln\left(\frac{\tau}{\tau^*}\right)\varepsilon_{i+1}^*\big(1+\varepsilon_1^*+\varepsilon_1^{*2}+\varepsilon_1^*\varepsilon_2^*\big)\\
    &\qquad\quad+\frac{1}{2}\left(\ln\left(\frac{\tau}{\tau^*}\right)\right)^2\varepsilon_{i+1}^*\big(\varepsilon_{i+1}^*+\varepsilon_{i+2}^*+\varepsilon_1^*\varepsilon_2^*+2\varepsilon_1^*\varepsilon_{i+1}^*+2\varepsilon_1^*\varepsilon_{i+2}^*\big)\\
    &\qquad\quad-\frac{1}{6}\left(\ln\left(\frac{\tau}{\tau^*}\right)\right)^3\varepsilon_{i+1}^*\big(\varepsilon_{i+1}^{*2}+3\varepsilon_{i+1}^*\varepsilon_{i+2}^*+\varepsilon_{i+2}^{*2}+\varepsilon_{i+2}^*\varepsilon_{i+3}^*\big)\Bigg]+\mathcal{O}\big(\varepsilon^{*5}\big),\label{eq:eistar}\end{split}\\
    \begin{split}\dot{\rho}&=\dot{\rho}^*\Bigg[1+\ln\left(\frac{\tau}{\tau^*}\right)\varepsilon_{1}^*\big(1+\varepsilon_1^*+\varepsilon_1^{*2}+\varepsilon_1^*\varepsilon_2^*\big)\\
    &\qquad\quad+\frac{1}{2}\left(\ln\left(\frac{\tau}{\tau^*}\right)\right)^2\varepsilon_{1}^*\big(\varepsilon_1^*-\varepsilon_2^*+2\varepsilon_1^{*2}-3\varepsilon_1^*\varepsilon_2^*\big)\\
    &\qquad\quad+\frac{1}{6}\left(\ln\left(\frac{\tau}{\tau^*}\right)\right)^3\varepsilon_{1}^*\big(\varepsilon_1^{*2}-3\varepsilon_1^*\varepsilon_2^*+\varepsilon_{2}^{*2}+\varepsilon_2^*\varepsilon_3^*\big)\Bigg]+\mathcal{O}\big(\varepsilon^{*4}\big).\label{eq:rhodotstar}\end{split}
\end{align}
Another useful quantity is the scale factor $e^\rho$. We can first expand $\rho$ in a similar fashion:
\begin{align}
\begin{split}
    \rho&=\rho^*+(t-t^*)H^*-\frac{1}{2}(t-t^*)^2H^{*2}\varepsilon_1^*+\frac{1}{6}(t-t^*)^3H^{*3}\varepsilon_1^*\big(2\varepsilon_1^*-\varepsilon_2^*\big)\\
    &\quad
-\frac{1}{24}(t-t^*)^4H^{*4}\varepsilon_1^*\big(6\varepsilon_1^{*2}-7\varepsilon_1^*\varepsilon_2^*+\varepsilon_2^{*2}+\varepsilon_2^*\varepsilon_3^*\big)+\mathcal{O}\big(\varepsilon^{*4}\big)\\
    &=\rho^*-\ln\left(\frac{\tau}{\tau^*}\right)-\varepsilon_1^*\Bigg(\ln\left(\frac{\tau}{\tau^*}\right)\big(1+\varepsilon_1^*+\varepsilon_2^*+\varepsilon_1^{*2}+3\varepsilon_1^*\varepsilon_2^*+\varepsilon_2^{*2}+\varepsilon_2^*\varepsilon_3^*\big)\\
    &\qquad\qquad\qquad\qquad\qquad
-\frac{1}{2}\left(\ln\left(\frac{\tau}{\tau^*}\right)\right)^2\vep_2^*\big(1+3\varepsilon_3^*+\varepsilon_2^*+\varepsilon_3^*\big)\\
    &\qquad\qquad\qquad\qquad\qquad
+\frac{1}{6}\left(\ln\left(\frac{\tau}{\tau^*}\right)\right)^3\varepsilon_2^{*}\big(\varepsilon_2^{*}+\varepsilon_3^*\big)\Bigg)+\mathcal{O}\big(\varepsilon^{*4}\big)
    \end{split}
\end{align}
and get the expansion of the scale factor $e^\rho$:
\begin{align}\begin{split}
    e^{\rho}&=e^{\rho^*}\frac{\tau^*}{\tau}\Bigg[1-\ln\left(\frac{\tau}{\tau^*}\right)\varepsilon_1^*\big(1+\varepsilon_1^*+\varepsilon_2^*+\varepsilon_1^{*2}+3\varepsilon_1^*\varepsilon_2^*+\varepsilon_2^{2*}+\varepsilon_2^*\varepsilon_3^*\big)\\
    &\qquad\qquad+\frac{1}{2}\left(\ln\left(\frac{\tau}{\tau^*}\right)\right)^2\varepsilon_1^*\big(\varepsilon_1^*+\varepsilon_2^*+2\varepsilon_1^{*2}+5\varepsilon_1^*\varepsilon_2^*+\varepsilon_2^{*2}+\varepsilon_2^*\varepsilon_3^*\big)\\
     &\qquad\qquad-\frac{1}{6}\left(\ln\left(\frac{\tau}{\tau^*}\right)\right)^3\varepsilon_1^*\big(\varepsilon_1^{*2}+3\varepsilon_1^*\varepsilon_2^*+\varepsilon_{2}^{*2}+\varepsilon_2^*\varepsilon_3^*\big)\Bigg]+\mathcal{O}\big(\varepsilon^{*4}\big).\end{split}
\end{align}
Removing $e^{\rho^*}\tau^*$ by using \eqref{eq:etaorder3}, we finally get
\begin{align}
    \begin{split}\label{eq:expstar}
        e^\rho&=-\frac{1}{\tau H^*}\Bigg[1+\varepsilon_1^*\big(1+\varepsilon_1^*+\varepsilon_2^*+\varepsilon_1^{*2}+3\varepsilon_1^*\varepsilon_2^*+\varepsilon_2^{*2}+\varepsilon_2^*\varepsilon_3^*\big)\\
        &\qquad\qquad\quad
-\ln\left(\frac{\tau}{\tau^*}\right)\varepsilon_1^*\big(1+2\varepsilon_1^*+\varepsilon_2^*+3\varepsilon_1^{*2}+5\varepsilon_1^*\varepsilon_2^*+\varepsilon_2^{*2}+\varepsilon_2^*\varepsilon_3^*\big)\\
        &\qquad\qquad\quad
+\frac{1}{2}\left(\ln\left(\frac{\tau}{\tau^*}\right)\right)^2\varepsilon_1^*\big(\varepsilon_1^*+\varepsilon_2^*+3\varepsilon_1^{*2}+6\varepsilon_1^*\varepsilon_2^*+\varepsilon_2^{*2}+\varepsilon_2^*\varepsilon_3^*\big)\\
        &\qquad\qquad\quad
-\frac{1}{6}\left(\ln\left(\frac{\tau}{\tau^*}\right)\right)^3\varepsilon_1^*\big(\varepsilon_1^{*2}+3\varepsilon_1^*\varepsilon_2^*+\varepsilon_2^{*2}+\varepsilon_2^*\varepsilon_3^*\big)\Bigg]+\mathcal{O}\big(\varepsilon^{*4}\big)
    \end{split}
\end{align}
and its inverse
\begin{align}\begin{split}
    e^{-\rho}&=-\tau H^*\Bigg[1-\varepsilon_1^*\big(1+\varepsilon_2^*+\varepsilon_1^*\varepsilon_2^*+\varepsilon_2^{*2}+\varepsilon_2^*\varepsilon_3^*\big)\\
    &\qquad\qquad\quad
+\ln\left(\frac{\tau}{\tau^*}\right)\varepsilon_1^*\big(1+\varepsilon_2^*+\varepsilon_1^*\varepsilon_2^*+\varepsilon_2^{*2}+\varepsilon_2^*\varepsilon_3^*\big)\\
    &\qquad\qquad\quad
-\frac{1}{2}\left(\ln\left(\frac{\tau}{\tau^*}\right)\right)^2\varepsilon_1^*\big(-\varepsilon_1^*+\varepsilon_2^*-\varepsilon_1^{*2}+\varepsilon_2^{*2}+\varepsilon_2^*\varepsilon_3^*\big)\\
    &\qquad\qquad\quad
+\frac{1}{6}\left(\ln\left(\frac{\tau}{\tau^*}\right)\right)^3\varepsilon_1^*\big(\varepsilon_1^{*2}-3\varepsilon_1^*\varepsilon_2^*+\varepsilon_2^{*2}+\varepsilon_2^*\varepsilon_3^*\big)\Bigg]+\mathcal{O}\big(\varepsilon^{*4}\big).\end{split}
\end{align}

\section{The Wightman functions\label{wightman}}
In this appendix, we derive the Wightman functions for the scalar perturbation $\zeta$ at tree-level at N3LO in Hubble flow parameters. In this appendix, we set $M_\Pl=1$.

\subsection{Wightman functions at N3LO around a pivot scale\label{N3LO}}
Let us first summarise the Wightman functions defined in Appendix~\ref{subsec:master}:
\begin{align}
G^>(t,\vec{x};t',\vec{y})
&=\int\frac{d^3\vec{k}}{(2\pi)^3}e^{i\vec{k}\cdot(\vec{x}-\vec{y})}G^>_k(t,t'),\label{N3LO:eq:Gright}\\
G^<(t,\vec{x};t',\vec{y})
&=\int\frac{d^3\vec{k}}{(2\pi)^3}e^{i\vec{k}\cdot(\vec{x}-\vec{y})}G^<_k(t,t'),\label{N3LO:eq:Gleft}
\end{align}
and the Fourier modes $G^>_k,G^<_k$ are given in terms of the mode function $\zt_k$ by
\begin{align}
G^>_k(t_1,t_2)&=\zt_k(t_1)\bar\zt_k(t_2), \\
G^<_k(t_1,t_2)&=\zt_k(t_2)\bar\zt_k(t_1),
\end{align} 
where $\zt_k$ satisfies the free equation of motion \eqref{ztk-eom} and the normalisation of the Wronskian \eqref{ztk-wronskian}.
They are consequently linked to each other by an exchange of their arguments and a complex conjugation:
\begin{align}
    G^<(t,\vec{x};t',\vec{y})&=G^>(t',\vec{y};t,\vec{x}),\\
    G^<_k(t,t')&=G^>_k(t',t), \\
    G^<_k(t,t')&=\bar{G}^>_k(t,t').
\end{align}
To compute these functions, we start with the equation of motion \eqref{ztk-eom}.
To solve this, we introduce $\vphi_k(\tau)$ with conformal time $\tau$ by rescaling $\zt_k(t)$:
\begin{align}
    \zeta_k(t)=\frac{e^{-\rho(t)}}{\sqrt{2\varepsilon_1(t)}}\varphi_k(\tau).
\end{align}
Substituting this into the equation of motion \eqref{ztk-eom}, we find that $\vphi_k$ satisfies a Klein-Gordon equation
\begin{align}
    \varphi_k''+\big(k^2+m^2\big)\varphi_k=0,\label{eq:phik}
\end{align}
with a \emph{time-dependent} mass $m^2$ given by ($H=\dot\rho$)
\begin{align}\label{eq:mass}
    m^2=-H^2e^{2\rho}\left[2-\varepsilon_1+\frac{3}{2}\varepsilon_2-\frac{1}{2}\varepsilon_1\varepsilon_2+\frac{1}{4}\varepsilon_2^2+\frac{1}{2}\varepsilon_2\varepsilon_3\right].
\end{align}
In order to solve this equation, we actually need to know precisely what is its dependence on time. This is why one has to introduce a pivot scale $t^*$ and to expand the different quantities around that scale. In this appendix, we will compute the Wightman functions at N3LO. Namely, we need to expand the mass \eqref{eq:mass} up to third order in Hubble flow parameters $\varepsilon_i^*$. Thanks to equations \eqref{eq:eistar}, \eqref{eq:rhodotstar} and \eqref{eq:expstar}, we get:
\begin{align}
    m^2=-\frac{2+\mu}{\tau^2}
\end{align}
with
\begin{align}
    \mu(\tau)=\alpha+\beta\ln\left(\frac{\tau}{\tau^*}\right)+\gamma\left(\ln\left(\frac{\tau}{\tau^*}\right)\right)^2+\mathcal{O}\big(\varepsilon^{*4}\big)\label{mu(tau)}
    \end{align}
    and
    \begin{align}
    \begin{split}    
        \alpha&=3\varepsilon_1^*+\frac{3}{2}\varepsilon_2^*+4\varepsilon_1^{*2}+\frac{13}{2}\varepsilon_1^*\varepsilon_2^*+\frac{1}{4}\varepsilon_2^{*2}+\frac{1}{2}\varepsilon_2^*\varepsilon_3^* \\
        &\qquad\quad
+5\varepsilon_1^{*3}+\frac{35}{2}\varepsilon_1^{*2}\varepsilon_2^*+\frac{15}{2}\varepsilon_1^*\varepsilon_2^{*2}+5\varepsilon_1^*\varepsilon_2^*\varepsilon_3^*, 
    \end{split} \\
        \beta&=-\varepsilon_2^*\left(3\varepsilon_1^*+\frac{3}{2}\varepsilon_3^*+11\varepsilon_1^{*2}+\frac{13}{2}\varepsilon_1^*\varepsilon_2^{*}+8\varepsilon_1^*\varepsilon_3^*+\frac{1}{2}\varepsilon_2^{*}\varepsilon_3^*+\frac{1}{2}\varepsilon_3^{*2}+\frac{1}{2}\varepsilon_3^*\varepsilon_4^*\right),\\
        \gamma&=\frac{1}{2}\varepsilon_2^*\left(3\varepsilon_1^*\varepsilon_2^*+3\varepsilon_1^*\varepsilon_3^*+\frac{3}{2}\varepsilon_3^{*2}+\frac{3}{2}\varepsilon_3^*\varepsilon_4^*\right).
\end{align}
We also define the dimensionless variable $x=-k\tau$ (and $x^*=-k\tau^*$) so that \eqref{eq:phik} becomes
\begin{align}\label{eq:eqdiffphik}
    \frac{d^2\varphi_k}{dx^2}+\left(1-\frac{2}{x^2}-\frac{\mu(x)}{x^2}\right)\varphi_k=0
\end{align}
with
\begin{align}
    \mu(x)&=A+B\ln(x)+C\big(\ln(x)\big)^2+\mathcal{O}\big(\varepsilon^{*4}\big)
\end{align}
and
\begin{align}
    A&=\alpha-\beta\ln(x^*)+\gamma\big(\ln(x^*)\big)^2,\label{eq:A}\\
    B&=\beta-2\gamma\ln(x^*),\label{eq:B}\\
    C&=\gamma.\label{eq:C}
\end{align}
The exact solution to this equation turns out to be
\begin{align}\label{vphik-sol}
    \varphi_k(x)=\frac{i}{2}\int_x^\infty\frac{\mu(y)}{y^2}\varphi_k(y)\Big(u(x)\Bar{u}(y)-\Bar{u}(x)u(y)\Big)dy+\varphi_{k,0}(x),
\end{align}
where
\begin{align}
    u(x)=\left(1+\frac{i}{x}\right)e^{ix},
\end{align}
$\Bar{u}(x)$ is its complex conjugate and $\varphi_{k,0}$ is a solution of the homogeneous equation, namely the zeroth-order solution when $\varepsilon_i^*=0$ and therefore $\mu(x)=0$. For $\varphi_{k,0}$, one can get
\begin{align}
    \varphi_{k,0}=c_1\left(1+\frac{i}{x}\right)\frac{e^{ix}}{\sqrt{2k}}+c_2\left(1-\frac{i}{x}\right)\frac{e^{-ix}}{\sqrt{2k}}
\end{align}
with two constants $c_1$ and $c_2$. These constants can be fixed as follows. We first impose the the Wronskian condition \eqref{ztk-wronskian}, which reads in terms of $\vphi_k$ as
\begin{align}
    \varphi_k\ol{\varphi}_k'-\ol{\varphi}_k\varphi_k'=i,
\end{align}
where $\ol{\vphi}_k$ is the complex conjugate of $\vphi_k$.
Since the Wronskian is time-independent, its value is fixed by giving $\vphi_k$ at the past infinity $x\sim\infty$. By construction of the solution \eqref{vphik-sol}, we have $\vphi_k(\infty)=\vphi_{k,0}(\infty)$. Therefore, it is enough to determine $\vphi_{k,0}(\infty)$ such that 
\begin{align}\label{Wronskian-vphik0}
    \varphi_{k,0}\ol{\varphi}_{k,0}'-\ol{\varphi}_{k,0}\varphi_{k,0}'=i.
\end{align}
Next, we require that $\vphi_k(\infty)$ should behave as the wavefunction in the Bunch-Davies vacuum. Since $\vphi_k(\infty)=\vphi_{k,0}(\infty)$, it is enough to require $\vphi_{k,0}$ to satisfy this condition. Combining this with the Wronskian condition \eqref{Wronskian-vphik0} then yields
\begin{align}
    \varphi_{k,0}\sim\frac{e^{ix}}{\sqrt{2k}} \quad \text{as} \quad x\sim\infty,
\end{align}
which means $c_1=1$ and $c_2=0$. The solution at zeroth-order is therefore given by  
\begin{align}
    \varphi_{k,0}(x)=\left(1+\frac{i}{x}\right)\frac{e^{ix}}{\sqrt{2k}}.
\end{align}
We can now compute the higher order parts recursively as
\begin{align}
    \varphi_{k,1}(x)=&\int_x^\infty\mu_1(y)\varphi_{k,0}(y)v(x,y)dy,\\
    \varphi_{k,2}(x)=&\int_x^\infty\Big(\mu_1(y)\varphi_{k,1}(y)+\mu_2(y)\varphi_{k,0}(y)\Big)v(x,y)dy,\\
    \varphi_{k,3}(x)=&\int_x^\infty\Big(\mu_1(y)\varphi_{k,2}(y)+\mu_2(y)\varphi_{k,1}(y)+\mu_3(y)\varphi_{k,0}(y)\Big)v(x,y)dy,\\
    \vdots& \nn
\end{align}
where $\mu_n$ is the $n^{\text{th}}$ order part of $\mu$ and we have expanded $\varphi_k$ order by order as
\begin{align}
    \varphi_k(x)=\varphi_{k,0}(x)+\varphi_{k,1}(x)+\varphi_{k,2}(x)+\varphi_{k,3}(x)+\mathcal{O}\big(\varepsilon^{*4}\big),
\end{align}
and where we have defined
\begin{align}
    v(x,y)=\frac{i}{2}\frac{u(x)\Bar{u}(y)-\Bar{u}(x)u(y)}{y^2}.
\end{align}
By defining the following integrals
\begin{align}
    I_1(x)=&\int_x^\infty v(x,y)u(y)dy,\\
    I_2(x)=&\int_x^\infty v(x,y)u(y)\ln(y)dy,\\
    I_3(x)=&\int_x^\infty v(x,y)I_1(y)dy,\\
    I_4(x)=&\int_x^\infty v(x,y)u(y)\ln^2(y)dy,\\
     I_5(x)=&\int_x^\infty v(x,y)\ln(y)I_1(y)dy,\\
      I_6(x)=&\int_x^\infty v(x,y)I_2(y)dy,\\
       I_7(x)=&\int_x^\infty v(x,y)I_3(y)dy,
\end{align}
one can finally get
\begin{align}
    \sqrt{2k}\varphi_{k,0}(x)&=u(x),\\
    \sqrt{2k}\varphi_{k,1}(x)&=A_1I_1(x),\\
    \sqrt{2k}\varphi_{k,2}(x)&=A_2I_1(x)+B_2I_2(x)+A_1^2I_3(x),\\
  \begin{split}  
    \sqrt{2k}\varphi_{k,3}(x)&=A_3I_1(x)+B_3I_2(x)+2A_1A_2I_3(x)+C_3I_4(x)\\
    &\quad +A_1B_2(I_5(x)+I_6(x))+A_1^3I_7(x),
\end{split}
\end{align}
where $A_i,\ B_i$ and $C_i$ are the $i^{\text{th}}$ order parts of $A,\ B$ and $C$ defined in \eqref{eq:A}, \eqref{eq:B} and \eqref{eq:C}. As suggested in \cite{Auclair:2022yxs}, one can introduce the integrals
\begin{align}
    F_n(x)=&\int_x^\infty\frac{e^{2iy}}{y}\ln^n(y)dy,\\
    F_{00}(x)=&\int_x^\infty\frac{e^{-2iy}}{y}F_0(y)dy,\\
    F_{01}(x)=&\int_x^\infty\frac{e^{-2iy}}{y}F_1(y)dy,\\
    F_{10}(x)=&\int_x^\infty\frac{e^{-2iy}}{y}\ln(y)F_0(y)dy,\\
    F_{000}(x)=&\int_x^\infty\frac{e^{2iy}}{y}F_{00}(y)dy,
\end{align}
so that, after some permutations of integrations and a huge number of integrations by part, the $I_n$ integrals can be expressed using the $F$ integrals only as:
\begin{align}
    I_1(x)&=\frac{e^{-ix}}{3x}\Big[2ie^{2ix}-(x-i)F_0(x)\Big],\\
    I_2(x)&=\frac{e^{-ix}}{9x}\Big[2ie^{2ix}\big(4+3\ln(x)\big)-(x-i)\big(7F_0(x)+3F_1(x)\big)\Big],\\
    I_3(x)&=\frac{e^{-ix}}{27x}\Big[e^{2ix}\big(-2i+3(x+i)F_{00}(x)\big)+(x+5i)F_0(x)\Big],\\
    \begin{split}I_4(x)&=\frac{e^{-ix}}{27x}\Big[2ie^{2ix}\big(26+3\ln(x)\big(8+3\ln(x)\big)\big)\\
    &\qquad\qquad-(x-i)\big(50F_0(x)+42F_1(x)+9F_2(x)\big)\Big],\end{split}\\
   \begin{split} I_5(x)&=\frac{e^{-ix}}{81x}\Big[e^{2ix}\big(-2i\big(14+3\ln(x)\big)+21(x+i)F_{00}(x)+9(x+i)F_{10}(x)\big)\\
    &\qquad\qquad+2\big(-i+13x+9i\ln(x)\big)F_0(x)+3(x-i)F_1(x)\Big],\end{split}\\
    \begin{split}I_6(x)&=\frac{e^{-ix}}{81x}\Big[e^{2ix}\big(-2i\big(14+3\ln(x)\big)+21(x+i)F_{00}(x)+9(x+i)F_{01}(x)\big)\\
    &\qquad\qquad+2(8i+13x)F_0(x)+3(5i+x)F_1(x)\Big],\end{split}\\
    I_7(x)&=\frac{e^{-ix}}{243x}\Big[2e^{2ix}\big(2i-3(x-2i)F_{00}(x)\big)-9(x-i)F_{000}(x)-2(x+5i)F_0(x)\Big].
\end{align}
We therefore have all the ingredients to compute $\varphi_k$ at N3LO. To compute the $\zeta_{k}$, with
\begin{align}
    \zeta_{k}(t)=\frac{e^{-\rho(t)}}{\sqrt{2\varepsilon_1(t)}}\varphi_{k}(\tau),
\end{align}
the final step is to expand the prefactor as
\begin{align}
    \frac{e^{-\rho}}{\sqrt{2\varepsilon_1}}=&-\frac{\tau H^*}{\sqrt{2\varepsilon_1^*}}\Bigg(\tilde{\alpha}+\tilde{\beta}\ln\left(\frac{\tau}{\tau^*}\right)+\tilde{\gamma}\ln\left(\frac{\tau}{\tau^*}\right)^2+\tilde{\delta}\ln\left(\frac{\tau}{\tau^*}\right)^3\Bigg)+\mathcal{O}\big(\varepsilon^{*4}\big)
\end{align}
where the coefficients are
\begin{align}
    \tilde{\alpha}&=1-\varepsilon_1^*-\varepsilon_1^*\varepsilon_2^*-\varepsilon_1^{*2}\varepsilon_2^*-\varepsilon_1^*\varepsilon_2^{*2}-\varepsilon_1^*\varepsilon_2^*\varepsilon_3^*,\\
    \tilde{\beta}&=\varepsilon_1^*+\frac{\varepsilon_2^*}{2}+\varepsilon_1^*\varepsilon_2^*+\varepsilon_1^{*2}\varepsilon_2^*+\varepsilon_1^*\varepsilon_2^{*2}+\varepsilon_1^*\varepsilon_2^*\varepsilon_3^*,\\
    \tilde{\gamma}&=\frac{1}{2}\left(\varepsilon_1^{*2}+\frac{\varepsilon_2^{*2}}{4}-\frac{\varepsilon_2^*\varepsilon_3^*}{2}+\varepsilon_1^{*3}+\varepsilon_1^{*2}\varepsilon_2^*-\frac{\varepsilon_1^*\varepsilon_2^{*2}}{4}-\frac{3}{2}\varepsilon_1^*\varepsilon_2^*\varepsilon_3^*\right),\\
    \tilde{\delta}&=\frac{1}{4}\left(\frac{2}{3}\varepsilon_1^{*3}-\varepsilon_1^{*2}\varepsilon_2^*+\frac{\varepsilon_1^*\varepsilon_2^{*2}}{6}-\frac{\varepsilon_1^*\varepsilon_2^*\varepsilon_3^*}{3}+\frac{\varepsilon_2^{*3}}{12}-\frac{1}{2}\varepsilon_2^{*2}\varepsilon_3^*+\frac{\varepsilon_2^*\varepsilon_3^{*2}}{3}+\frac{\varepsilon_2^*\varepsilon_3^*\varepsilon_4^*}{3}\right),
\end{align}
or as a function of $x$:
\begin{align}
    \frac{e^{-\rho}}{\sqrt{2\varepsilon_1}}=\frac{xH^*}{k\sqrt{2\varepsilon_1^*}}\Big(\tilde{A}+\tilde{B}\ln(x)+\tilde{C}\big(\ln(x)\big)^2+\tilde{D}\big(\ln(x)\big)^3\Big)+\mathcal{O}\big(\varepsilon^{*4}\big)
\end{align}
with
\begin{align}
    \tilde{A}&=\tilde{\alpha}-\tilde{\beta}\ln(x^*)+\tilde{\gamma}\big(\ln(x^*)\big)^2-\tilde{\delta}\big(\ln(x^*)\big)^3,\\
    \tilde{B}&=\tilde{\beta}-2\tilde{\gamma}\ln(x^*)+3\tilde{\delta}\big(\ln(x^*)\big)^2,\\
    \tilde{C}&=\tilde{\gamma}-3\tilde{\delta}\ln(x^*),\\
    \tilde{D}&=\tilde{\delta}.
\end{align}
Finally we have:
\begin{align}
    \zeta_{k,0}&=\frac{xH^*}{2k\sqrt{k\varepsilon_1^*}}\tilde{A}_0 u(x),\\
    \zeta_{k,1}&=\frac{xH^*}{2k\sqrt{k\varepsilon_1^*}}\Big[\Big(\tilde{A}_1 +\tilde{B}_1\ln(x)\Big)u(x)+\tilde{A}_0A_1I_1(x)\Big],\\
    \begin{split}\zeta_{k,2}&=\frac{xH^*}{2k\sqrt{k\varepsilon_1^*}}\Big[\Big(\tilde{A}_2+\tilde{B}_2\ln(x)+\tilde{C}_2\big(\ln(x)\big)^2\Big)u(x)\\
    &\quad+\Big(\tilde{A}_0A_2+A_1\big(\tilde{A}_1+\tilde{B}_1\ln(x)\big)\Big)I_1(x)+\tilde{A}_0B_2I_2(x)+\tilde{A}_0A_1^2I_3(x)\Big],\end{split}\\
    \begin{split}\zeta_{k,3}&=\frac{xH^*}{2k\sqrt{k\varepsilon_1^*}}\Big[\Big(\tilde{A}_3+\tilde{B}_3\ln(x)+\tilde{C}_3\big(\ln(x)\big)^2+\tilde{D}_3\big(\ln(x)\big)^3\Big)u(x)\\
    &\quad+\Big(\tilde{A}_0A_3+\tilde{A}_1A_2+\tilde{A}_2A_1+\big(A_1\tilde{B}_2+A_2\tilde{B}_1\big)\ln(x)+A_1\tilde{C}_2\big(\ln(x)\big)^2\Big)I_1(x)\\
    &\quad+\Big(\tilde{A}_0B_3+\tilde{A}_1B_2+\tilde{B}_1B_2\ln(x)\big)I_2(x)\\
    &\quad+\Big(2\tilde{A}_0A_1A_2+\tilde{A}_1A_1^2+\tilde{B}_1A_1^2\ln(x)\Big)I_3(x)\\
    &\quad+\tilde{A}_0C_3I_4(x)+\tilde{A}_0A_1B_2\big(I_5(x)+I_6(x)\big)+\tilde{A}_0A_1^3I_7(x)\Big].\end{split}
\end{align}
One can now construct the Wightman functions up to N3LO as
\begin{align}
     G_k^>(t,t')=\zeta_{k}(t)\bar{\zeta}_{k}(t')
\end{align}
and its derivatives defined by
\begin{align}
    \dot{G}^>_k(t,t')&\equiv\frac{\partial}{\partial t'}G_k^>(t,t')=\zeta_{k}(t)\dot{\bar{\zeta}}_{k}(t'),\\
    \d{$G$}_k^>(t,t')&\equiv\frac{\partial}{\partial t}G_k^>(t,t')=\dot{\zeta}_{k}(t)\bar{\zeta}_{k}(t').
\end{align}
The other ordering, $G_k^<,\ \dot{G}_k^<$ and \d{$G$}$_k^<$ are just their complex conjugates. In Section \ref{sec:shift}, we are actually interested in those functions at equal time $t=t'$ in the late-time limit only. One can therefore compute the late-time limits of the $F$ functions as in \cite{Auclair:2022yxs}\footnote{$F_{01}$ has been computed in a similar way by defining a generating function for the $F_{0n}$ hierarchy and $F_{10}$ has been computed by using the relation: $F_{10}=F_0\bar{F}_1-\bar{F}_{01}$.}:
\begin{align}
    F_0(x)&=-\ln(x)-\kappa+\mathcal{O}(x),\\
    F_{1}(x)&=-\frac{1}{2}\big(\ln(x)\big)^2+\frac{\kappa^2}{2}+\frac{\pi^2}{12}+\mathcal{O}(x),\\
    F_2(x)&=-\frac{1}{3}\big(\ln(x)\big)^3-\frac{\kappa^3}{3}-\frac{\pi^2}{6}\kappa-\frac{2}{3}\zeta(3)+\mathcal{O}(x),\\
    F_{00}(x)&=\frac{\pi^4}{4}+\frac{\kappa^2}{2}+\kappa\ln(x)+\frac{1}{2}\big(\ln(x)\big)^2+\mathcal{O}(x),\\
    F_{01}(x)&=-\frac{1}{3}\zeta(3)-\frac{\pi^2}{4}\kappa-\frac{1}{6}\kappa^3+\frac{1}{2}\kappa\big(\ln(x)\big)^2+\frac{1}{3}\big(\ln(x)\big)^3+\mathcal{O}(x)\\
   \begin{split} F_{10}(x)&=\frac{1}{3}\zeta(3)+\frac{i\pi^3}{12}+\frac{\pi^2}{6}\kappa-\frac{i\pi}{2}\kappa^2-\frac{\kappa^3}{3}+\left(\frac{5\pi^2}{12}-i\pi\kappa-\frac{\kappa^2}{2}\right)\ln(x)\\
    &\quad-\frac{i\pi}{2}\big(\ln(x)\big)^2+\frac{1}{6}\big(\ln(x)\big)^3+\mathcal{O}(x),\end{split}\\
    \begin{split}F_{000}(x)&=-\frac{7}{3}\zeta(3)-\frac{\pi^2}{4}\kappa-\frac{\kappa^3}{6}-\left(\frac{\pi^2}{4}+\frac{\kappa^2}{2}\right)\ln(x)
    -\frac{\kappa}{2}\big(\ln(x)\big)^2-\frac{1}{6}\big(\ln(x)\big)^3+\mathcal{O}(x),\end{split}
\end{align}
where we have introduced the constant
\begin{align}
    \kappa=\gamma_E+\ln(2)-\frac{i\pi}{2}.
\end{align}
Finally, we get:
\begin{align}
G^>_k(t,t)&=\frac{H^{*2}}{4k^3\varepsilon_1^*}\Big[\mathcal{A}+\mathcal{B}\ln(-k\tau^*)+\mathcal{C}\big(\ln(-k\tau^*)\big)^2+\mathcal{D}\big(\ln(-k\tau^*)\big)^3\Big]+\mathcal{O}\big(\varepsilon^{*3},\tau\big),\\
\dot{G}^>_k(t,t)&=i\pi\frac{H^{*2}}{16k^3\varepsilon_1^*}\varepsilon_2^*\big(2\varepsilon_1^*+\varepsilon_2^*\big)\big(2\varepsilon_1^*+\varepsilon_3^*\big)\big(2+\lambda+\ln(-k\tau)\big)+\mathcal{O}\big(\varepsilon^{*3},\tau\big),\\
\d{$G$}_k^>(t,t)&=-i\pi\frac{H^{*2}}{16k^3\varepsilon_1^*}\varepsilon_2^*\big(2\varepsilon_1^*+\varepsilon_2^*\big)\big(2\varepsilon_1^*+\varepsilon_3^*\big)\big(2+\lambda+\ln(-k\tau)\big)+\mathcal{O}\big(\varepsilon^{*3},\tau\big),\\
    \d{$\dot{G}$}^>(t,t)&=\mathcal{O}\big(\varepsilon^{*3},\tau\big),
\end{align}
with
\begin{align}
    \lambda=\kappa-2+\frac{i\pi}{2}=\gamma_E+\ln(2)-2
\end{align}
and
\begin{align}
    \begin{split}\mathcal{A}&=1-2(\lambda+1)\varepsilon_1^*-\lambda\varepsilon_2^*+\frac{1}{2}\big(\pi^2+4\lambda^2+4\lambda-6\big)\varepsilon_1^{*2}\\
    &\quad+\frac{1}{12}\big(7\pi^2+12\lambda^2-12\lambda-72\big)\varepsilon_1^*\varepsilon_2^*+\frac{1}{8}\big(\pi^2+4\lambda^2-8\big)\vep_2^{*2}+\frac{1}{24}\big(\pi^2-12\lambda^2\big)\varepsilon_2^*\varepsilon_3^*\\
    &\quad-\frac{1}{24}\big(4\lambda^3+3(\pi^2-8)\lambda+14\zeta(3)-16\big)\big(8\varepsilon_1^{*3}+\varepsilon_2^{*3}\big)\\
    &\quad+\frac{1}{12}\big(13\pi^2-8(\pi^2-9)\lambda+36\lambda^2-84\zeta(3)\big)\varepsilon_1^{*2}\varepsilon_2^*\\
    &\quad-\frac{1}{24}\big(8\lambda^3-15\pi^2+6(\pi^2-4)\lambda-12\lambda^2+100\zeta(3)+16\big)\varepsilon_1^*\varepsilon_2^{*2}\\
    &\quad+\frac{1}{12}\big(8\lambda^3+\pi^2+6(\pi^2-12)\lambda-12\lambda^2-8\zeta(3)-8\big)\varepsilon_1^*\varepsilon_2^*\varepsilon_3^*\\
    &\quad+\frac{1}{24}\big(12\lambda^3+(5\pi^2-48)\lambda\big)\varepsilon_2^{*2}\varepsilon_3^*\\
    &\quad+\frac{1}{24}\big(\pi^2\lambda-4\lambda^3-8\zeta(3)+16\big)\big(\varepsilon_2^*\varepsilon_3^{*2}+\varepsilon_2^*\varepsilon_3^*\varepsilon_4^*\big),\end{split}\\
    \begin{split}\mathcal{B}&=-2\varepsilon_1^*-\varepsilon_2^*+(4\lambda+2)\varepsilon_1^{*2}+(2\lambda-1)\varepsilon_1^*\varepsilon_2^*+\lambda\big(\varepsilon_2^{*2}-\varepsilon_2^*\varepsilon_3^*\big)\\
    &\quad-\frac{1}{8}\big(\pi^2+4\lambda^2-8\big)\big(8\varepsilon_1^{*3}+\varepsilon_2^{*3}\big)\\
    &\quad-\frac{2}{3}\big(\pi^2-9\lambda-9\big)\varepsilon_1^{*2}\varepsilon_2^*-\frac{1}{4}\big(\pi^2+4\lambda^2-4\lambda-4\big)\varepsilon_1^*\varepsilon_2^{*2}\\
    &\quad+\frac{1}{2}\big(\pi^2+4\lambda^2-4\lambda-12\big)\varepsilon_1^*\varepsilon_2^*\varepsilon_3^*+\frac{1}{24}\big(5\pi^2+36\lambda^2-48\big)\varepsilon_2^{*2}\varepsilon_3^*\\
    &\quad+\frac{1}{24}\big(\pi^2-12\lambda^2\big)\big(\varepsilon_2^*\varepsilon_3^{*2}+\varepsilon_2^*\varepsilon_3^*\varepsilon_4^*\big),\end{split}\\
    \begin{split}
        \mathcal{C}&=2\varepsilon_1^{*2}+\varepsilon_1^*\varepsilon_2^*+\frac{1}{2}\varepsilon_2^{*2}-\frac{1}{2}\varepsilon_2^*\varepsilon_3^*-4\lambda\varepsilon_1^{*3}+3\varepsilon_1^{*2}\varepsilon_2^*-\frac{1}{2}(2\lambda-1)\big(\varepsilon_1^*\varepsilon_2^{*2}-2\varepsilon_1^*\varepsilon_2^*\varepsilon_3^*\big)\\
        &\quad+\frac{1}{2}\lambda\big(-\varepsilon_2^{*3}+3\varepsilon_2^{*2}\varepsilon_3^*-\varepsilon_2^*\varepsilon_3^{*2}-\varepsilon_2^*\varepsilon_3^*\varepsilon_4^*\big),\end{split}\\
        \mathcal{D}&=-\frac{1}{4}\varepsilon_1^{*3}-\frac{1}{3}\varepsilon_1^*\varepsilon_2^{*2}+\frac{2}{3}\varepsilon_1^*\varepsilon_2^*\varepsilon_3^*-\frac{1}{6}\varepsilon_2^{*3}+\frac{1}{2}\varepsilon_2^{*2}\varepsilon_3^*-\frac{1}{6}\varepsilon_2^*\varepsilon_3^{*2}-\frac{1}{6}\varepsilon_2^*\varepsilon_3^*\varepsilon_4^*.
\end{align}
The computations of Section \ref{sec:shift} actually involve those quantities up to N2LO. One can therefore conclude that up to that order, the derivatives of the Wightman functions at equal time indeed vanish in the late-time limit. Moreover, by defining the power spectrum as
\begin{align}
    \mathcal{P}_\zeta(k)=\frac{k^3}{2\pi^2}G^>(t,t),
\end{align}
we recover here the power spectrum at N3LO computed in \cite{Auclair:2022yxs,Bianchi:2024qyp}.

\subsection{Wightman functions at NLO}
Actually, if one is interested in the Wightman functions at NLO only, it is not necessary to expand everything around an arbitrary pivot scale. This is because at NLO, the Hubble flow functions $\varepsilon_i$ can actually be considered as \emph{constants} as their variations appear at next order. Therefore, the mass in \eqref{eq:mass} is just
\begin{align}
    m^2&=-H^2e^{2\rho}\left(2-\varepsilon_1+\frac{3}{2}\varepsilon_2\right)+\mathcal{O}\big(\varepsilon^2\big)\\
    &=-\frac{1}{\tau^2}\left(2+3\varepsilon_1+\frac{3}{2}\varepsilon_2\right)+\mathcal{O}\big(\varepsilon^2\big),
\end{align}
so $\mu$ given by \eqref{mu(tau)}:
\begin{align}\label{mu1}
    \mu=3\varepsilon_1+\frac{3}{2}\varepsilon_2+\mathcal{O}\big(\varepsilon^2\big)
\end{align}
is now considered as a constant. 
Computing the Wightman functions order by order as previously explained  and expanding the functions of $t'$ around $t$ as a pivot scale then gives us
\begin{align}\begin{split}\label{eq:grightNLO}
    G^>_k(t,t')&=\frac{H^2}{4k^3\varepsilon_1}e^{-ik(\tau-\tau')}(1+ik\tau)(1-ik\tau')\\
    &\quad\times\Bigg\{1-2\varepsilon_1+\left(\varepsilon_1+\frac{\varepsilon_2}{2}\right)\Bigg[\frac{2}{1+ik\tau}+\frac{2}{1-ik\tau'}\\
    &\qquad\quad -e^{2ik\tau}\frac{1-ik\tau}{1+ik\tau}\Big(\text{Ei}(-2ik\tau)-i\pi\Big)\\
    &\qquad\quad -e^{-2ik\tau'}\frac{1+ik\tau'}{1-ik\tau'}\Big(\text{Ei}(2ik\tau')+i\pi\Big)+\ln\left(\frac{\tau'}{\tau}\right)\Bigg]\Bigg\}+\mathcal{O}(\varepsilon),\end{split}\\
    \begin{split}G^<_k(t,t')&=\frac{H^2}{4k^3\varepsilon_1}e^{ik(\tau-\tau')}(1-ik\tau)(1+ik\tau')\\
    &\quad\times\Bigg\{1-2\varepsilon_1+\left(\varepsilon_1+\frac{\varepsilon_2}{2}\right)\Bigg[\frac{2}{1-ik\tau}+\frac{2}{1+ik\tau'}\\
    &\qquad\quad -e^{-2ik\tau}\frac{1+ik\tau}{1-ik\tau}\Big(\text{Ei}(2ik\tau)+i\pi\Big)\\
    &\qquad\quad -e^{2ik\tau'}\frac{1-ik\tau'}{1+ik\tau'}\Big(\text{Ei}(-2ik\tau')-i\pi\Big)+\ln\left(\frac{\tau'}{\tau}\right)\Bigg]\Bigg\}+\mathcal{O}(\varepsilon),\label{eq:GleftNLO}\end{split}
\end{align}
and
\begin{align}
\begin{split}
    \dot{G}_k^>(t,t')&=-\frac{H^3}{4k\varepsilon_1}e^{-ik(\tau-\tau')}(1+ik\tau)\tau'^2\\
    &\quad
\times\Bigg\{1-3\varepsilon_1+\left(\varepsilon_1+\frac{\varepsilon_2}{2}\right)\Bigg[\frac{2}{1+ik\tau}-e^{-2ik\tau'}\big(\text{Ei}(2ik\tau')+i\pi\big)\\
    &\quad\qquad
-e^{2ik\tau}\frac{1-ik\tau}{1+ik\tau}\big(\text{Ei}(-2ik\tau)-i\pi\big)\Bigg]+\left(2\varepsilon_1+\frac{\varepsilon_2}{2}\right)\ln\left(\frac{\tau'}{\tau}\right)\Bigg\}+\mathcal{O}(\varepsilon),
    \end{split}\\
    \begin{split}
    \dot{G}_k^<(t,t')&=-\frac{H^3}{4k\varepsilon_1}e^{ik(\tau-\tau')}(1-ik\tau)\tau'^2\\
    &\quad
\times\Bigg\{1-3\varepsilon_1+\left(\varepsilon_1+\frac{\varepsilon_2}{2}\right)\Bigg[\frac{2}{1-ik\tau}-e^{2ik\tau'}\big(\text{Ei}(-2ik\tau')-i\pi\big)\\
    &\quad\qquad
-e^{-2ik\eta}\frac{1+ik\tau}{1-ik\tau}\big(\text{Ei}(2ik\tau)+i\pi\big)\Bigg]+\left(2\varepsilon_1+\frac{\varepsilon_2}{2}\right)\ln\left(\frac{\tau'}{\tau}\right)\Bigg\}+\mathcal{O}(\varepsilon).
    \end{split}
\end{align}
Actually, as $\mu$ is constant at NLO, equation \eqref{eq:eqdiffphik} can be solved exactly, leading to the usual form of the Wightman functions:
\begin{align}
    G_{k}^>(t,t')&=\frac{\pi}{8}\frac{e^{-\rho(t)}e^{-\rho(t')}}{\sqrt{\varepsilon_1(t)\varepsilon_1(t')}}
{\sqrt{\tau\tau'}}H_\nu^{(1)}(-k\tau)H_{\nu}^{(2)}(-k\tau'),\\
    G_{k}^<(t,t')&=\frac{\pi}{8}\frac{e^{-\rho(t)}e^{-\rho(t')}}{\sqrt{\varepsilon_1(t)\varepsilon_1(t')}}
{\sqrt{\tau\tau'}}H_\nu^{(2)}(-k\tau)H_{\nu}^{(1)}(-k\tau'),
\end{align}
where the functions $H_{\nu}^{(i)}$ are the Hankel functions and the index $\nu$ is given by
\begin{align}
    \nu=\frac{1}{2}\sqrt{9+4\mu}=\frac{3}{2}\sqrt{1+\frac{4}{3}\left(\varepsilon_1+\frac{\varepsilon_2}{2}\right)},
\end{align}
where we used the linear part of \eqref{mu1}.
Note however that this result is true only up to NLO. To render that explicit, one can do a Taylor expansion of $\nu$ around $3/2$ up to first order. This will introduce derivatives of the Hankel functions with respect to their parameter. In order to compute it, we use the expression 
\begin{align}
    H_{\nu}^{(1)}(z)=\sqrt{\frac{2}{\pi z}}\frac{\exp\left[i\left(z-\frac{\pi}{2}\nu-\frac{\pi}{4}\right)\right]}{\Gamma\left(\nu+\frac{1}{2}\right)}\int_{0}^{\infty}\left(1+\frac{it}{2z}\right)^{\nu-\frac{1}{2}}t^{\nu-\frac{1}{2}}e^{-t}dt
\end{align}
valid for $\text{Re}(\nu)>-1/2$ and $-\pi/2<\text{arg}(z)<3\pi/2$, to show that
\begin{align}
    \partial_\nu H_{\nu=\frac{3}{2}}^{(1)}(z)=\frac{1}{\sqrt{2\pi}z^{\frac{3}{2}}}\left[\big(\pi(iz-1)-4i\big)e^{iz}+2(1+iz)\big(\pi+i\text{Ei}(2iz)\big)e^{-iz}\right]
\end{align}
and identically 
\begin{align}
    \partial_\nu H_{\nu=\frac{3}{2}}^{(2)}(z)=\frac{1}{\sqrt{2\pi}z^{\frac{3}{2}}}\left[\big(4i-\pi(iz+1)\big)e^{-iz}+2(1-iz)\big(\pi-i\text{Ei}(-2iz)\big)e^{iz}\right].
\end{align}
Going through this procedure will give again, as expected, the previous results \eqref{eq:grightNLO} and \eqref{eq:GleftNLO}.

\section{The $f_{\text{NL}}$ parameters\label{sec:fNLs}}

In this appendix, we examine in greater detail the $f_{\text{NL}}$ parameter-function, defined as
\begin{align}
    f_{\text{NL}}=\frac{5}{12}\frac{\mathscr{A}}{k_1^3+k_2^3+k_3^3}.
\end{align}
To this end, we decompose it at leading and next-to-leading order as
\begin{align}
    f_{\text{NL,LO}}&=f_1\varepsilon_1+f_2\varepsilon_2,\\
    f_{\text{NL,NLO}}&=f_{11}\varepsilon_1^2+f_{12}\varepsilon_1\varepsilon_2+f_{22}\varepsilon_2^2+f_{23}\varepsilon_2\varepsilon_3,
\end{align}
where the coefficients $f_i$ are functions of the three momenta. In fact, the functions $f_1$ and $f_2$ coincide with the previously introduced quantities $\tilde{K}_1$ and $\tilde{K}_2$ in \eqref{Ktilde}, yielding
\begin{equation}\label{eq:fiLO}
    \begin{array}{ccccccccccc}
    0.83&\lesssim&f_{1}&\lesssim&1.53&\qquad\quad&\text{and}&\qquad\quad&f_2&\simeq&0.42.
    \end{array}
\end{equation}
Since the functions $f_i$ are dimensionless, they can be expressed in terms of rescaled momenta. 
By normalising all momenta to the largest one $k_{i,\max}$, they depend only on two independent variables, $x$ and $y$, 
corresponding to the ratios of the other two momenta $k_j/k_{i,\max}$. 
Introducing the functions
\begin{align}
    \mathscr{F}_n(x,y)=\frac{1+x^4+y^4-10\big(x^2y^2+x^2+y^2\big)}{1+x+y}-nxy,
\end{align}
and evaluating them at the last horizon exit time\footnote{The functions $f_i$ can actually be expressed solely in terms of $x$ and $y$ for all contributions except $\ln\big(-\tau(k_1+k_2+k_3)\big)$, which at the last horizon exit reduces to $\ln(1+x+y) + \mathcal{O}(\varepsilon)$, thus depending only on $x$ and $y$ as well.} $-\tau=1/k_{i,\max}+\mathcal{O}(\varepsilon)$, we find:
\begin{align}
    f_1&=-\frac{5}{12}\frac{\mathscr{F}_2}{1+x^3+y^3},\\
    f_2&=\frac{5}{12},\\
    f_{11}&=\frac{5}{12}\frac{1}{1+x^3+y^3}\Bigg\{-4\mathscr{F}_{5/2}+4\mathscr{F}_2\big(\gamma_E+\ln2\big) 
    +2\frac{1+y}{1-x+y}\mathscr{F}_0\ln x+2\frac{1+x}{1+x-y}\mathscr{F}_0\ln y \nn\\
    &\quad+16\frac{x^2+y^2+x^2y^2}{1+x+y}\left(\frac{x+y}{x+y-1}+\frac{1+x}{1+x-y}+\frac{1+y}{1-x+y}-2\right)\ln\left(\frac{1+x+y}{2}\right)\Bigg\}, \\
    \begin{split}f_{12}&=\frac{5}{12}\frac{1}{1+x^3+y^3}\Bigg\{(1+x+y)(x+y+xy)+2\big(1+x^3+y^3\big)-3\mathscr{F}_0\\
    &\quad+\Big(\mathscr{F}_2-4\big(1+x^3+y^3\big)\Big)\big(\gamma_E+\ln 2\big)\\
    &\quad+\left(\frac{1+y}{1-x+y}\mathscr{F}_0-2\big(1+y^3\big)\right)\ln x+\left(\frac{1+x}{1+x-y}\mathscr{F}_0-2\big(1+x^3\big)\right)\ln y\\
    &\quad+\left(\mathscr{F}_2-\left(\frac{x+y}{x+y-1}+\frac{1+x}{1+x-y}+\frac{1+y}{1-x+y}\right)\mathscr{F}_0\right)\ln\left(\frac{1+x+y}{2}\right)\Bigg\},\end{split}\\
    f_{22}&=\frac{5}{12}\Bigg\{4-2\big(\gamma_E+\ln 2\big)-\frac{1+y^3}{1+x^3+y^3}\ln x-\frac{1+x^3}{1+x^3+y^3}\ln y\Bigg\},\\
    f_{23}&=\frac{5}{12}\Bigg\{\frac{4xy-(x+y+xy)(1+x+y)}{1+x^3+y^3}-1+\big(\gamma_E+\ln 2\big)+\ln\left(\frac{1+x+y}{2}\right)\Bigg\}.
\end{align}
The range of numerical values of the LO contributions are given in \eqref{eq:fiLO}, while those of the NLO ones are found to be:
\begin{equation}
    \begin{array}{ccccccccccc}
    2.23&\lesssim&f_{11}&\lesssim&\infty,&\qquad\qquad&3.61&\lesssim&f_{12}&\lesssim&\infty,\\
    0.61&\lesssim&f_{22}&\lesssim&\infty,&\qquad\qquad&-0.45&\lesssim&f_{23}&\lesssim&-0.30.
    \end{array}
\end{equation}
The functions $f_{11}$, $f_{12}$, and $f_{22}$ diverge in the squeezed limit, where $x$ or $y$ approaches zero. However, as discussed in Section~\ref{sec:numest}, only a finite range of momenta is observationally accessible. Consequently, the ratios $x$ and $y$ cannot be arbitrarily small and are bounded below by $k_\text{min}/k_\text{max}$, with $\ln\big(k_\text{min}/k_\text{max}\big)\simeq -7$. This leads to:
\begin{align}
    f_{11,\max}\simeq12.43, \qquad\quad   f_{12,\max}\simeq13.07 \qquad\quad\text{and}\quad\qquad   f_{22,\max}\simeq3.52.
\end{align}

We could also introduce a time-independent $\widehat{f}_{\text{NL}}$ parameter as
\begin{align}
    \widehat{f}=\frac{5}{6}\frac{\mathcal{B}\big(k_1,k_2,k_3\big)}{\mathcal{P}_{k_1}\mathcal{P}_{k_2}+\mathcal{P}_{k_1}\mathcal{P}_{k_3}+\mathcal{P}_{k_2}\mathcal{P}_{k_3}}.
\end{align}
Writing again
\begin{align}
    \widehat{f}_{\text{NL,LO}}&=\widehat{f}_1\varepsilon_1+\widehat{f}_2\varepsilon_2,\\
    \widehat{f}_{\text{NL,NLO}}&=\widehat{f}_{11}\varepsilon_1^2+\widehat{f}_{12}\varepsilon_1\varepsilon_2+\widehat{f}_{22}\varepsilon_2^2+\widehat{f}_{23}\varepsilon_2\varepsilon_3,
\end{align}
and evaluating them at the last horizon exit time $-\tau=1/k_{i,\max}+\mathcal{O}(\varepsilon)$, we find:
\begin{align}
    \widehat{f}_1&=f_1,\\
    \widehat{f}_2&=f_2,\\
    \begin{split}
    \widehat{f}_{11}&=\frac{5}{12}\frac{1}{1+x^3+y^3}\Bigg\{\mathscr{F}_0\\
    &~~+2\frac{x}{1+x^3+y^3}\left(\frac{1+y}{1-x+y}\big(1-y+x^2+y^2\big)\mathscr{F}_0+2y\big(1+y^3\big)\right)\ln x\\
    &~~+2\frac{y}{1+x^3+y^3}\left(\frac{1+x}{1+x-y}\big(1-x+x^2+y^2\big)\mathscr{F}_0+2x\big(1+x^3\big)\right)\ln y\\
    &~~+16\frac{x^2+y^2+x^2y^2}{1+x+y}\left(\frac{x+y}{x+y-1}+\frac{1+x}{1+x-y}+\frac{1+y}{1-x+y}-2\right)\ln\left(\frac{1+x+y}{2}\right)\Bigg\},
    \end{split}\\
    \begin{split}
    \widehat{f}_{12}&=\frac{5}{12}\frac{1}{1+x^3+y^3}\Bigg\{(1+x+y)(x+y+xy)-2\big(1+x^3+y^3\big)+\mathscr{F}_8-\mathscr{F}_2\big(\gamma_E+\ln 2\big) \\
    &\quad+\frac{x}{1+x^3+y^3}\left(\frac{1+y}{1-x+y}\big(1-y+x^2+y^2\big)\mathscr{F}_0+2y\big(1+y^3\big)\right)\ln x\\
    &\quad+\frac{y}{1+x^3+y^3}\left(\frac{1+x}{1+x-y}\big(1-x+x^2+y^2\big)\mathscr{F}_0+2x\big(1+x^3\big)\right)\ln y \\
    &\quad+\left(\mathscr{F}_2-\left(\frac{x+y}{x+y-1}+\frac{1+x}{1+x-y}+\frac{1+y}{1-x+y}\right)\mathscr{F}_0\right)\ln\left(\frac{1+x+y}{2}\right)\Bigg\},
    \end{split}\\
    \widehat{f}_{22}&=0,\\
    \widehat{f}_{23}&={f}_{23}.
\end{align}
We thus identify two key differences compared to $f_{\text{NL}}$: $\widehat{f}_{22}$ vanishes, and the divergences are absent, yielding
\begin{equation}
    \begin{array}{ccccccccccccc}
    0.82&\lesssim&\widehat{f}_{11}&\lesssim&2.35&\qquad\quad&\text{and}&\qquad\quad&0.54&\lesssim&\widehat{f}_{12}&\lesssim&2.42.
    \end{array}
\end{equation}

\begin{figure}[h!]
    \begin{subfigure}[t]{0.5\textwidth}
    \centering\includegraphics[width=0.65\linewidth]{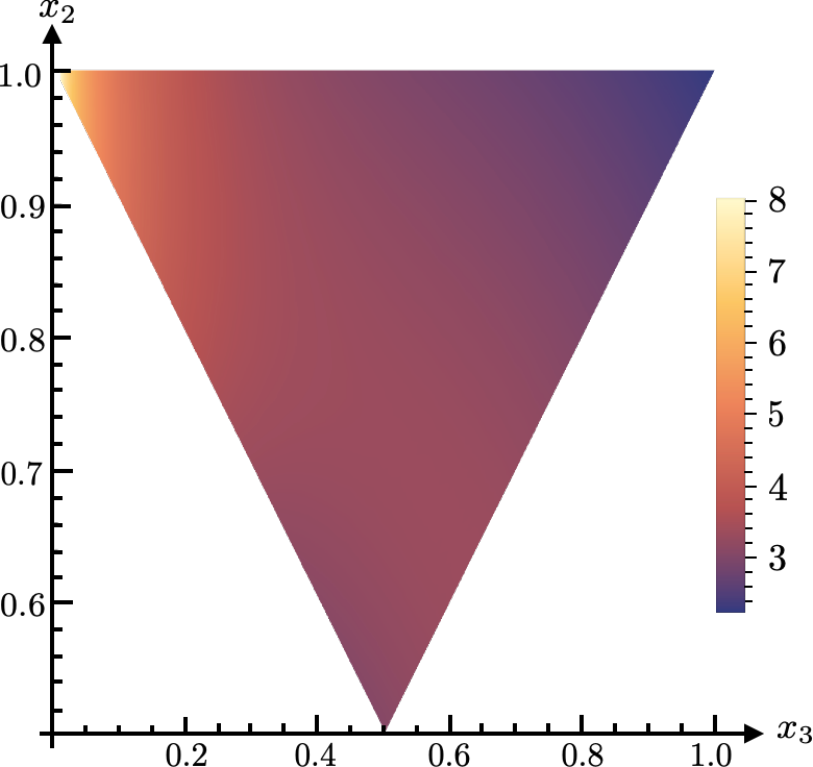}
    \subcaption{$f_{11}(1,x_2,x_3)$}
    \label{fig:f11}
    \end{subfigure}
    \begin{subfigure}[t]{0.5\textwidth}
    \centering\includegraphics[width=0.65\linewidth]{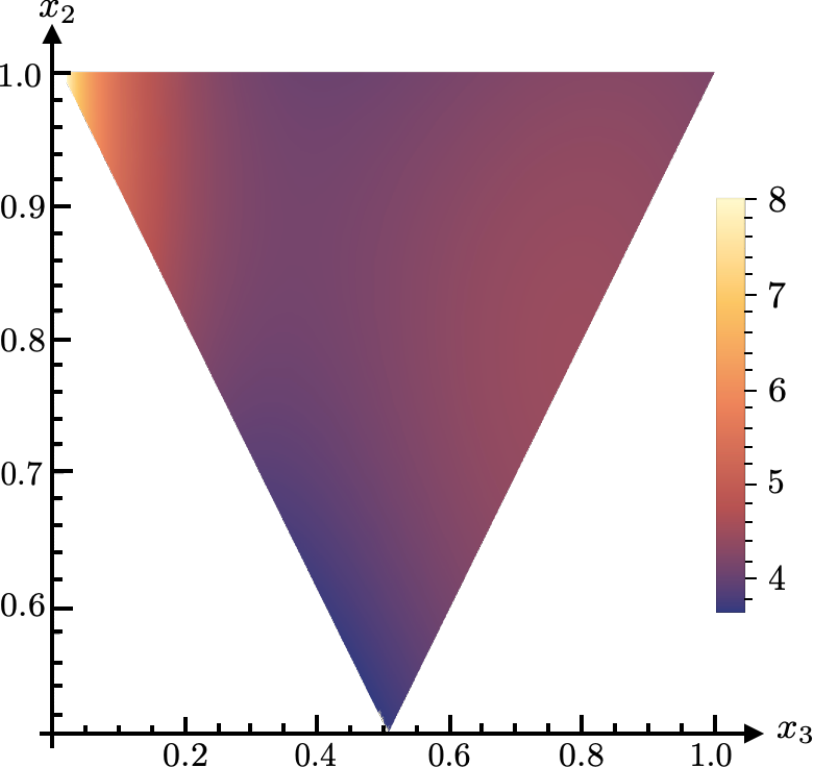}
    \subcaption{$f_{12}(1,x_2,x_3)$}
    \label{fig:f12}
    \end{subfigure}
    \begin{subfigure}[t]{0.5\textwidth}
    \centering\includegraphics[width=0.65\linewidth]{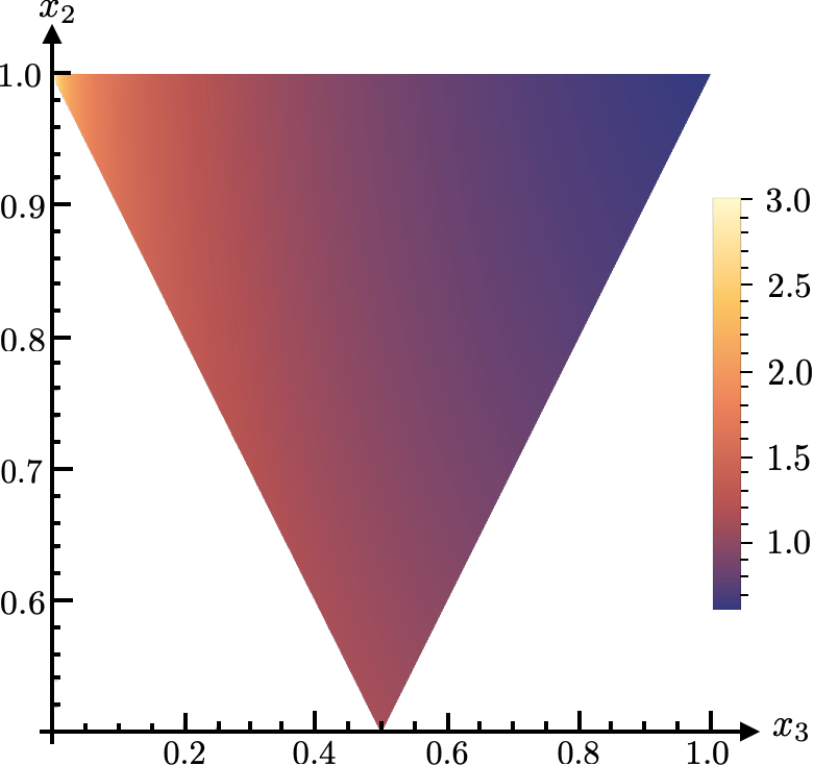}
    \subcaption{$f_{22}(1,x_2,x_3)$}
    \label{fig:f22}
    \end{subfigure}
    \begin{subfigure}[t]{0.5\textwidth}
    \centering\includegraphics[width=0.65\linewidth]{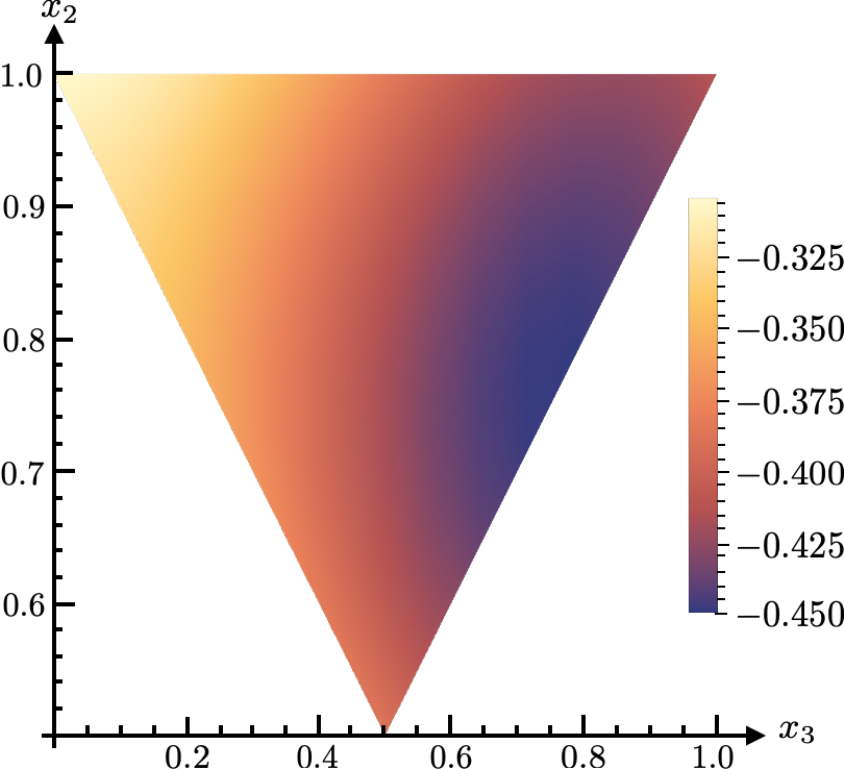}
    \subcaption{$f_{23}(1,x_2,x_3)$}
    \label{fig:f23}
    \end{subfigure}
    \begin{subfigure}[t]{0.5\textwidth}
    \centering\includegraphics[width=0.65\linewidth]{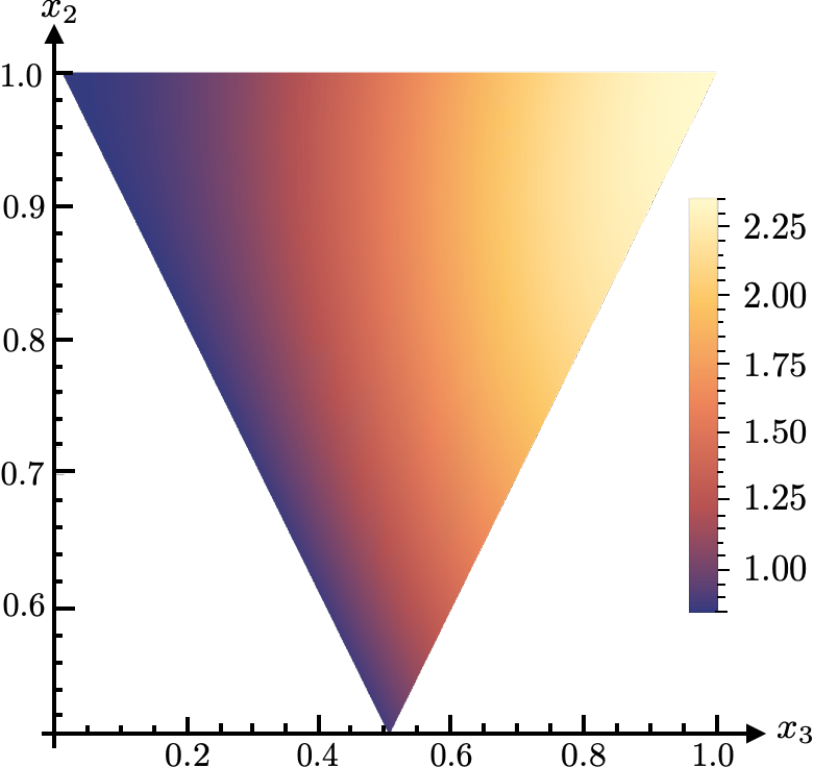}
    \subcaption{$\widehat{f}_{11}(1,x_2,x_3)$}
    \label{fig:fhat11}
    \end{subfigure}
    \begin{subfigure}[t]{0.5\textwidth}
    \centering\includegraphics[width=0.65\linewidth]{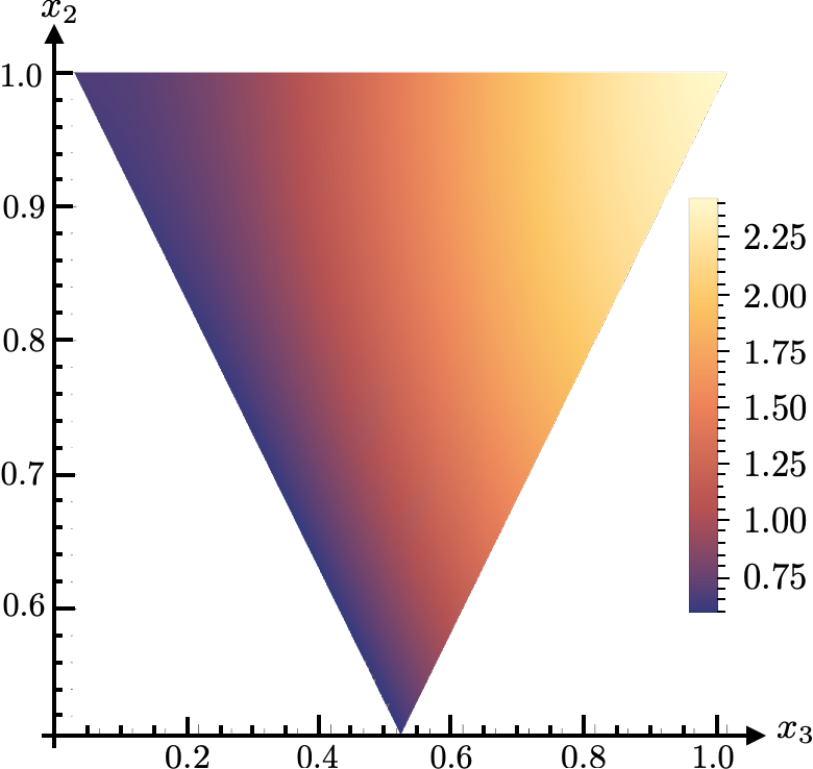}
    \subcaption{$\widehat{f}_{12}(1,x_2,x_3)$}
    \label{fig:fhat12}
    \end{subfigure}
  \caption{Values of the functions $f_i(k_1,k_2,k_3)$ and $\widehat{f}_i(k_1,k_2,k_3)$ plotted for the rescaled momenta $x_1=1,\ x_2=k_2/k_1$ and $x_3=k_3/k_1$. The momenta are chosen to be ordered as $x_3\leq x_2\leq x_1$ and obey the triangle inequality $x_2+x_3\geq 1$.}
  \label{fig:fi}
\end{figure}

All independent shape functions are plotted in Figure~\ref{fig:fi}, whereas $f_1 = \widehat{f}_1 = \tilde{K}_1$ was already shown in Figure~\ref{fig:K1}.
\providecommand{\href}[2]{#2}\begingroup\raggedright\endgroup



\end{document}